\documentclass{aastex}
\usepackage{spr-astr-addons}
\usepackage{url}
\usepackage{float}
\usepackage{subfloat}
\usepackage{adjustbox}
\usepackage[utopia]{mathdesign}
\usepackage{ragged2e}
\usepackage{soul}

\RequirePackage{color}

\newcommand{\emaila}{authors@email.com}

\begin{document}

\title{Low resolution spectroscopy of selected Algol systems}

\shorttitle{Low resolution spectroscopy of selected Algol systems}
\shortauthors{Shanti et al.}

\author{D. Shanti Priya\altaffilmark{1*}} \and \author{J. Rukmini\altaffilmark{1}}
\and \author{M. Parthasarathy\altaffilmark{2}}
\and \author{D.K. Sahu\altaffilmark{2}}
\and \author{Vijay Mohan\altaffilmark{3}}
\and \author{B.C. Bhatt\altaffilmark{2}}
\and \author{Vineet S. Thomas\altaffilmark{4}}

\altaffiltext{1}{Department of Astronomy, Osmania University, Hyderabad, Telangana 500007, India. \email{\emaila}{*Corresponding author email: astroshanti@gmail.com}}
\altaffiltext{2}{Indian Institute of Astrophysics, Koramangala, Bengaluru, Karnataka 560034, India.}
\altaffiltext{3}{Inter-University Centre for Astronomy and Astrophysics, Pune University Campus, Pune 411007, India.}
\altaffiltext{4}{The University of Akron, Akron, Ohio 44304, USA.}

\begin{abstract}
The analysis of  spectroscopic data for 30 Algol-type binaries is presented.  All these systems are short period Algols having primaries with spectral types B and A. Dominant spectral lines were identified for the spectra collected and their equivalent widths were calculated. All the spectra were examined to understand presence of mass transfer, a disk or circumstellar matter and chromospheric emission. We also present first spectroscopic and period study for few Algols and conclude that high resolution spectra within and outside the primary minimum are needed for better understanding of these Algol type close binaries. 
\end{abstract}

\section{Introduction}

Algol systems are semi-detached close interacting binaries, which consist of a hot and more massive B-A type main-sequence primary star and a cool, giant or subgiant secondary star of F-K spectral type that transfers mass and angular momentum to the primary via Roche lobe overflow (RLOF) (\cite{kopal1955, giuricin1983}). They are an important source to study various phenomena, such as mass transfer and accretion, angular momentum, magnetic activity in the late-type companion and orbit evolution. Studies of these binaries resulting in reliable fundamental parameters has significantly increased due to the availability of high-resolution spectroscopic data.
Detailed studies of Algols, both detached and semidetached are important in understanding and developing theoretical models representing the formation and evolution of binary systems as well as a single stars. 
The chemical composition of stellar photospheres in mass-transferring binary systems like Algols is an important diagnostic to study the nucleosynthesis processes which occur deep inside the component stars. This study also provides the information on the components history. The evolutionary process in these interacting binaries cause many observable effects (changes in orbital period, erratic light variability, distorted radial velocity curves, etc.), with the most important being mass transfer. Up to 80\% of the more massive star's initial mass can be lost, exposing deep inner layers of the star that have been changed by thermonuclear fusion during its main sequence evolution. There are also changes happening in some of the material transferred to the companion. The abundance pattern in Algol-type binaries could reveal their past, and would be strong evidence for postulated mass transfer between the components (\cite{sarna1996}).
The current work highlights the results of low-resolution spectroscopy carried out over several nights to investigate the changes in emission and absorption line profiles of the star systems. The orbital period variations were investigated for selected Algol-type eclipsing binaries in this study. The O-C diagrams of 13 of these Algols show some measurable period variation, whereas for the other 17 Algols the changes are undefined due to insufficient number or non-availability of data. 

\section{Observations, Data Reduction and Analysis}

Low resolution spectral observations were carried out using the 2-m Himalayan Chandra Telescope (HCT) of Indian Astronomical Observatory equipped with a Himalaya Faint Object Spectrograph Camera (HFOSC). The telescope has a 2K$\times$4K CCD and a central strip of 500 x 3500 pixels was used.
A total of 31 Algols (as listed in Table \ref{T1}) were observed on various nights in the years 2013 and 2014 with exposures varying between 380s-1200s as listed in Table \ref{T1}. The observed data covers wavelength region of 3500-9100 \AA. This is achieved using two grism Gr7 and Gr8. For Gr7 the wavelength coverage is 3500-7800 \AA \hspace{0.5mm} and for Gr8 it is 5200-9200 \AA. This gives a dispersion of 1.5 \AA/pixel and a resolution of $\sim$11 \AA. The FeAr arc lamp was used for wavelength calibration of spectrum taken with Gr7 and FeNe arc lamp spectrum was used for wavelength calibration of spectrum taken with Gr8. The IRAF\footnotemark  package and different sub-packages like ONEDSPEC were used for reducing the spectra after bias and flat-field corrections. Later the spectra were normalized for further studies. 

\footnotetext{IRAF is distributed by the National Optical Astronomy Observatory, which is operated by the Association of Universities for Research in Astronomy (AURA) under cooperative agreement with the National Science Foundation.}

\begin{table*}
\caption{Observed Algols: their periods and UT date of observations}
\begin{center}
\begin{adjustbox}{width=1\textwidth}
\begin{tabular}{@{}cccccccccc@{}}
\hline
S.No. & Algol Name & Period &  & Date of Observation & Epoch & Phase & Exposure time(secs) \\
\hline
\\
1.	&	 V0616  Aql &	1.6906&		&	Apr 23, 2014	& 2456770.486 & 0.973  & 1380 &\\
2.	&	 V0769 Aql    &	4.5623	&		&	Nov 20, 2013 &  2452500.270  & 0.977 & 1200 &\\
3.	&	 V1340 Aql  &	1.5969&		&	Mar 18, 2013	&  2452501.200   & 0.349 & 1380 &\\
4.	&	  HN Cas      &	             2.6594&		&	Nov 12, 2013 &  2456619.306   & 0.147	&1200 &\\
5.	&	 V0380 Cas    &	1.3577	&		&	Nov 20, 2013 & 255479.656  & 0.075 &	600 &\\
6.	&	 AV Cep       &	2.9584	&		&	Oct 13, 2013	&   255643.5478  &  0.020 & 720 &\\
7.	&	 XY Cet       &	           2.7807     &		&	Nov 20, 2013 & 2456617.251  & 0.018 & 360 &\\
8.	&	 RS CMi       &	5.0278&		&	Feb 20, 2013 &  2448659.708  &  0.412  & 1200 &\\
	&		          &		  &		&	 Feb 20, 2014 & 2448659.708   &  0.004   & 1200 &\\
9.	&	RS CVn        &	4.7979&		&	Apr 23, 2014	& 2448230.537 & 0.101 & 1320 &\\
10.	&	RR Dra         &	2.8313&		&	Apr 23, 2014 &  2454200.534    &  0.021	& 1320 & \\  
11.	&	TZ Eri	      &	2.6061&		&	Nov 20, 2013 & 255478.764    & 0.972 &	720 &\\
12.	&	AN Gem       &	2.0325&		&	Nov 12, 2013  &  2448394.577 & 0.911 & 1200 &\\
13.	&	SX Gem        &	1.3669&		&	Feb 20, 2013 & 2448656.914  & 0.050 & 1200 &\\
	&		          &		  &		&	Feb 20, 2014 &   2448656.914     & 0.065 & 480 &\\
14.	&	TW Lac         &	3.0374 &		&	Dec 21, 2013  &  2456647.435  & 0.293 &	1200 &\\ 
15.	 &	FG Lyr	           &	2.8718  &	&	Apr 23, 2014	&   2456770.916    & 0.028 &  900 &\\
16.	&	BZ Mon         &	3.4518 &		&	Feb 20, 2013 & 2448660.263 & 0.070 & 900 &	\\
	&		          &		  &		&	Feb 20, 2014	&  2448660.263  &  0.807 & 900 &\\
17.	&	CH Mon       &	6.9223&		&	Nov 12, 2013	&  2456619.442  & 0.014 & 1200 &\\
18.	&	FW Mon       &	3.8736 &		&	Mar 18, 2013  & 2448630.672 & 0.003 & 1200 & \\
19.	&	HP Mon       &     	1.4547   &		&	Feb 20, 2013 & 2456344.174 & 0.342 & 1200 &\\
	&		          &		               &		&	Feb 20, 2014 & 2456344.174 & 0.261 & 1200 & \\
20.	 &	RV Oph      &	1.3577	&		&	Nov 20, 2013	& 2448634.730 & 0.324 & 1200 &\\
21.	&	FH Ori &	2.1512&		&	Oct 14, 2013	&  255581.721  & 0.249  & 900 &\\
22.	&	Z Ori   &	5.2033	&		&	Mar 21, 2014	&  254516.431     &  0.991 & 900 & \\
23.	&	V0640 Ori &	2.0207&		&	Oct 13, 2013	&   253008.704    &     0.013 & 1200 &\\
24.	&	CK Per	 &	2.3728&		&	Oct 13, 2013 & 2456617.245  &  0.995	& 1800 & \\
	&		&		&		&	Nov 20, 2013 & 2456617.245    &  0.015	& 1200 &\\
25.	&	Z Per	&	3.0563	&		&	Feb 20, 2013 & 2456343.109   & 0.015 & 720 &\\
	&		&		&		&	Oct 13, 2013 &  2456343.109  &  0.016 &	900 &\\
	&		&		&		&	Feb 20, 2014 & 2456343.109  & 0.443 &	600 &\\
26.	&	XY Pup  &	14.7783 &		&	Feb 20, 2013 &   2456709.207   &0.196 & 720 & \\
27.	&	AC Tau   & 2.0434 &		&	Dec 21, 2013 &  2456648.145  &  0.938 & 1200 &\\
28.	&	RW Tau &	2.7688 &		&	Oct 16, 2013 &  2456582.117  & 0.115 & 480 &\\
29.	&	AF UMa &	5.2576 &		&	Mar 18, 2013 & 2448631.425 & 0.942 &	1200 &\\
	&		    &		   &		&	Mar 21, 2014 &  2448631.425  &  0.940 & 1200 &\\
30.	&	VV Vul &	3.4114	&		&	Nov 20, 2013 &  2454410.390   &  0.868 & 720 & \\

\hline
\end{tabular}
\end{adjustbox}
\end{center}
\label{T1}
\end{table*}

\section{Period Analysis and Spectroscopic study}

In order to carry out period analysis for the Algols in this study, times of minimum available in the literature were collected and the new phases listed in Table \ref{T1} were calculated. Fig 1 show the (O-C) diagrams that are constructed from all times of primary minima available. Only 13 Algols have data which were used to investigate period variations. The O-C diagrams show significant period variations which can be fit to polynomial equation. The period changes of some of the systems are tilted sinusoidal variations superimposed on either upward or downward parabolic forms suggesting presence of tertiary component.

Algols are close interacting binaries in which mass transfer occurs as a result of RLOF from the evolved late-type secondary star. The stream, impact point and other accretion structures arising from mass transfer often give rise to emission and absorption lines. In this study we also present the observed spectral profiles of 30 Algol binaries. The dominant spectral lines identified in spectral range 3000-7000\AA \hspace{0.5mm} and their profiles are presented and discussed. The equivalent widths of identified spectral lines with absorption feature for all the Algols and emission feature for XY Pup are being reported for the first time. The equivalent widths obtained for the dominant spectral lines are given in Table \ref{T2}. 
To derive the spectral class and thus the effective temperatures of the components of Algols in this study, the observed spectra were first compared with standard spectra catalogued in the library of Stellar Spectra (\cite{jacoby1984}). Based on chi square minima and visual inspections, the best fit spectral model was selected. It was also observed that the dominant spectral lines were normal and free of emission (except in XY Pup) while atomic absorption features matched very well with the corresponding atomic absorption features of the standard spectra.  

\subsection{V0616 Aql}


V0616 Aql (= GSC 04147-01115 = TYC 4147-1115-1, V = 14.55) is one of the least studied Algol binary which was first catalogued by \cite{hoffmeister1943} and the linear elements were derived by \cite{zedja2002, kreiner2004}. This is one of least observed Algol with only 6 times of minima(ToM) in the literature. The O-C residuals were obtained from the recent epoch available in the literature and the same is plotted in the Figure \ref{fig2}. The quadratic fit for the O-C variation resulted in dp/dt = -0.001481 days/yr.  The significant change observed from the best fit curve is of that of a decreasing trend extended over a period of 8 years with 200 orbital cycles. The ephemeris obtained in the current study is HJD (Min I) = 2456770.486 + 1$^{d}$.690532$\times$E. However, better conclusions can be laid on only with further observations.

\begin{figure}[H]
\centering
\includegraphics[scale=0.11, angle=0 ]{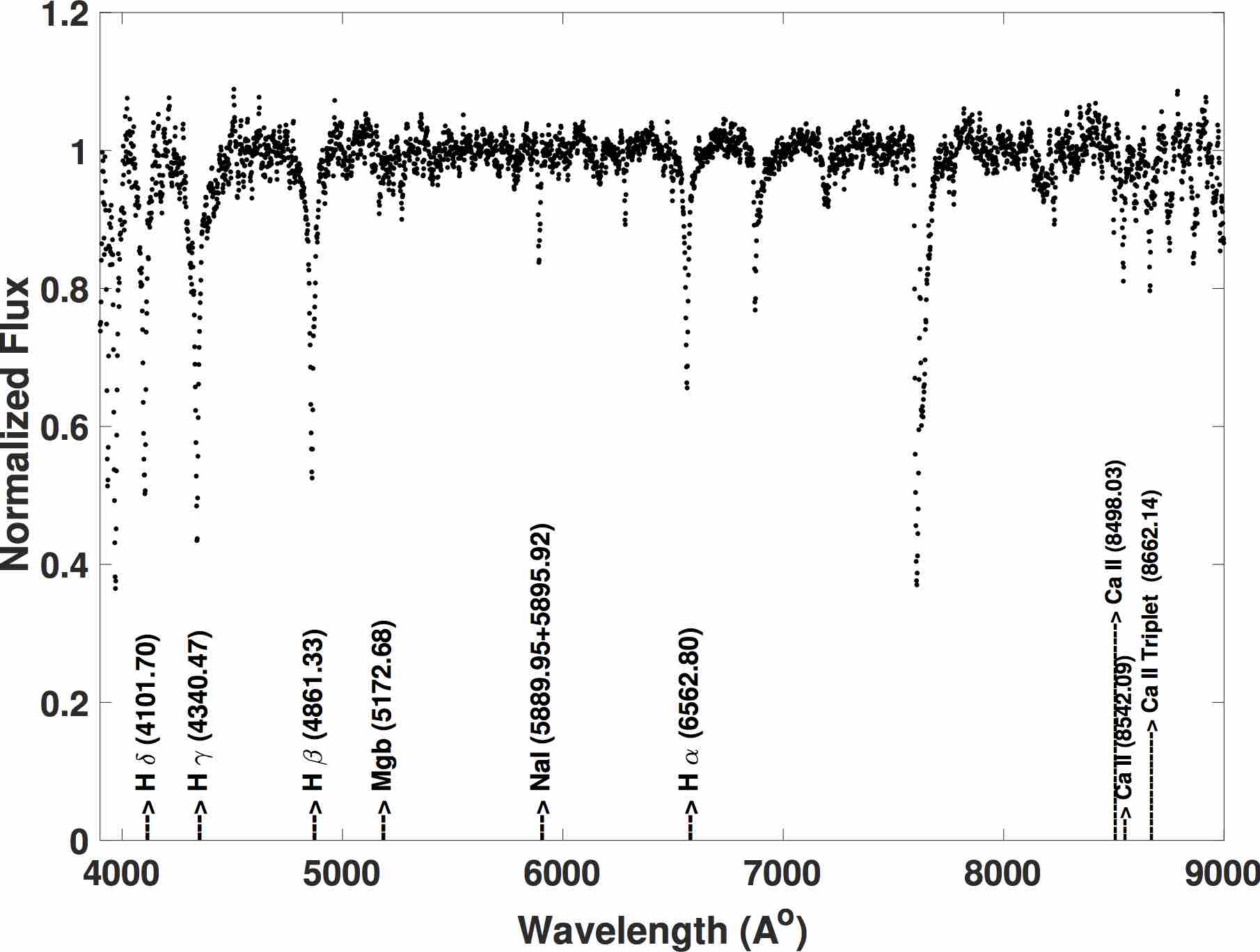}
\caption{Spectrum of V0616 Aql.}
\label{fig1}
\end{figure}

\begin{table*}
\tabletypesize{\scriptsize}
\caption{The equivalent widths of the spectral lines obtained for the Algols that were observed with HCT}
\begin{adjustbox}{width=1\textwidth}
\small
\begin{tabular}{@{}ccccccccccc@{}}
\hline
S.No. & Algol Name & Phase & H$\delta$ & H$\gamma$ & H$\beta$ & H$\alpha$ & NaI & Ca II & Ca II & Ca triplet  \\
 &  &   &(4101.70\AA)& (4340.47\AA)&(4861.33\AA) &(6562.80\AA) & (5889.95\AA + & (8498.03\AA) & (8542.09\AA) & (8662.14\AA) \\
  &  &   & & & & & 5895.92\AA)& & &  \\
\hline
\\
1.	 &	 V0616 Aql & 0.878 & 10.970 & 7.162 & 3.988 & 3.761 & 1.916 &  0.261 & 1.018 & 1.339 \\
2.	 &	 V0769 Aql & 0.977 & 2.541 & 3.876 & 4.664 & 3.411 & 0.968 &  1.095 & 1.885 & 2.591 \\
3.	 &	 V1340 Aql & 0.349 & 0.284 & 5.909 & 3.124 & 3.067 & 0.428 &  0.167 & - & 0.136 \\
4.	 &	  HN Cas	& 0.147 & 12.110 & 9.818 & 8.806 & 2.638 & 1.935 &  0.618 & 0.263 & 0.843 \\
5.	 &	 V0380 Cas & 0.706 & 11.820 & 11.730 & 13.460 & 7.816 & 1.108 &  0.910 & 1.655 & 5.379 \\
6.	 &	 AV Cep	&  0.384  & 8.494 & 10.600 & 7.414 & 1.475 & 1.636 &  1.598 & 0.941 & 10.130 \\
7.	 &	 XY Cet	& 0.018 & 6.202 & 4.409 & 7.159 & 6.144 & 0.301 &  1.003 & 1.227 & 1.953 \\
8.	 &	 RS CMi	& 0.412 & 11.010 & 13.370 & 5.012 & 4.095 & 1.256 &  0.765 & 0.080 & 0.393 \\
9.	 &	 RS CVn	& 0.101 & 2.877 & 1.744 & 2.745 & 1.043 & 1.354 &  0.991 & 1.973 & 2.060 \\
10.	 &	RR Dra	& 0.021 & 11.060 & 10.500 & 12.360 & 0.974 & 1.398 &  1.486 & 2.241 & 1.807 \\
11.	&	TZ Eri	& 0.071 & 6.888 & 4.737 & 7.622 & 3.716 & 1.766 &  1.913 & 3.642 & 3.362 \\
12.	 &	AN Gem	& 0.911 & 4.346 & 19.600 & 7.176 & 5.424 & 0.504 &  0.216 & 1.208 & 9.990 \\
13.	&	SX Gem	& 0.050 & 9.676 & 10.140 & 8.992 & 5.754 & 1.345 &  0.591 & 1.094 & 1.201 \\ 
	&		        & 0.065 & 10.200 & 10.410 & 8.790 & 6.112 & 1.322 &  0.501 & 1.146 & 2.525 \\
14.	&	TW Lac	& 0.293 & 8.559 & 9.155 & 7.726 & 5.362 & 1.322 &  0.501 & 1.146 & 2.525 \\
15.	&	FG Lyr	& 0.028 & 2.437 & 2.283 & 2.616 & 1.887 & 2.772 &  0.946 & 2.192 & 1.071 \\
16.	&	BZ Mon	& 0.070 & 0.663 & 4.252 & 2.315 & 1.263 &  0.117 &  - & - & - 	\\
&				& 0.806 & 13.010 & 9.118 & 6.619 & 0.111 & 0.071 &  - & - & - 	\\
17.	&	CH Mon	& 0.014 & 8.302 & 8.190 & 5.493 & 1.029 & 2.051 &  0.731 & 2.008 & 0.328\\
18.	&	FW Mon	& 0.022 & 16.680 & 4.690 & 4.228 & 2.196 & 1.296 &  0.247 & 0.346 & 3.520 \\
19.	&	HP Mon	& 0.342 & 1.471 & 14.820 & 3.136 & 5.877 & 1.628 &  0.086 & 1.076 & 0.659 \\
&		            	& 0.260 & 6.594 & 7.184 & 12.250 & 6.640 & 0.992 &  0.298 & 1.123 & 0.694 \\ 
20.	&	RV Oph	& 0.324 & 11.620 & 11.130 & 9.907 & 6.107 & 1.354 &  0.991 & 1.973 & 2.060 \\
21.	&	FH Ori	& 0.249 & 11.350 & 5.984 & 3.323 & 7.141 & 1.405 &  1.077 & 9.734 & 3.037 \\
22.	&	Z Ori	 	& 0.991 & 4.714 & 5.648 & 4.773 & 3.511 &  0.437  & 0.479 & 1.723 & 0.646 \\
23.	 &	V0640 Ori	& 0.013 & 2.546 & 4.816 & 5.433 & 4.061 & 1.445 &  1.195 & 1.577 & 1.719 \\
24.	&	CK Per	& 0.995 & 1.100 & 2.762 & 3.517 & 2.105 & 0.556 &  0.304 & 0.429 & 0.323 \\
25.	&	Z Per	& 0.015 & 11.650 & 11.420 & 10.950 & 6.736 & 0.515 &  0.925 & 1.879 & 3.797 \\	
&		        		& 0.016 & 11.300 & 11.840 & 11.580 & 7.032 & 1.190 & 0.355  & 2.105 & 1.953 \\	
& 		        		& 0.443 & 11.640 & 11.970 & 12.650 & 7.358 & 0.979 & 0.953  & 2.260 & 2.295 \\
26.	&	XY Pup\footnotemark	& 0.195 & -4.605 & 0.744 & -1.891 & -3,585 & 3.872 &  1.316 & 2.208 & 1.655 \\
27.	&	AC Tau   	& 0.278& 6.723 & 7.445 & 5.691 & 4.464 & 0.323 &  2.920 & 1.360 & 0.995 \\	
                                 	&  &0.938 & 3.571 & 5.073 & 5.391 & 4.285 & 0.396 &  0.449 & 1.652 & 3.397 \\
28.	&	RW Tau	& 0.116 & 9.082 & 9.684 & 7.869 & 5.805 & 1.333 &  0.311 & 1.068 & 1.187 \\
29.	&	AF UMa	& 0.942 & 9.127 & 12.370 & 12.470 & 2.092 & 0.101 &  0.577 & 0.209 & 6.369 \\
	&		        & 0.940 & 13.050 & 15.850 & 14.300 & 8.766 & 0.358 &  6.112 & 0.760 & 9.880 \\
30.	&	VV Vul	& 0.869 & 11.070 & 10.560 & 9.785 & 6.715 & 1.644 &  0.084 & 2.097 & 2.417 \\
\hline

\end{tabular}
\end{adjustbox}
\label{T2}
\end{table*}
\footnotetext{Negative sign indicates emission line strength.} 

\begin{figure}[H]
\centering
\includegraphics[scale=0.11]{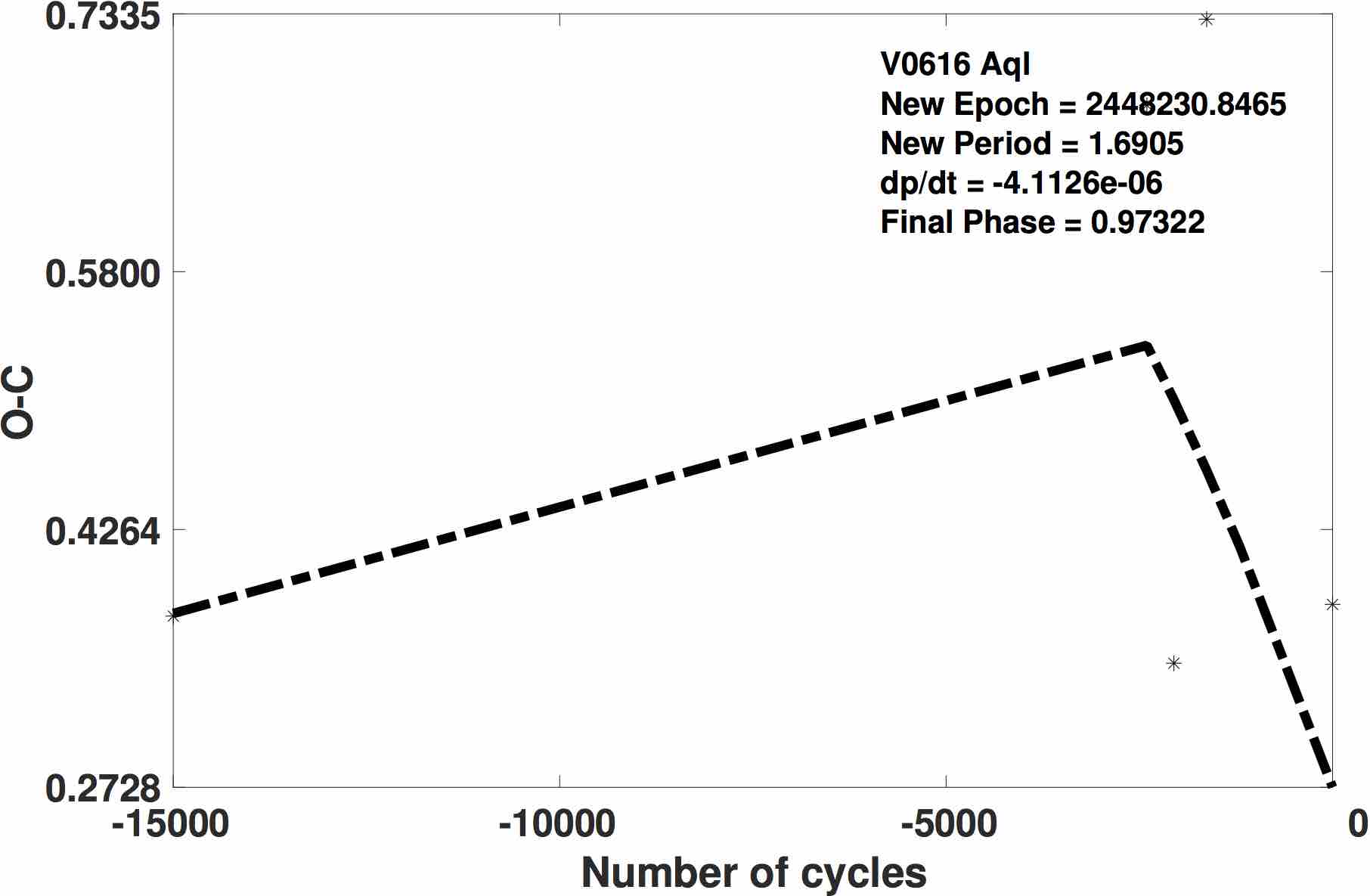}
\caption{O-C diagram of V0616 Aql.}
\label{fig2}
\end{figure}

One spectra for V0616 Aql was obtained on Apr 23, 2014 and the phase of observation calculated from the derived epoch is 0.9732. Spectral lines and their dominant profiles obtained can be seen in Figure \ref{fig1}.  All the Balmer lines in the selected wavelength range show good absorption profiles with equivalent widths as shown in Table \ref{T2}. It is observed that H$\alpha$ and H$\beta$  lines show a greater fill in the absorption profiles than other Balmer lines.

\subsection{V0769 Aql}

\begin{figure}[H]
\centering
\includegraphics[scale=0.11, angle=0 ]{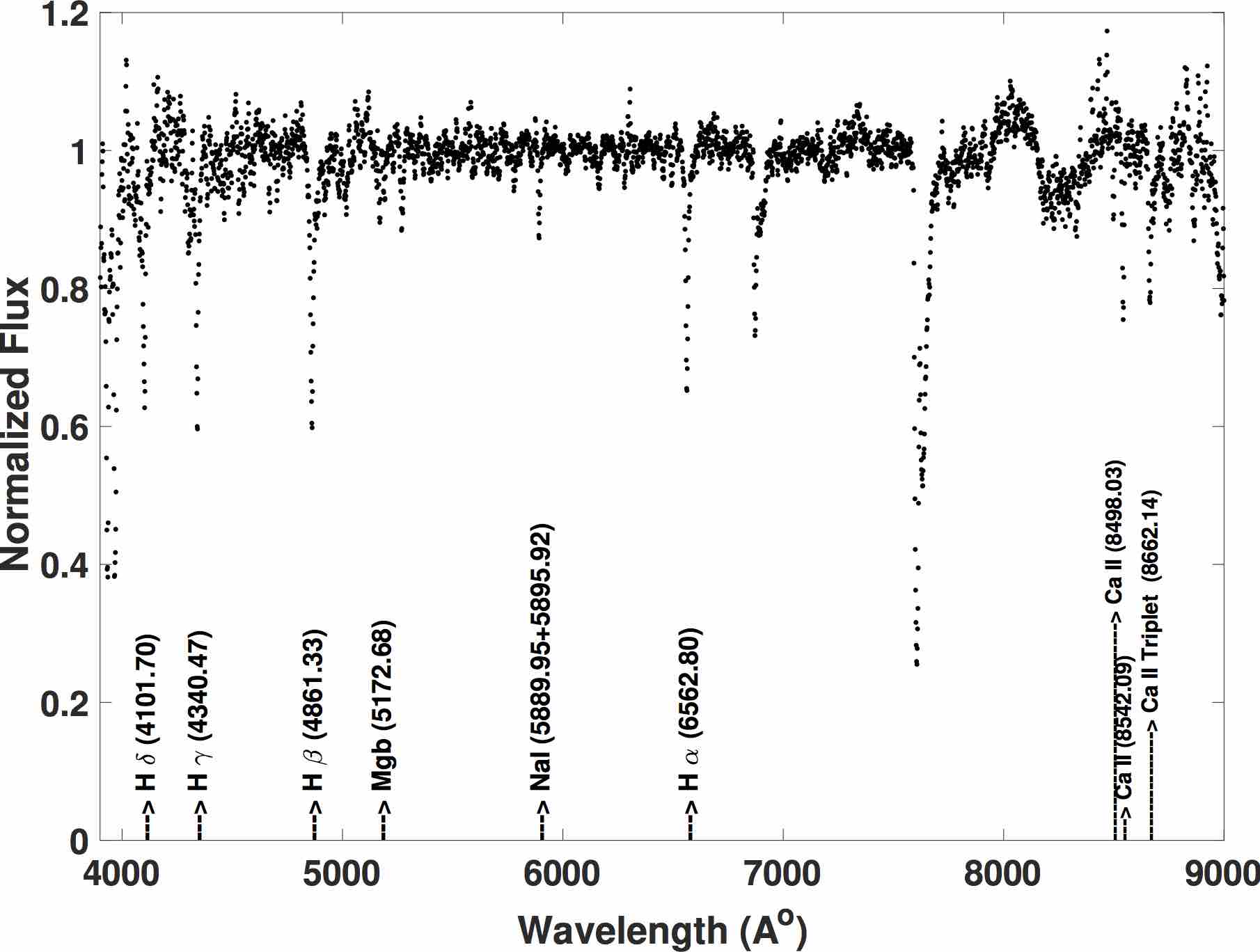}
\caption{Spectrum of V0769 Aql.}
\label{fig3}
\end{figure}

V0769 Aql (= GSC 05164-01535, V=15.40) was catalogued as an Algol type close binary by \cite{budding2004} and \cite{malkov2006}. This is least studied Algol binary with period derived as 4$^d$.5623 \cite{kukarkin1971}. The spectral type derived to be as G7V using the 2MASS magnitudes by \cite{cutri2003}. Due to lack of latest data, the available epoch in the literature was taken to calculate the phase. No period study was done due to unavailability of data. One spectrum for the variable was obtained on Nov 20, 2013 at phase 0.9773.  The dominant Balmer line profiles in the spectrum are presented for the first time in this paper (Figure \ref{fig3}) and the equivalent widths obtained are given in Table \ref{T2}. Due to the observed phase lying close to primary minima, the obtained spectra can be representing the cooler evolved secondary and its environment. However, the best fit spectral model couldn't be obtained using Jacoby spectral library.

\subsection{V1340 Aql}
\begin{figure}[H]
\centering
\includegraphics[scale=0.11, angle=0 ]{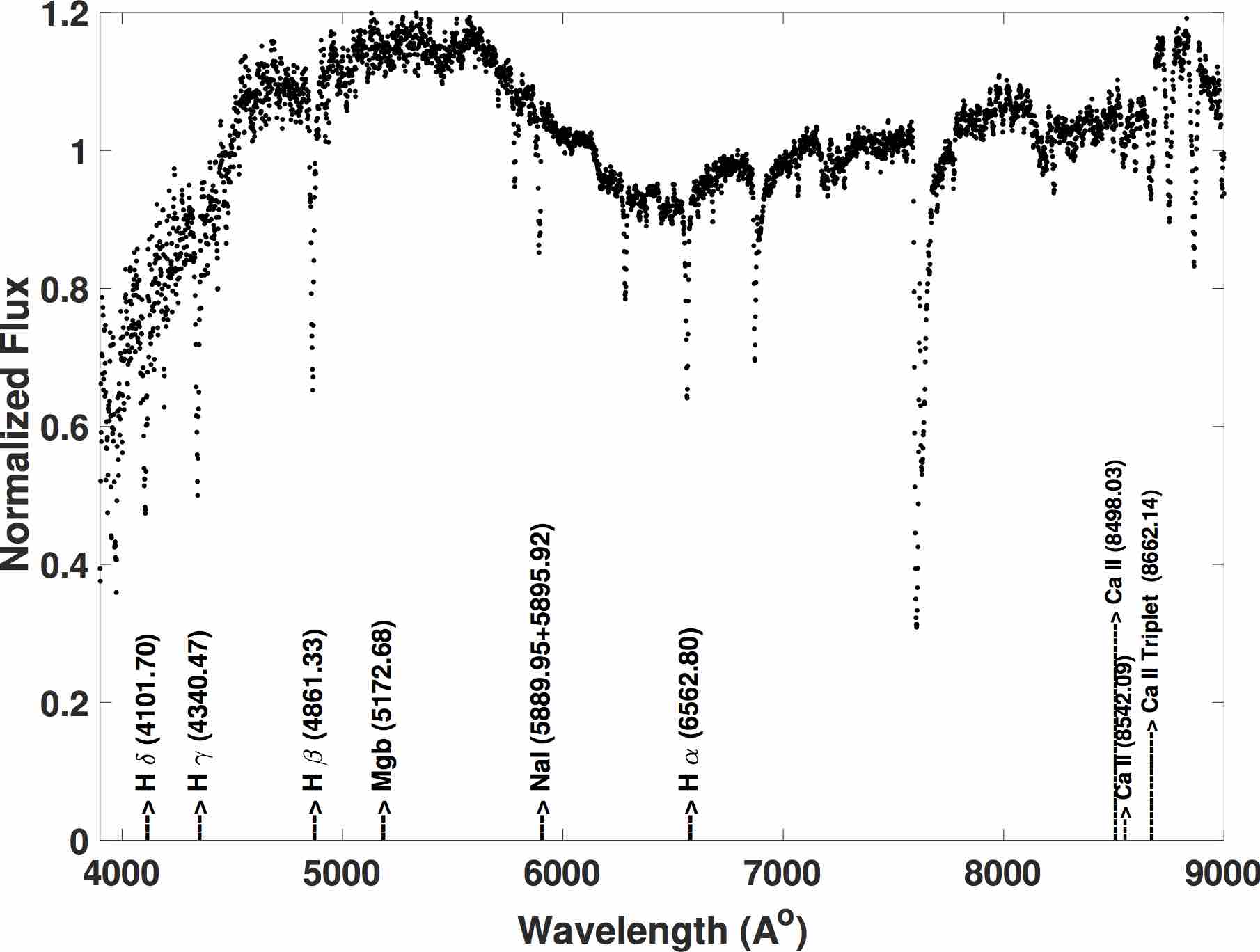}
\caption{Spectrum of V1340 Aql.}
\label{fig4}
\end{figure}

V1340 Aql(=2MASS J18441601-0330147, B=14.3) is also one of the least studied Algol with a period of 1$^d$.596940 (\cite{avvakumova2013}). It was identified as a variable star by \cite{kurochkin1954, kholpov1981} and later was catalogued as Algol type by \cite{budding2004} and \cite{malkov2006}. No period study was done on this variable due to lack of data. One spectrum for V1340 Aql at phase 0.3485 as calculated from available epoch, was obtained on Mar 18, 2013. We report the dominant Balmer lines in the spectrum for the first time. The spectrum is shown in Figure \ref{fig4} and corresponding equivalent widths are given in Table \ref{T2}. All the spectral lines are in absorption. H$\delta$ shows a weaker absorption compared to other Balmer lines.

\subsection{HN Cas}
\begin{figure}[H]
\centering
\includegraphics[scale=0.11, angle=0 ]{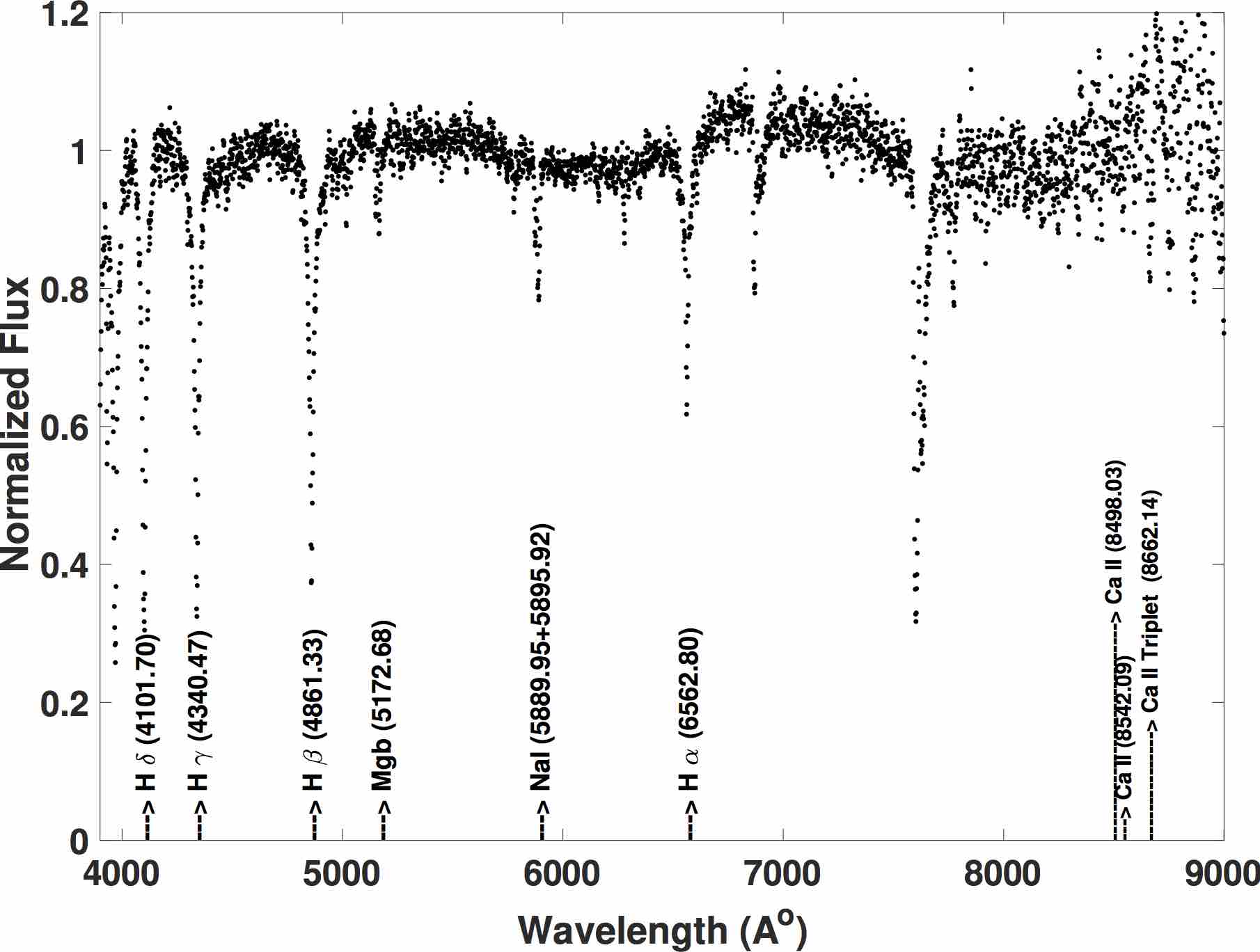}
\caption{Spectrum of HN Cas.}
\label{fig5}
\end{figure}

HN Cas (GSC 03672-01509, 2MASS J01005435 + 5554178, V=15.45) was classified as a variable star by \cite{hoffmeister1943, ahnert1947} and the period of the variable is 2$^d$.6594 (\cite{kukarkin1971}). Later it was catalogued as Algol by \cite{kinnunen2000},\cite{budding2004} and \cite{malkov2006}. Few times of minima (ToM) were given by \cite{brat2007} which is insufficient to carry out period study. Further no detailed study was carried out for the variable so far. One spectra for HN Cas at phase 0.1466 (as calculated from the available epoch) was obtained on Nov 12, 2013. The dominant Balmer spectral lines can be seen in Figure \ref{fig5} and equivalent widths are given in Table \ref{T2}. It is observed that H$\alpha$ absorption profile is relatively less prominent than the other Balmer lines which could be due to fill in effect. The spectral class determined in the current study using best fit spectral model (\cite{jacoby1984}) is A1 V.

\subsection{V0380 Cas}

\begin{figure}[H]
\centering
\includegraphics[scale=0.11, angle=0 ]{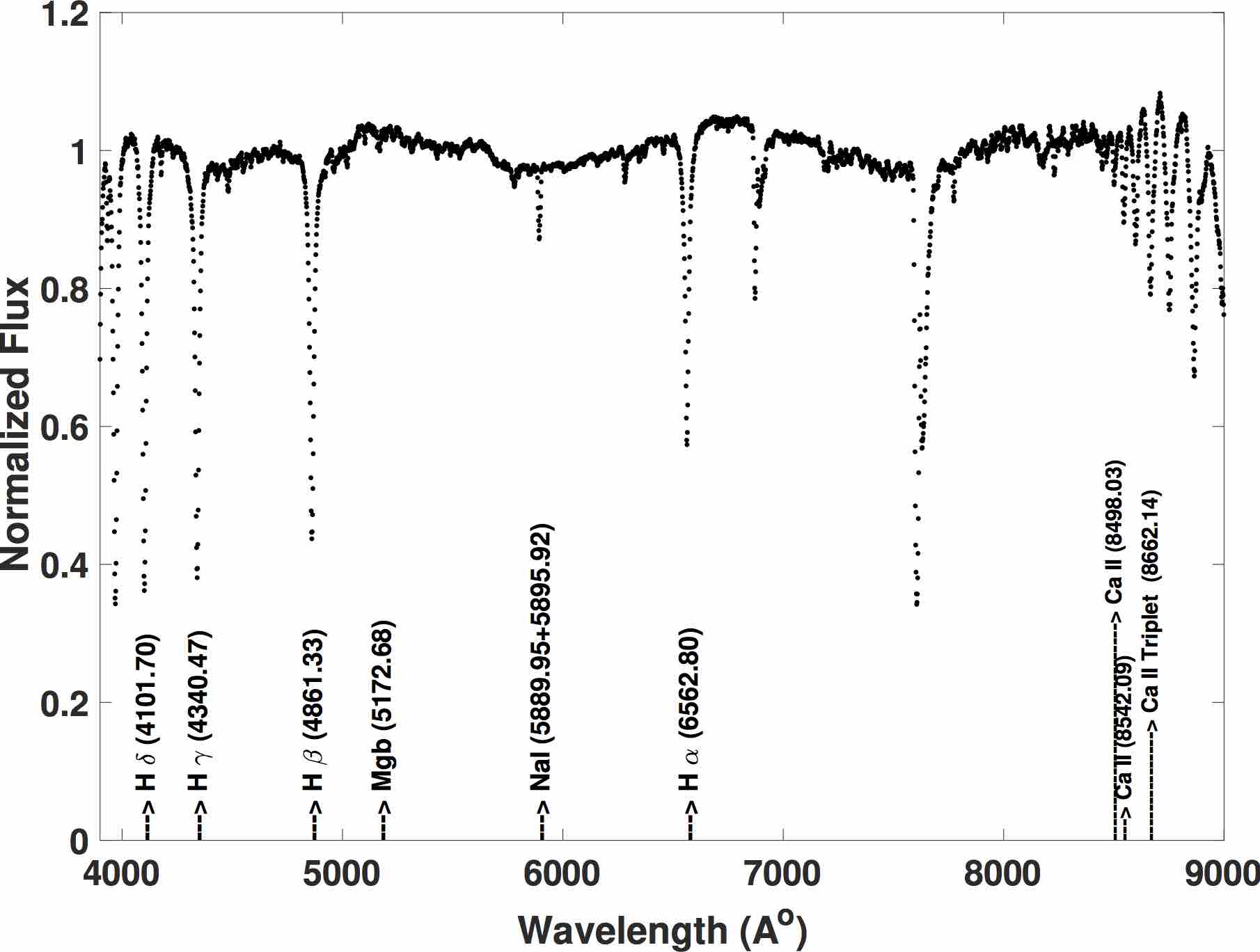}
\caption{Spectrum of V0380 Cas.}
\label{fig6}
\end{figure}

V0380 Cas (=GSC 04307-01121= TYC 4307-1121-1) was first discovered by Bauernfeind in 1899 and its eclipsing nature was studied by \cite{strohmeier1968}. \cite{meinunger1965} has classified its spectral type as A0 and orbital elements were first given by \cite{brancewicz1980}. This is a well studied Algol binary with current period given as 1$^d$.357270 by \cite{christopoulou2011} . They carried out first photometric study and found it to be a well-detached system with moderately evolved main-sequence components and a high resolution spectroscopy is recommended for this system to determine its evolutionary status. They also concluded from the period study that there is no indication of period change, eliminating the possibility of third companion. Hence no period period study was done in the current work. One spectrum for V0380 Cas was obtained on Nov 20, 2013 at phase 0.0749. The dominant Balmer line profiles can be seen in Figure \ref{fig6} and the derived equivalent widths are given in Table \ref{T2}. 

\subsection{AV Cep}
\begin{figure}[H]
\centering
\includegraphics[scale=0.11, angle=0 ]{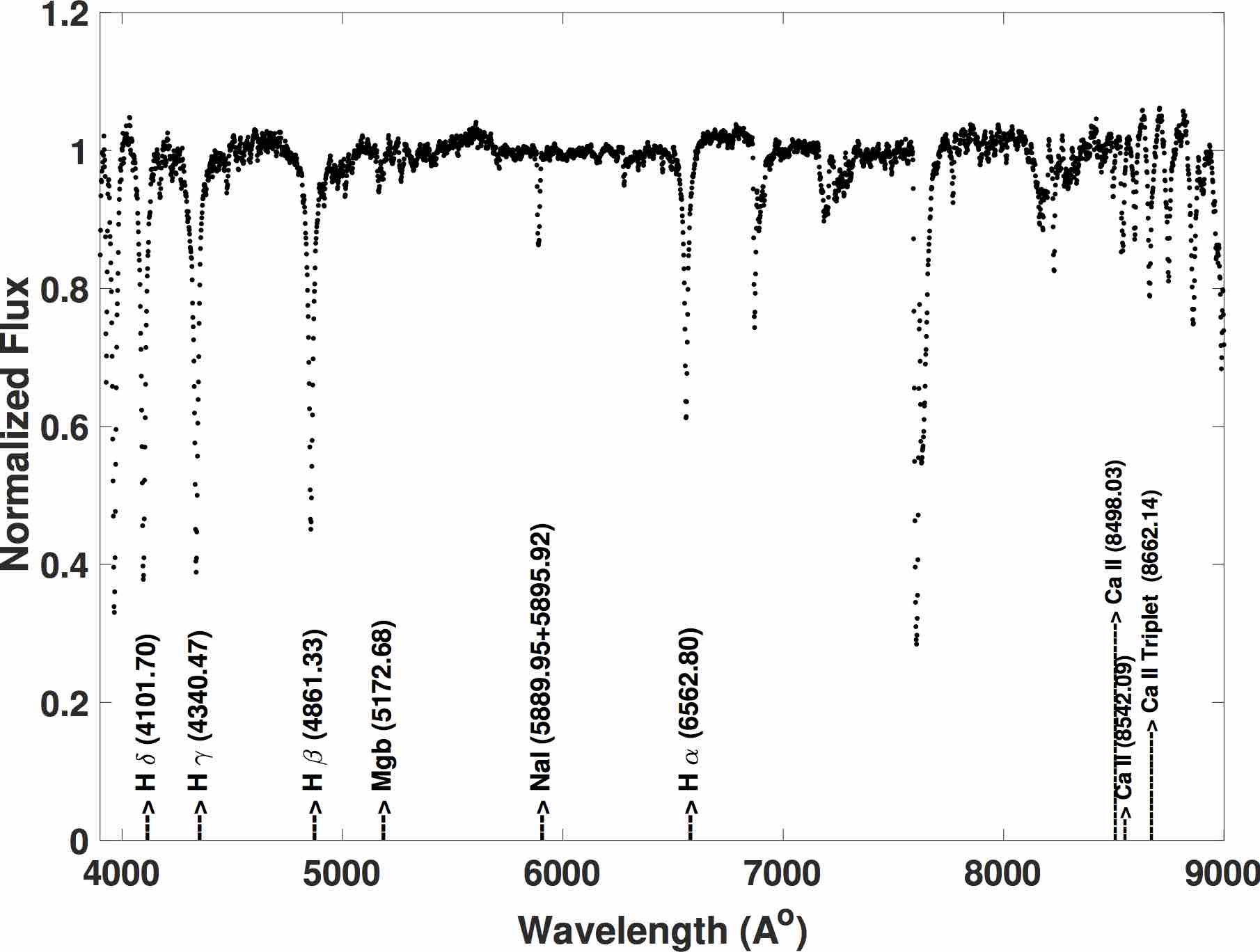}
\caption{Spectrum of AV Cep.}
\label{fig7}
\end{figure}
AV Cep (=	2MASS J05542644+8601206,V=12.18) was first identified and catalogued as eclipsing binary by \cite{kukarkin1971}.  It was later classified as an Algol type by \cite{budding2004} and the updated period of the variable was given as 2$^d$.958100 (\cite{malkov2006}). Photoelectric minima (\cite{agerer2003, hubscher2007, hubscher2012}) and few ToM (\cite{borovicka1993, kreiner2004}) are available in the literature. However, the ToM in the literature are very few to carry out O-C studies. One spectrum for AV Cep was obtained on Oct 13, 2013 at phase 0.0199 (calculated from the latest epoch available in the literature). Dominant spectral lines are shown in Figure \ref{fig7} and the calculated equivalent widths are given in Table \ref{T2}. From the observed phase the spectra can be attributed to the secondary component and it is also observed that H$\alpha$ absorption profile is relatively less prominent than the other Balmer lines which could be due to fill in effect .The spectral class determined in this current study using best fit spectral model (\cite{jacoby1984}) is F5 II. 

\subsection{XY Cet}

\begin{figure}[H]
\centering
\includegraphics[scale=0.11, angle=0 ]{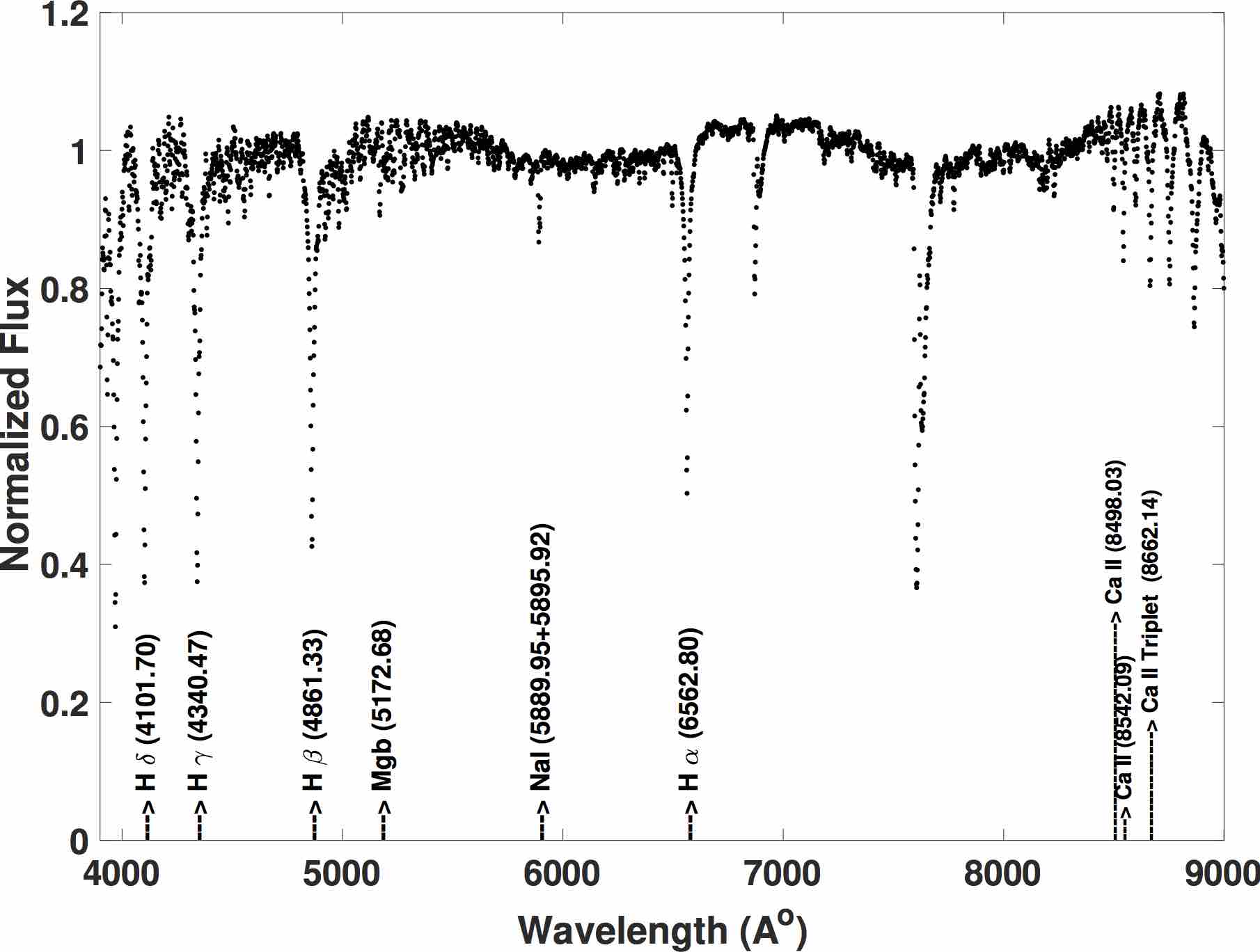}
\caption{Spectrum of XY Cet.}
\label{fig8}
\end{figure}

XY Cet(=TYC 51-832-1, V=8.75) is a well studied Algol type binary. It was first discovered by \cite{strohmeier1961} and was photoelectrically observed by \cite{morrison1968}. Detailed analysis of this binary was extensively done by many authors. The current period as derived by \cite{smalley2014} is 2$^{d}$.780710. Period studies were carried out by many authors beginning with \cite{srivastava1988} to \cite{southworth2011} however, no period variations were reported so far. Hence no period study was done in the present work. \cite{popper1971} obtained first spectroscopic data of this Algol and proposed its spectral type as A2 and F0 by inspecting metallic lines. This was further validated and refined using photometry data from superWASP study (\cite{southworth2011}). \cite{eker2014, eker2015} have deduced the double-lined binary nature and spectral type by disentangling previously observed data for the variable. In the current work one spectrum for XY Cet was obtained on Nov 20, 2013. From the latest epoch available the phase calculated is 0.0179.  The dominant profiles in the spectra are shown in Figure \ref{fig8} and the equivalent widths calculated are given in Table \ref{T2}. The spectral type determined using (\cite{jacoby1984}) is F0 III which is in agreement with that deduced from the work done by \cite{eker2015}.
 
\begin{figure}[H]
\centering
\includegraphics[scale=0.11]{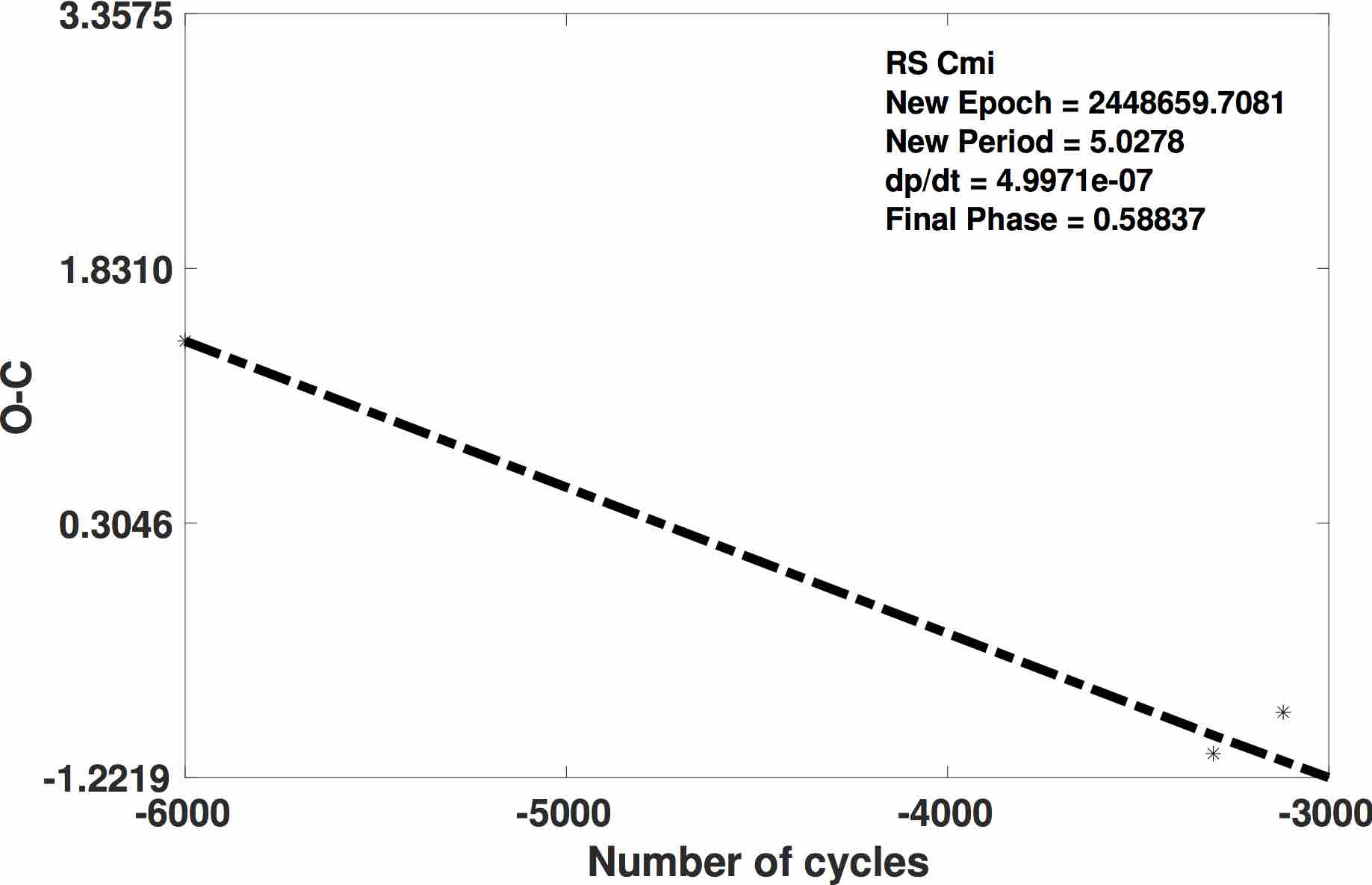}
\caption{O-C diagram of RS CMi.}
\label{fig9}
\end{figure}

\subsection{RS CMi}
RS CMi (= AN 154.1928, B=13.8) is an Algol-type eclipsing binary with an orbital period of 5$^{d}$.027800 (\cite{avvakumova2013}). It was discovered by \cite{hoffmeister1928}. \cite{budding2004} and \cite{malkov2006} have presented this variable in the catalogue of Algol type binary stars. Many authors \cite{diethelm2005, hubscher2007, hubscher2009} have presented ToM for this binary. It is one of the least studied Algols with only 5 ToM available of which a group of 4 are spread around 7 yrs and one observation separated from the group by about 60 years. We present the first period and spectroscopic study. The best fit for the O-C plot in Figure \ref{fig9} shows only a decreasing trend of period at the rate of dp/dt = 4.9971x10$^{-7}$ days/year. The new epoch obtained from current study is HJD (Min I) = 2448659.708 + 5$^{d}$.027755 $\times$ E. Two spectra were obtained for RS CMi on Feb 20, 2013 and on Feb 20, 2014 at phases 0.4117 \& 0.0044, respectively calculated from the new epoch. The spectral lines identified are shown in Figure \ref{fig10} and the equivalent widths of the dominant lines are given in Table \ref{T2}. From the spectral study it is observed that the absorption profile for the H$\alpha$ line is relatively less prominent when compared to other Balmer lines indicating fill in effect for the primary component. At the phase outside eclipse both H$\alpha$ \& H$\beta$ show fill in effect in the absorption profile relative to other Balmer lines. This can be related to the evolution of disk around the current primary, which can be confirmed with further observations.  Based on minimum res$^{2}$ and visual inspections, the spectral class is derived to be A3 III (\cite{jacoby1984}).

\begin{subfigures}
\begin{figure}[H]
\centering
\includegraphics[scale=0.11, angle=0 ]{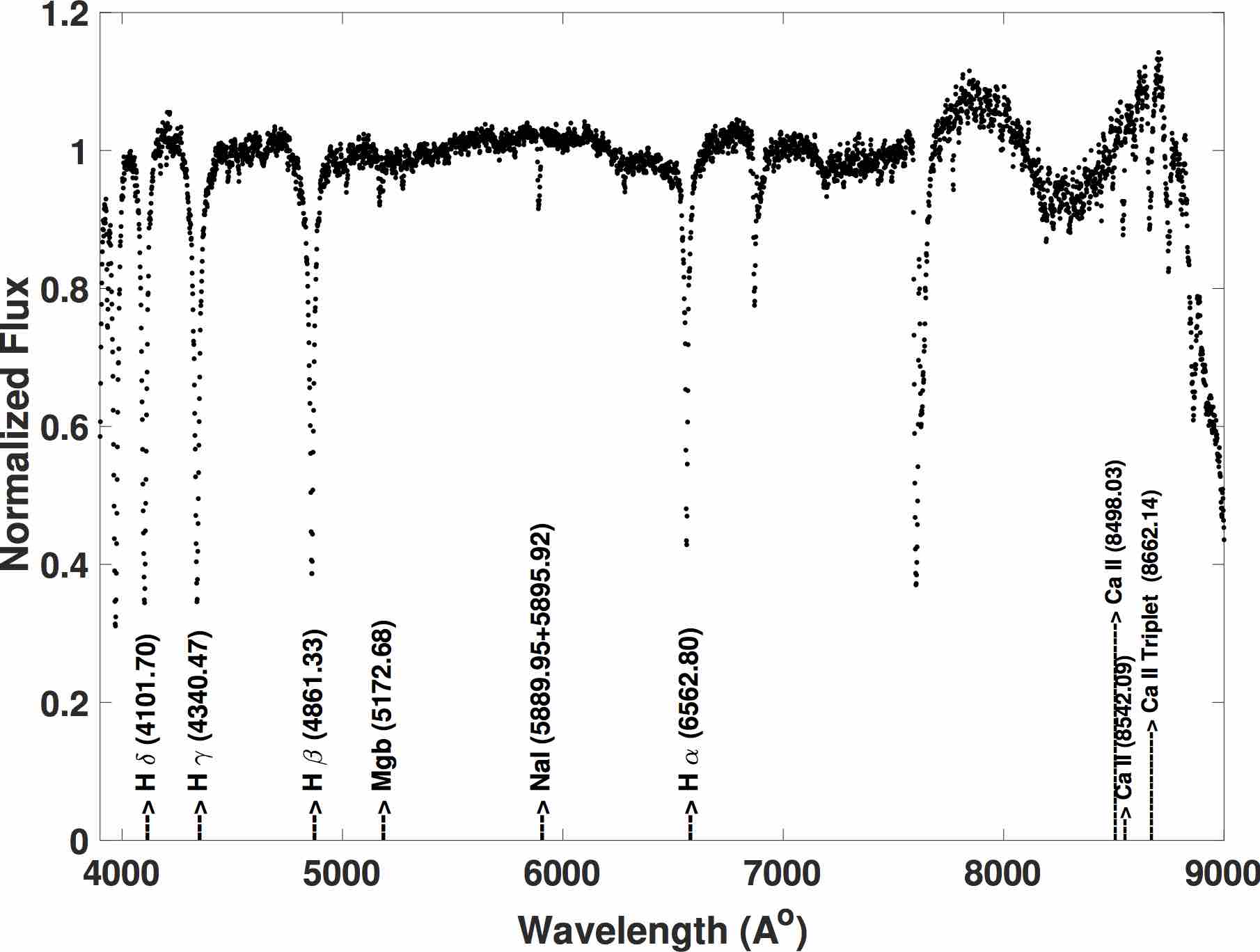}
\caption{Spectrum of RS CMi (2013).}
\end{figure}

\begin{figure}[H]
\centering
\includegraphics[scale=0.11, angle=0 ]{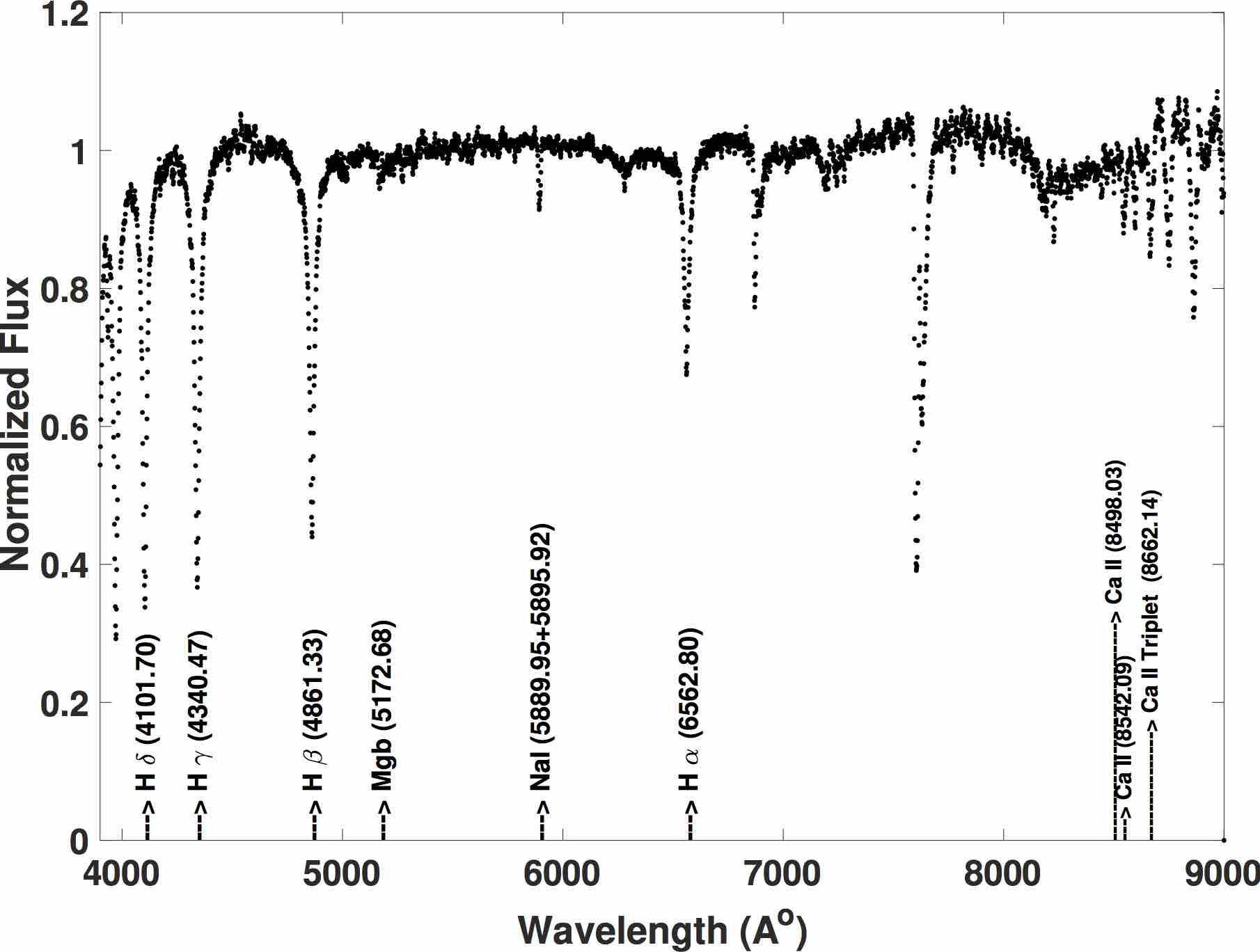}
\caption{Spectrum of RS CMi (2014).}
\end{figure}
\label{fig10}
\end{subfigures}
  
\subsection{RS CVn}
\begin{figure}[H]
\centering
\includegraphics[scale=0.11]{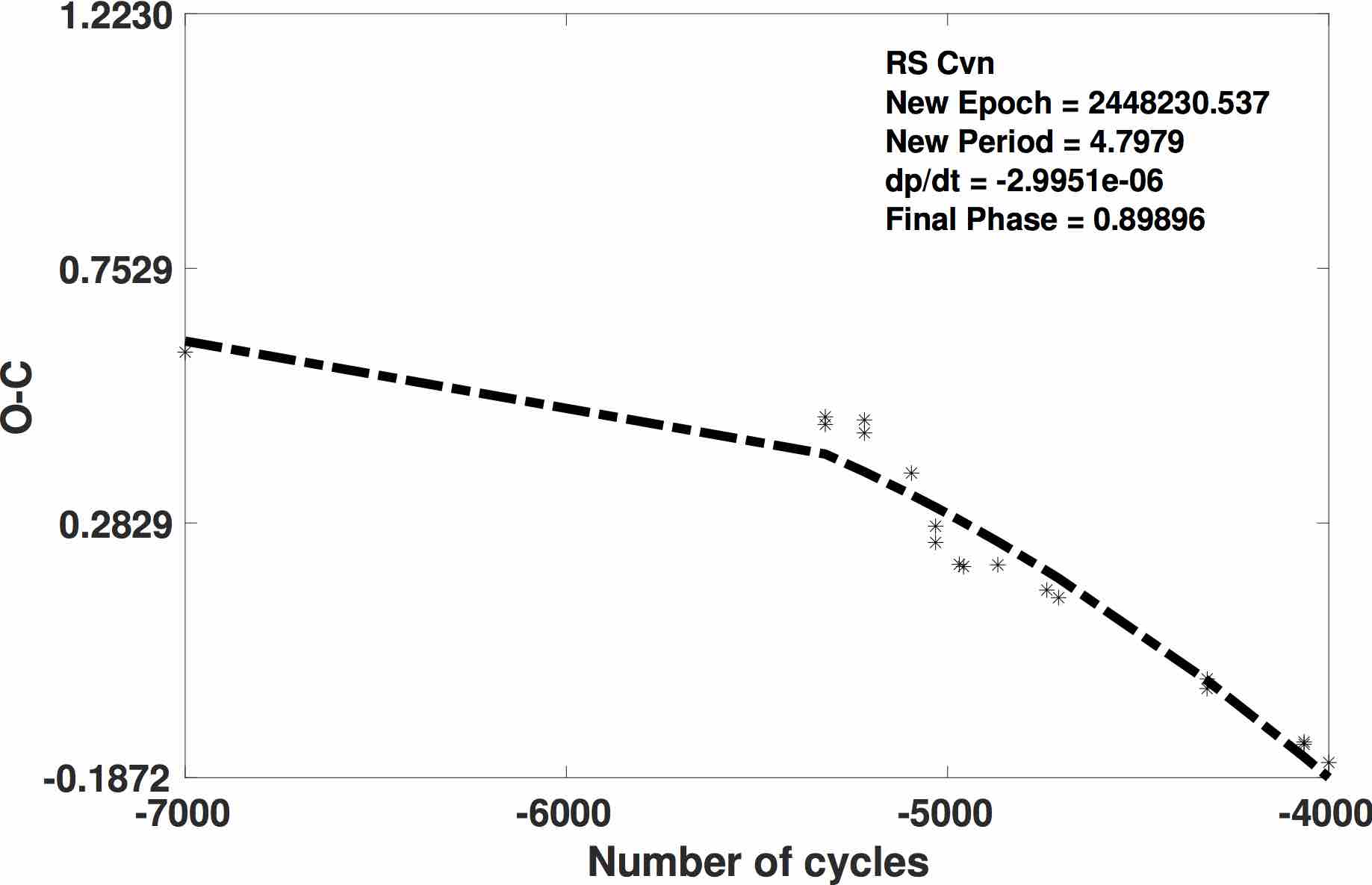}
\caption{O-C diagram of RS CVn.}
\label{fig11}
\end{figure}

\begin{figure}[H]
\centering
\includegraphics[scale=0.11]{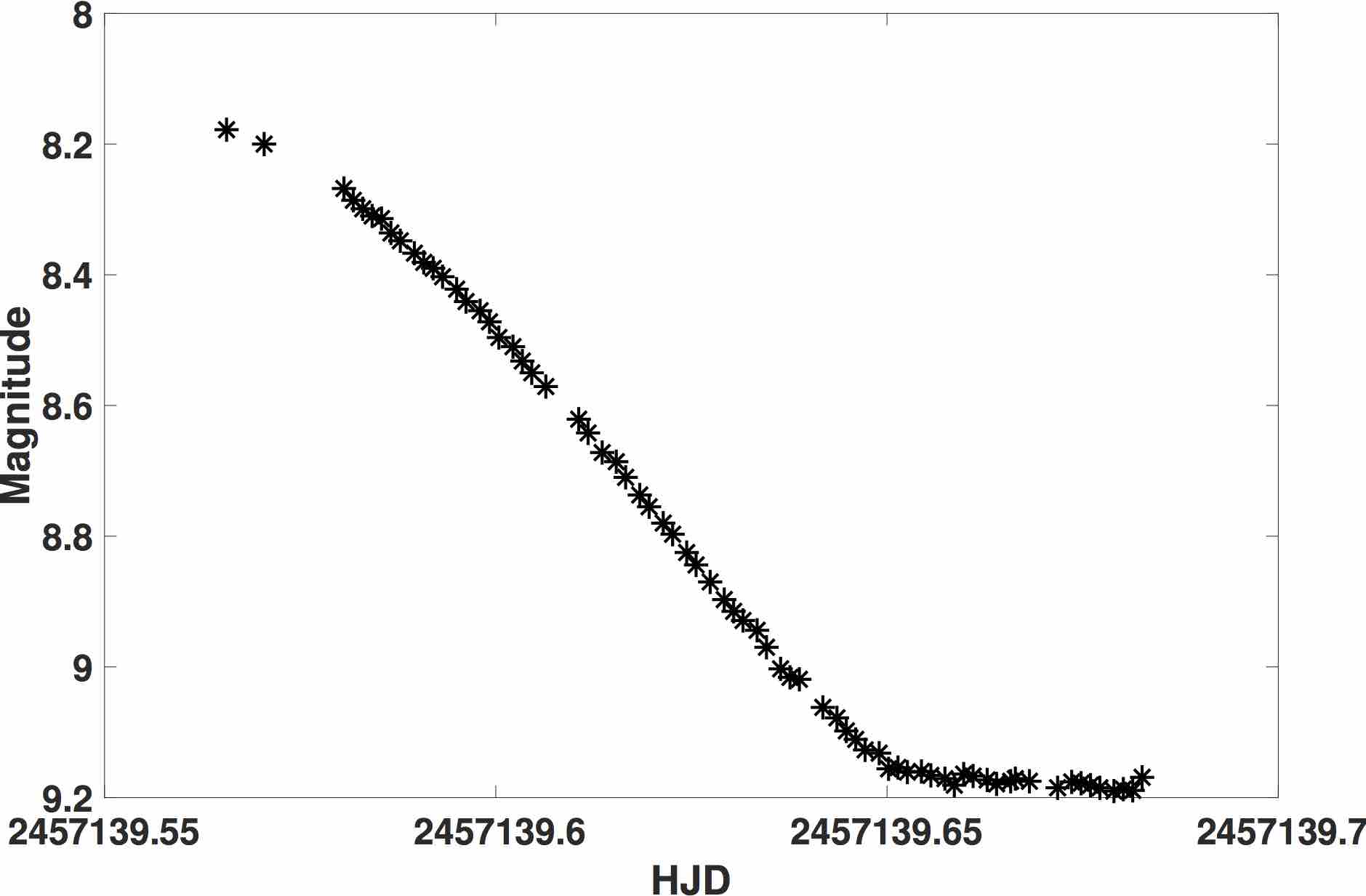}
\caption{Light curve of RS CVn.}
\label{fig12}
\end{figure}

\begin{figure}[H]
\centering
\includegraphics[scale=0.11, angle=0 ]{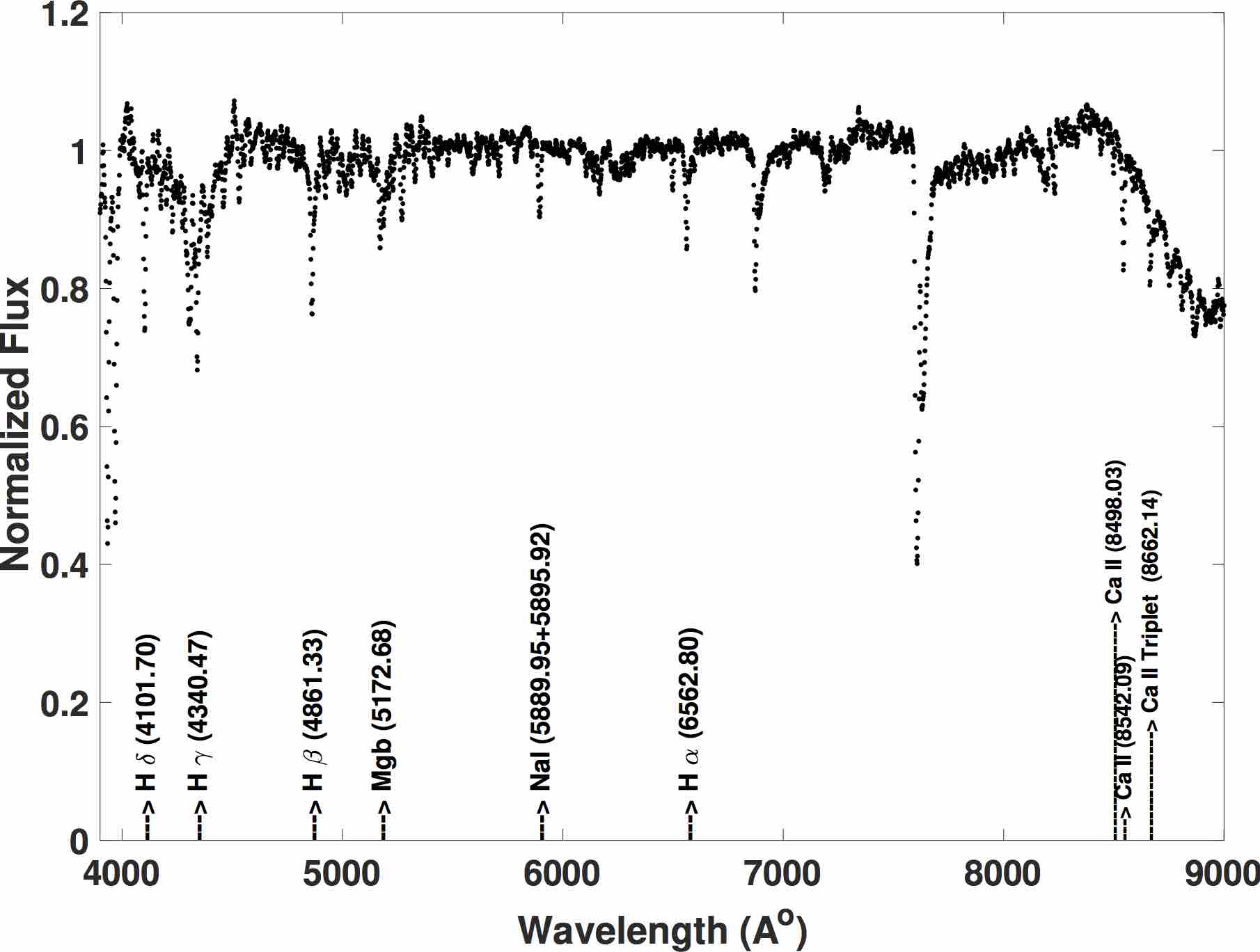}
\caption{Spectrum of RS CVn.}
\label{fig13}
\end{figure}

RS CVn (= GSC 02534-01642 = AN 10.1914, V = 7.93) is one of the most interesting and well studied Algol type binary since its discovery by \cite{hoffmeister1915} along with 5 times of minima. It has an orbital period of 4$^{d}$.797900 (\cite{kukarkin1971}) and spectral class F6IV+G8IV  (\cite{strassmeier1990}). Photometric and spectroscopic investigations revealed the orbital elements and spectral types of the components as F3 \& K0 by \cite{popper1980}. It was observed that there is a cyclic fluctuation in the primary period which was mostly due to mass loss  during continuous ejection of particles from migrating active regions like star spots on the cooler star. It was also hypothesized that the variable is in pre-main sequence contraction and the cooler star shows a T-Tauri star characteristics (\cite{sitterly1921, joy1922, keller1951, plavec1959, hall1972, arnold1973, catalano1974, rhombs1976}). In addition, the spectral studies carried out by \cite{fernandez1994} in the region of Ca II H \& K for the variable confirmed emission flux variations with the orbital phases. The ultraviolet excess characterizing RS CVn spectra was attributed to the free free emission from hot circumstellar gas by \cite{rhombs1976} . In the current work this is an Algol type with 18 observations of ToM is plotted for the period study. 17 of the observations are grouped around 40 years and this group is separated by one of the observations by about 46 years. However the data shows a sinusoidal component of period variations superimposed on the decreasing trend (Figure \ref{fig11}) of the best fit curve. A quadratic ephemeris to fit the O-C variation is derived to be dp/dt = -2.9951x10$^{-6}$ days/year. The new epoch deduced is HJD (Min I) = 2448230.537 + 4$^{d}$.797852 $\times$ E.  Mass exchange or mass loss or presence of third body can be excluded (\cite{lanza2004, frasca2005}) for variables showing magnetic activity. The magnetic activity is evident in RS CVn  (\cite{fernandez1994}) attributing to the sinusoidal period changes. One spectrum for RS CVn was obtained on Apr 23, 2014 at phase 0.1010 calculated from the derived epoch.  No spectroscopy work was carried out for the past three decades. In the current study we present the spectrum with dominant spectral lines observed as shown in Figure \ref{fig13}  and the equivalent widths calculated in Table \ref{T2}, the light curve obtained using data available in the literature is shown in Figure \ref{fig12}. The Balmer lines in the spectra show weak absorption profiles which could be due to fill in effect caused by activity on or near the primary component.

\subsection{RR Dra}
\begin{figure}[H]
\centering
\includegraphics[scale=0.11]{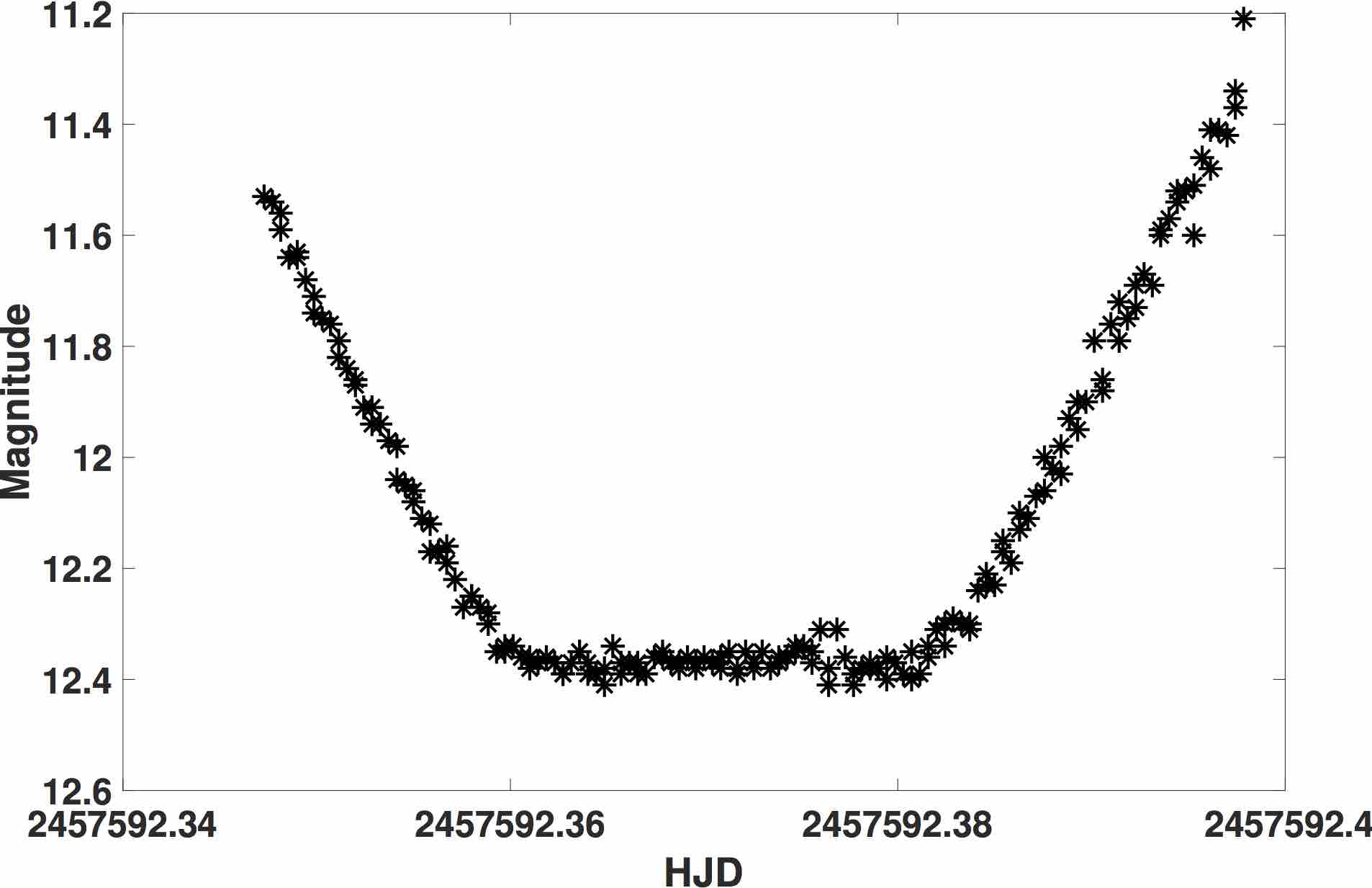}
\caption{Light curve of RR Dra.}
\label{fig14}
\end{figure}

\begin{figure}[H]
\centering
\includegraphics[scale=0.11, angle=0 ]{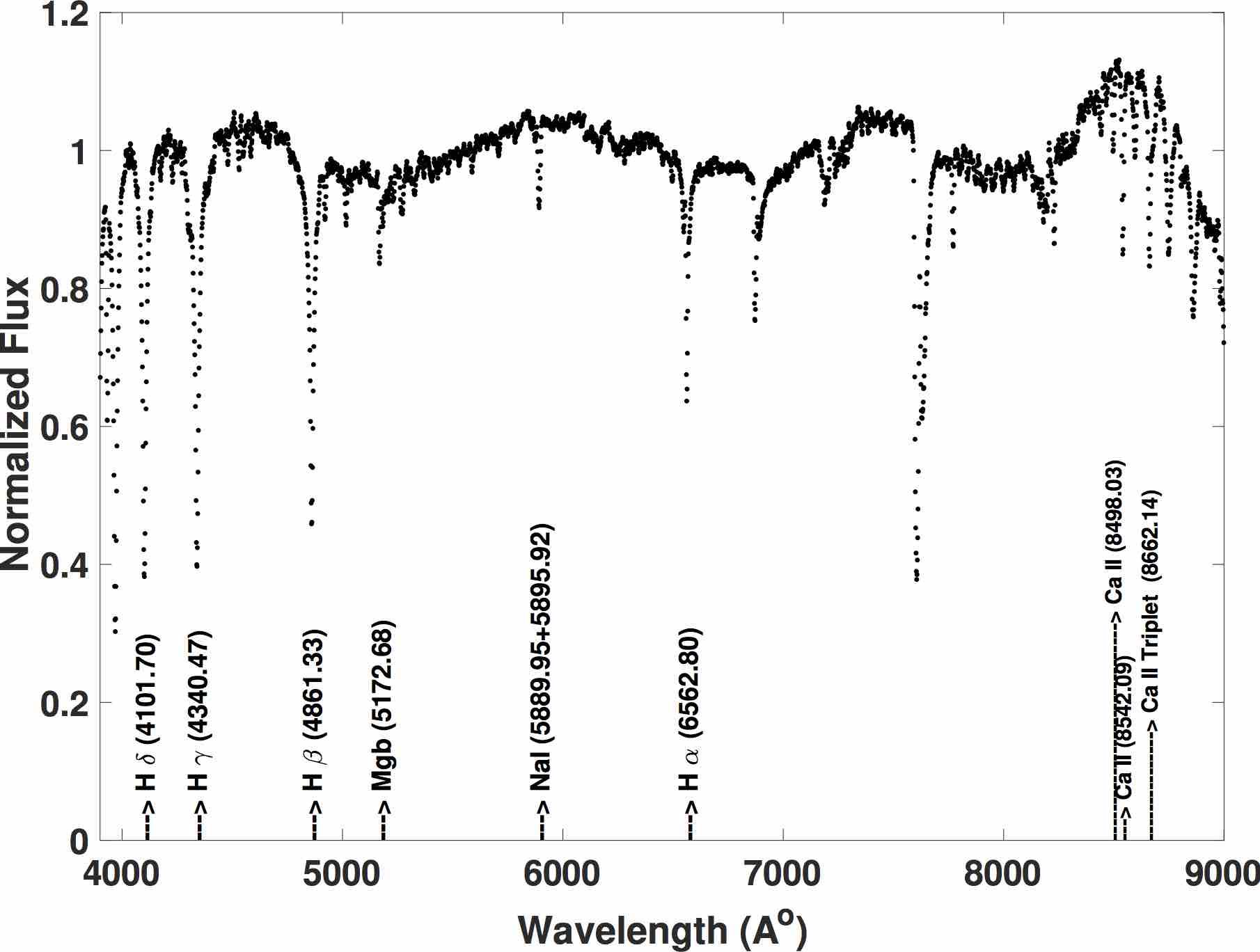}
\caption{Spectrum of RRDra.}
\label{fig15}
\end{figure}

RR Dra (= AN 188.1904, V = 9.831) is an Algol-type eclipsing binary with an orbital period of 2$^{d}$.831200 (\cite{kukarkin1971}). It contains an A2 type primary and a K0-type secondary (\cite{yoon1994}). Many times of light minimum of RR Dra were collected from previous work (\cite{dugan1939, szczepanowska1956, szczepanowska1959, whitney1957}) and extensive period study was carried out for this variable showing a secular change with rapid period increase rate of dp/dt = +4.24 × 10$^{-6}$ days/year (\cite{qian2002}).  A spectrum was obtained at phase 0.0210 (calculated using latest epoch available in the literature) on April 23, 2014. The dominant spectral lines identified are shown in Figure \ref{fig15}, their equivalent widths are given in Table \ref{T2} and the light curve is shown in Figure \ref{fig14}. The equivalent width of the H$\alpha$ line shows a remarkably high fill in effect compared to other Balmer lines of all the Algols in current study.   The work done by \cite{qian2002} suggests that the rapid period changes superimposed on the long-term increases could be due to the structural variation of the cool mass-loser through the instabilities in the convective outer layer (COL) or through cyclic magnetic activity of the K0-type components.  Since for this variable the period jumps are not observed in a short-term alternating way, the most possible explanation for period changes could be the structural change caused by instabilities in the COL via a dynamical mass loss from the cool subgiant. When the parameters were plotted (Figure \ref{fig55}) on the available models  (\cite{lubow1975, kaichuck1985}), the position of RR Dra is observed to be above the $\omega_{disk}$ implying that it is similar to those systems with permanent double peaked emission lines that may vary in strength and shape. It also indicates that stable disk could not form but circumstellar bulges may be possible, thus validating Qian's study.

\subsection{TZ Eri}
TZ Eri (= GSC 04147-01115 = TYC 4147-1115-1, V = 10.60) is an Algol-type eclipsing binary with an orbital period 2$^{d}$.6061 (\cite{kreiner2004}). Its variability was discovered by \cite{hoffmeister1929} and spectral class was first classified by \cite{cannon1934} as F type. Many authors have (\cite{kaitchuck1982, kaitchuck1988, vesper2001}) have studied the variable and given the parameters predicting the presence of an accretion disc in the system. The variable's possible link between the orbital and pulsational periods was investigated by \cite{soydugan2006}. Observed minima were given by many authors \cite{kordylewski1963, mallama1980, faulkner1983, samolyk2008, samolyk2010, samolyk2011} and \cite{harmanec1988} detected a third body in both light curve solutions as well as spectra. \cite{liakos2009} has published frequency analysis and the period study reveals a long-term period increase attributed to mass transfer from secondary component to primary component (\cite{zasche2008}). The spectral types of the components are given as A5/6 for primary and K0/1 III for secondary by \cite{barblan1998} and its pulsational behaviour was detected by \cite{mkrtichian2005}.   

\begin{figure}[H]
\centering
\includegraphics[scale=0.11]{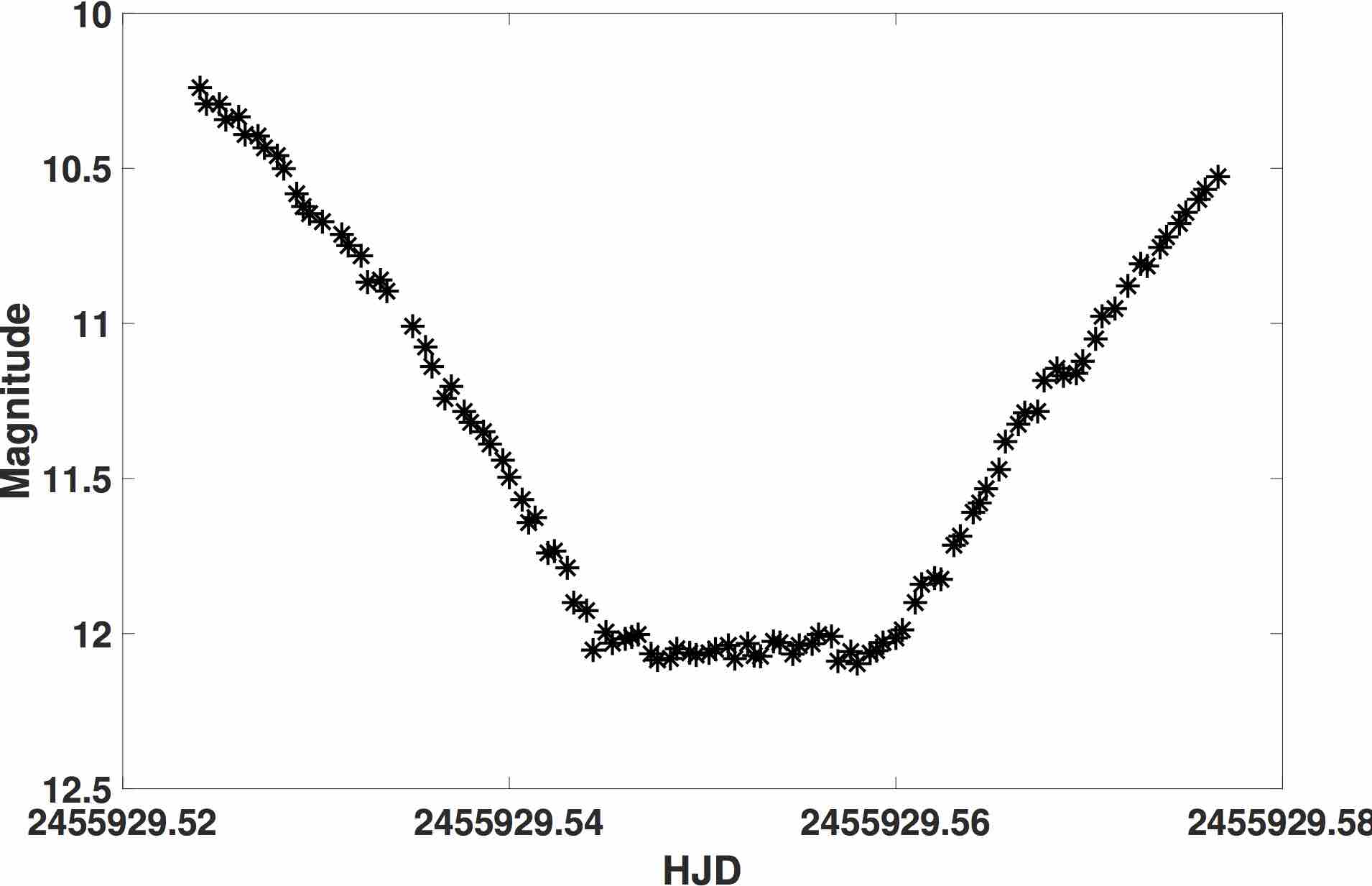}
\caption{Light curve of TZ Eri.}
\label{fig16}
\end{figure}

\begin{figure}[H]
\centering
\includegraphics[scale=0.11, angle=0 ]{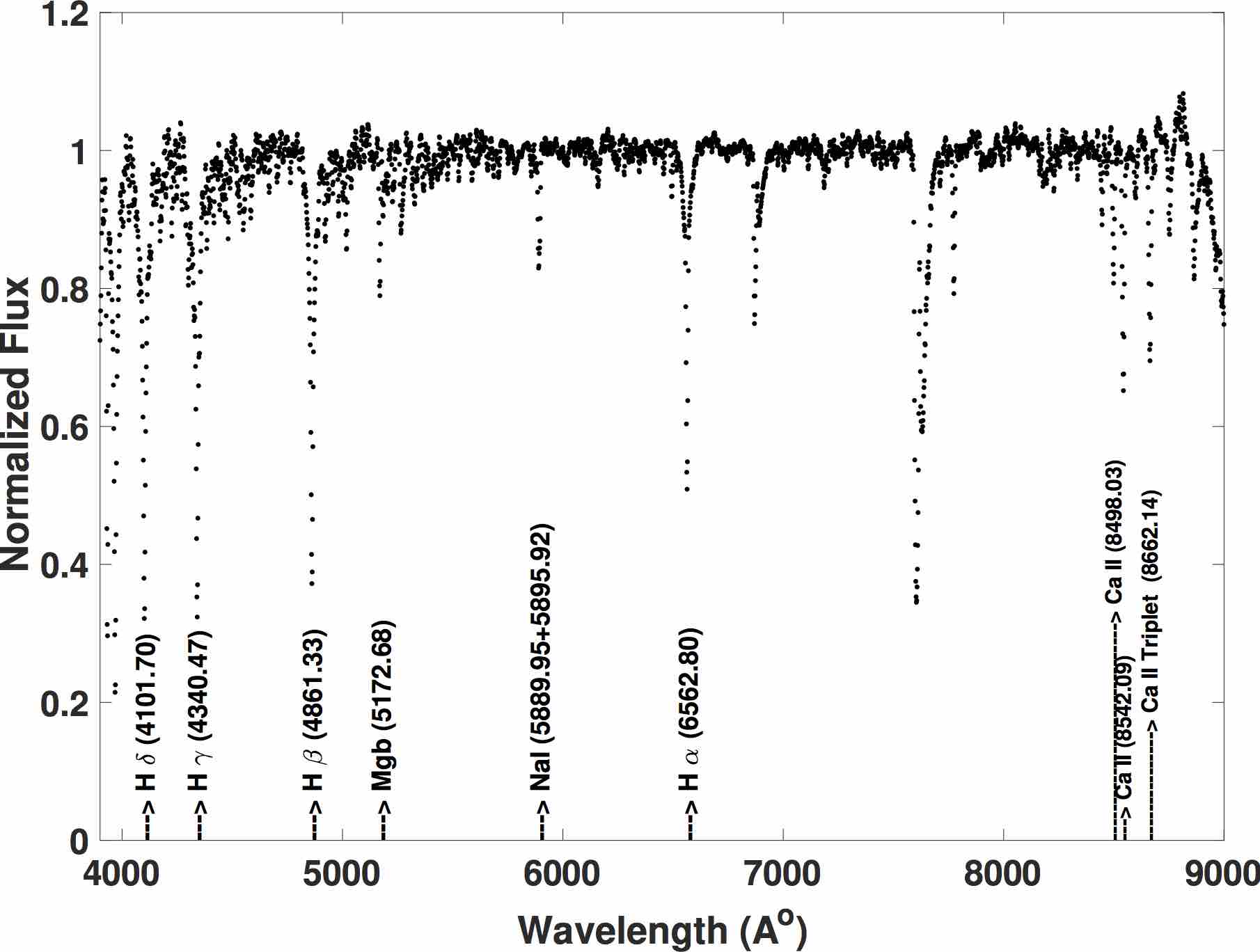}
\caption{Spectrum of TZ Eri.}
\label{fig18}
\end{figure}

 A spectrum of TZ Eri was obtained at phase 0.071 on Nov 20, 2013. The dominant spectral lines identified are given in Figure \ref{fig18}. The light curve is shown in Figure \ref{fig16} and the equivalent widths in Table \ref{T2}. The H$\alpha$ absorption profile is observed to be relatively less prominent than the other Balmer lines. The new phase obtained using derived ephemeris is  0.97171. From the available models \cite{lubow1975} and \cite{kaichuck1985}, the position of TZ Eri is below the $\omega_{min}$ line showing primary radius larger than $\omega_{disk}$. This implies that a stable accretion disk could not form but circumstellar bulges may prevail.
 
 \subsection{AN Gem}

\begin{figure}[H]
\centering
\includegraphics[scale=0.11]{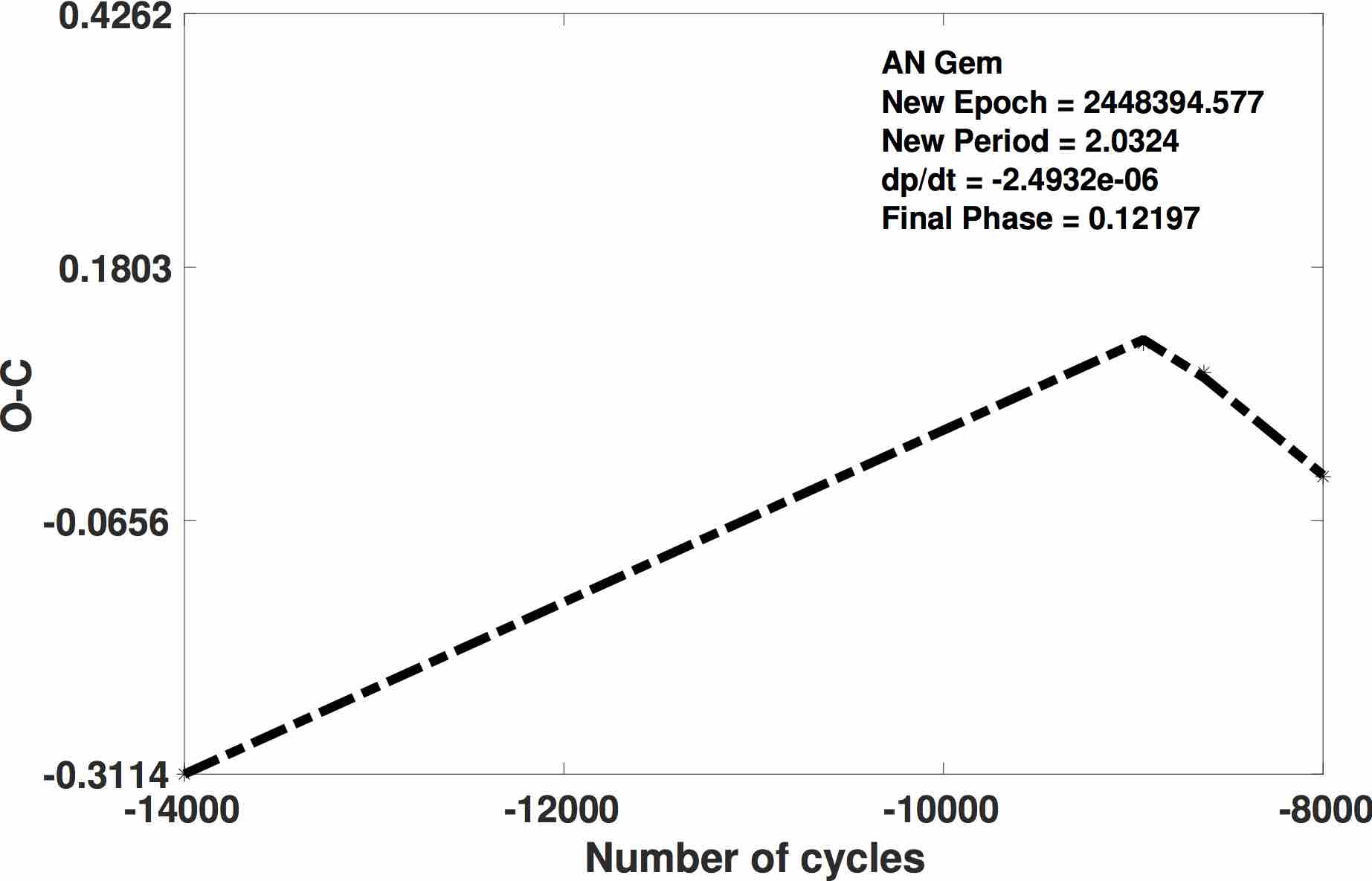}
\caption{O-C diagram of AN Gem.}
\label{fig19}
\end{figure}

\begin{figure}[H]
\centering
\includegraphics[scale=0.11, angle=0 ]{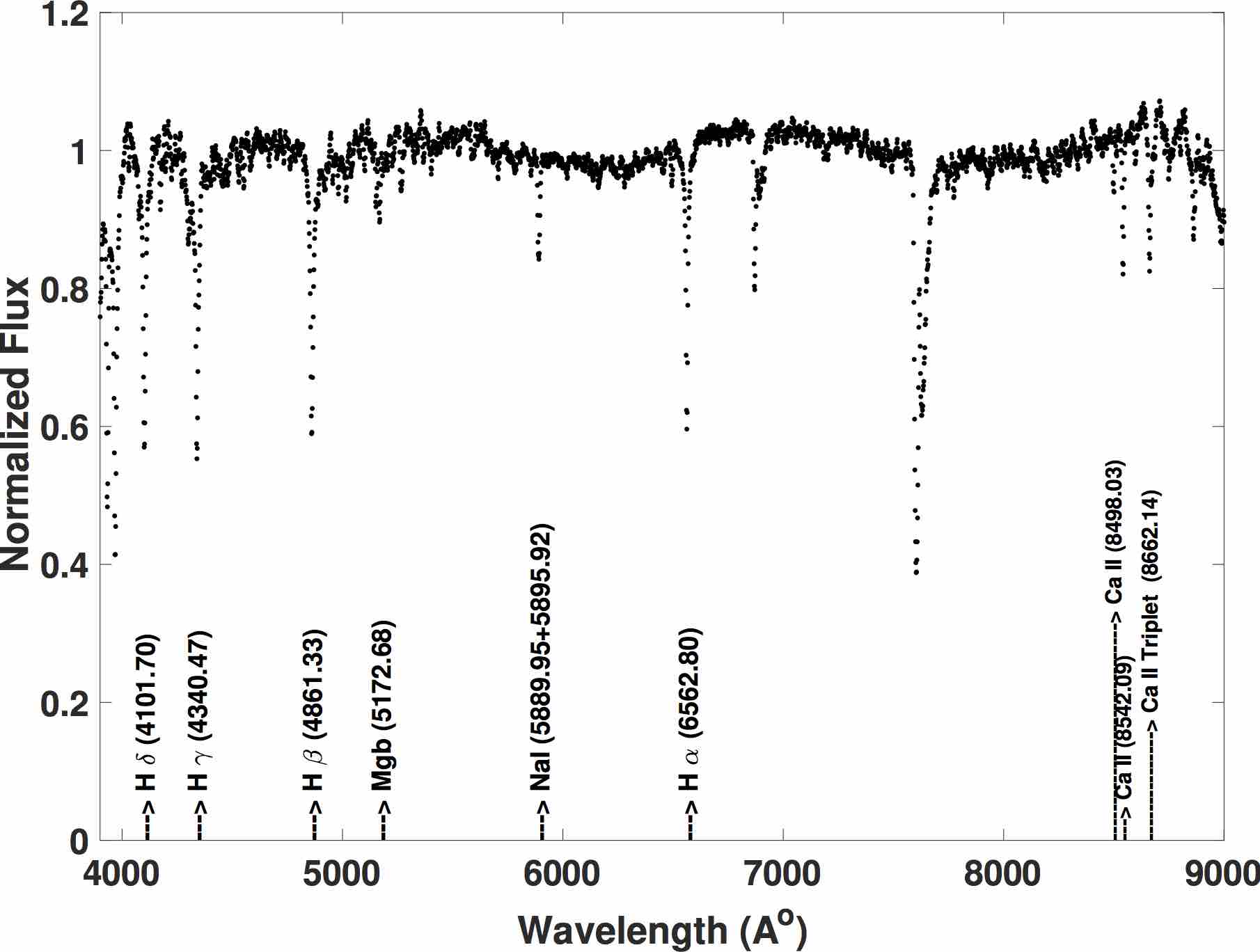}
\caption{Spectrum of AN Gem.}
\label{fig20}
\end{figure}

AN Gem (2MASS J07085752+1948136, V = 13.20) is an Algol-type eclipsing binary with an orbital period 2$^{d}$ .032300 (\cite{kreiner2004}). The linear elements for this variable were given by authors \cite{borovicka1993},\cite{hubscher2005} \& \cite{kreiner2004}. No further work was carried out on this variable so far. First period study is presented in the current work and from the quadratic ephemeris the O-C variation is calculated. The O-C diagram is shown in Figure \ref{fig19} has been plotted for only 3 data points spanning over 11 years and one point separated by 66 years. The fit suggests a decreasing period. The decreasing rate of period obtained is dp/dt = -2.4932x10$^{-6}$  days/year,  however additional observations are required to understand the change clearly. The new ephemeris obtained is as follows
 HJD (Min I) = 2448394.577 + 2$^{d}$.03246 $\times$ E. 
A spectrum for AN Gem was obtained on Nov 12, 2013 at phase 0.911 and the spectral lines identified are shown in Figure \ref{fig20} and the equivalent widths in Table \ref{T2}.  Based on minimum res$^{2}$ and visual inspections, the best fit spectral model (\cite{jacoby1984}) was selected to determine the spectral class as F3 III. 

 \subsection{SX Gem}
 
SX Gem (= AN 56.1908, V = 11.00) is an Algol-type eclipsing binary with an orbital period 1$^{d}$.366900 with spectral class A0+A9 (\cite{giuricin1984}). It was first listed in the catalogue of variable stars by \cite{gaposchkin1932}. Absolute parameters for this variable were deduced by \cite{brancewicz1980}. Many authors have recorded ToM but no period variation studies were done. \cite{zasche2011} has derived basic parameters for SX Gem and it is observed that the mass ratio is different in his study when compared to that derived by \cite{brancewicz1980}. 
\begin{figure}[H]
\centering
\includegraphics[scale=0.11]{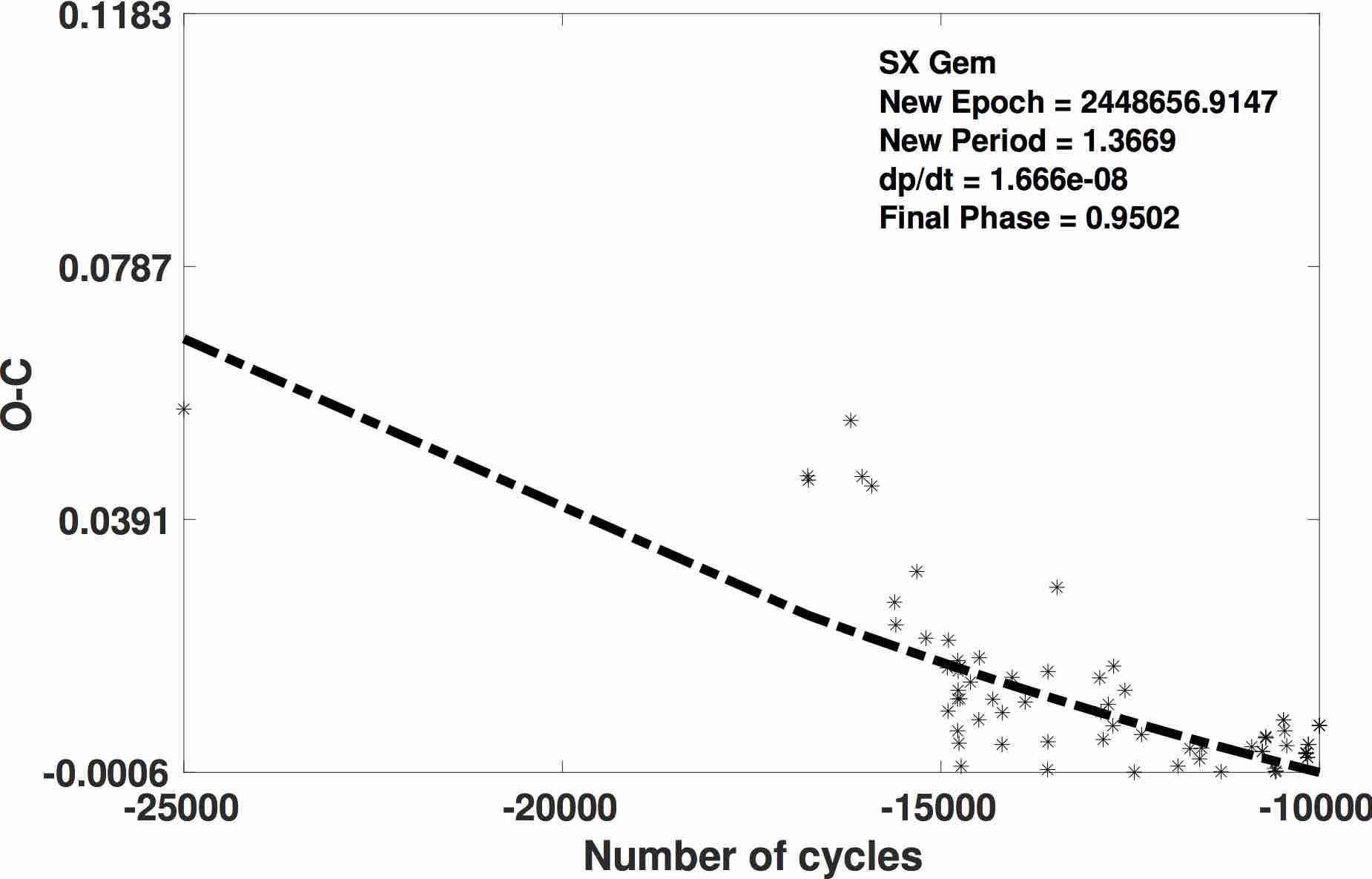}
\caption{O-C diagram of SX Gem.}
\label{fig21}
\end{figure}

\begin{figure}[H]
\centering
\includegraphics[scale=0.11]{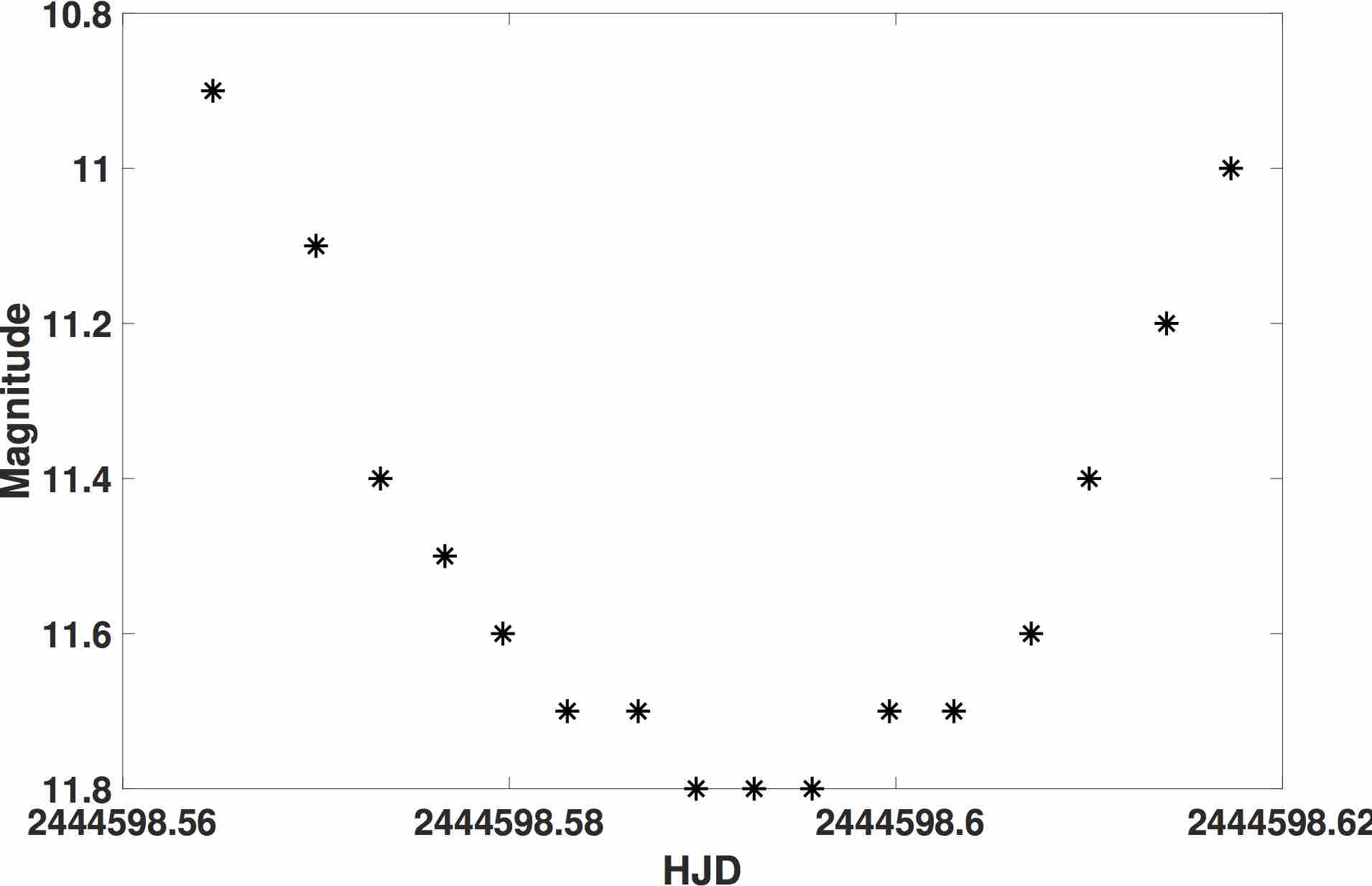}
\caption{Light curve of SX Gem.}
\label{fig22}
\end{figure}

\begin{subfigures}
\begin{figure}[H]
\centering
\includegraphics[scale=0.11, angle=0 ]{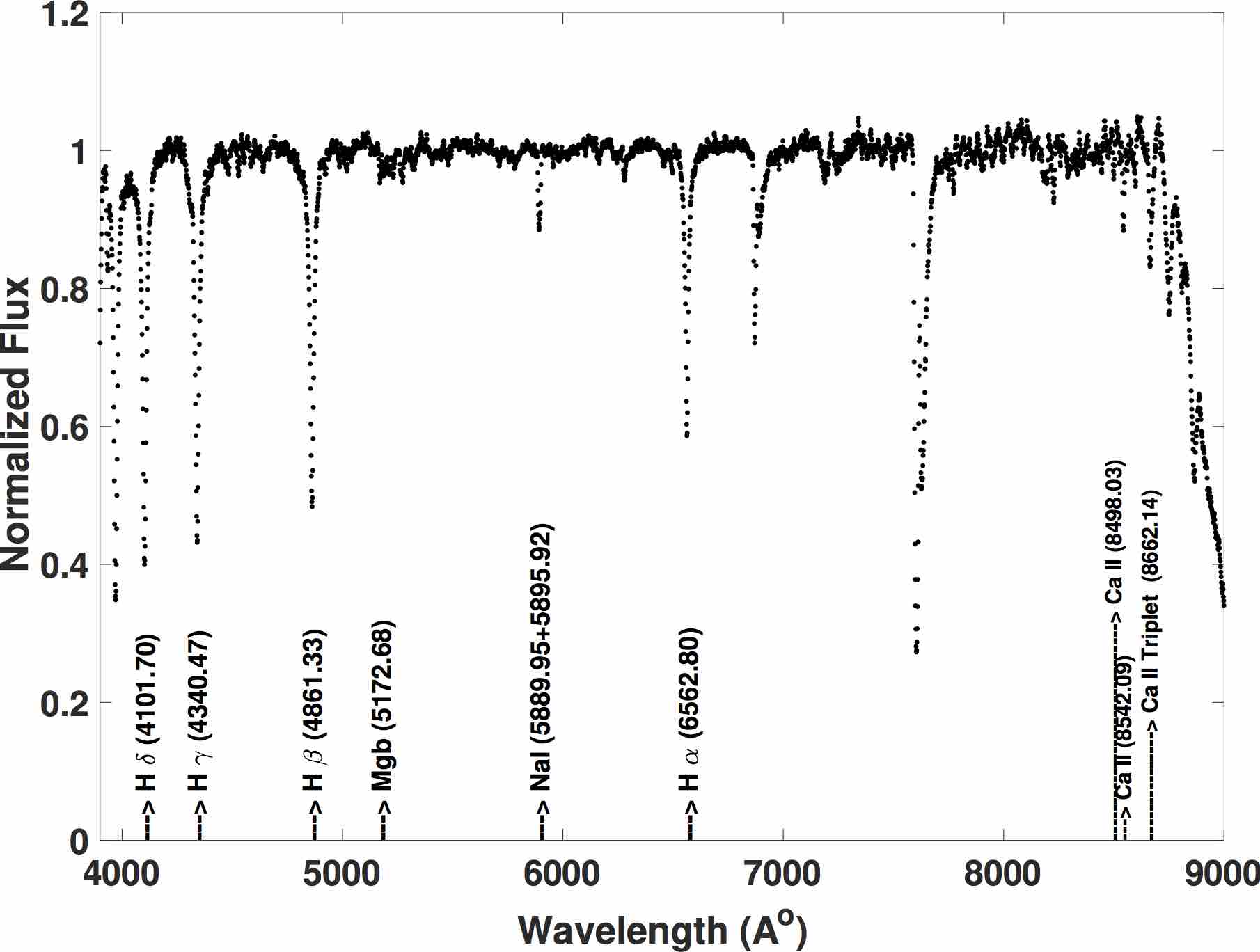}
\caption{Spectrum of SX Gem (2013).}
\end{figure}

The (O-C) is plotted (Figure \ref{fig21}) for 86 observations of ToM with a group of 85 data points scattered around 45 years and separated from one observation by another 45 years. The observations show a large scatter around the best fit curve, which is clearly showing a decreasing trend of the period and the rate of period decrease obtained is dp/dt = 1.666x10$^{-8}$ days/year. The decreasing trend of the period is in agreement with the change in mass ratio of the variable. The new ephemeris obtained is as follows HJD (Min I) = 2448656.9147 + 1$^{d}$.366877 $\times$ E. The light curve obtained from the data available in the literature is shown in Figure \ref{fig22}. 
Two spectra for SX Gem were observed on Feb 20, 2013 and Feb 20, 2014 at phases 0.050 and 0.065, obtained using latest epoch in the literature, respectively as shown in Figure \ref{fig23}. 
\begin{figure}[H]
\centering
\includegraphics[scale=0.11, angle=0 ]{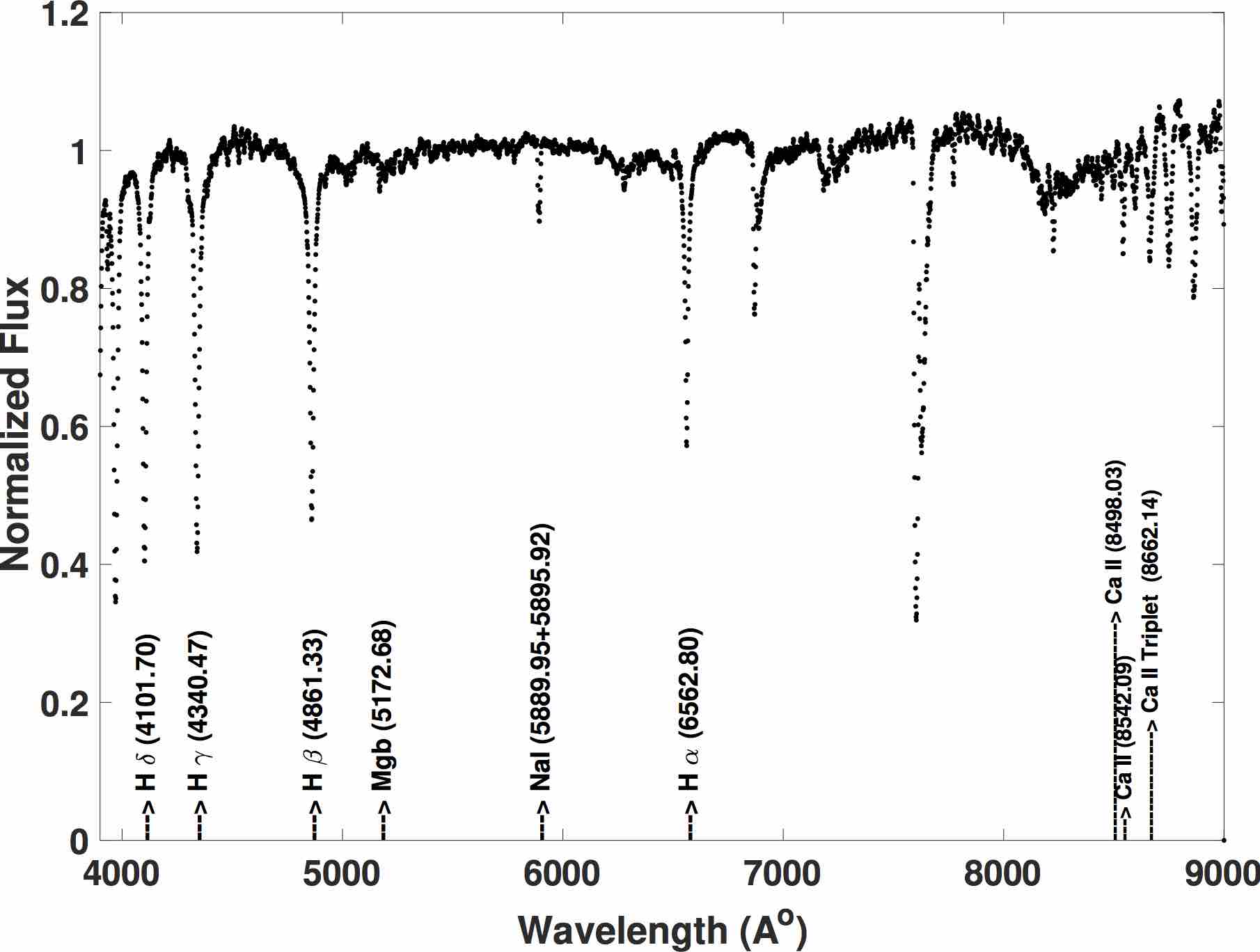}
\caption{Spectrum of SX Gem (2014).}
\end{figure}
\label{fig23}
\end{subfigures}

\subsection{TW Lac}
\begin{figure}[H]
\centering
\includegraphics[scale=0.11]{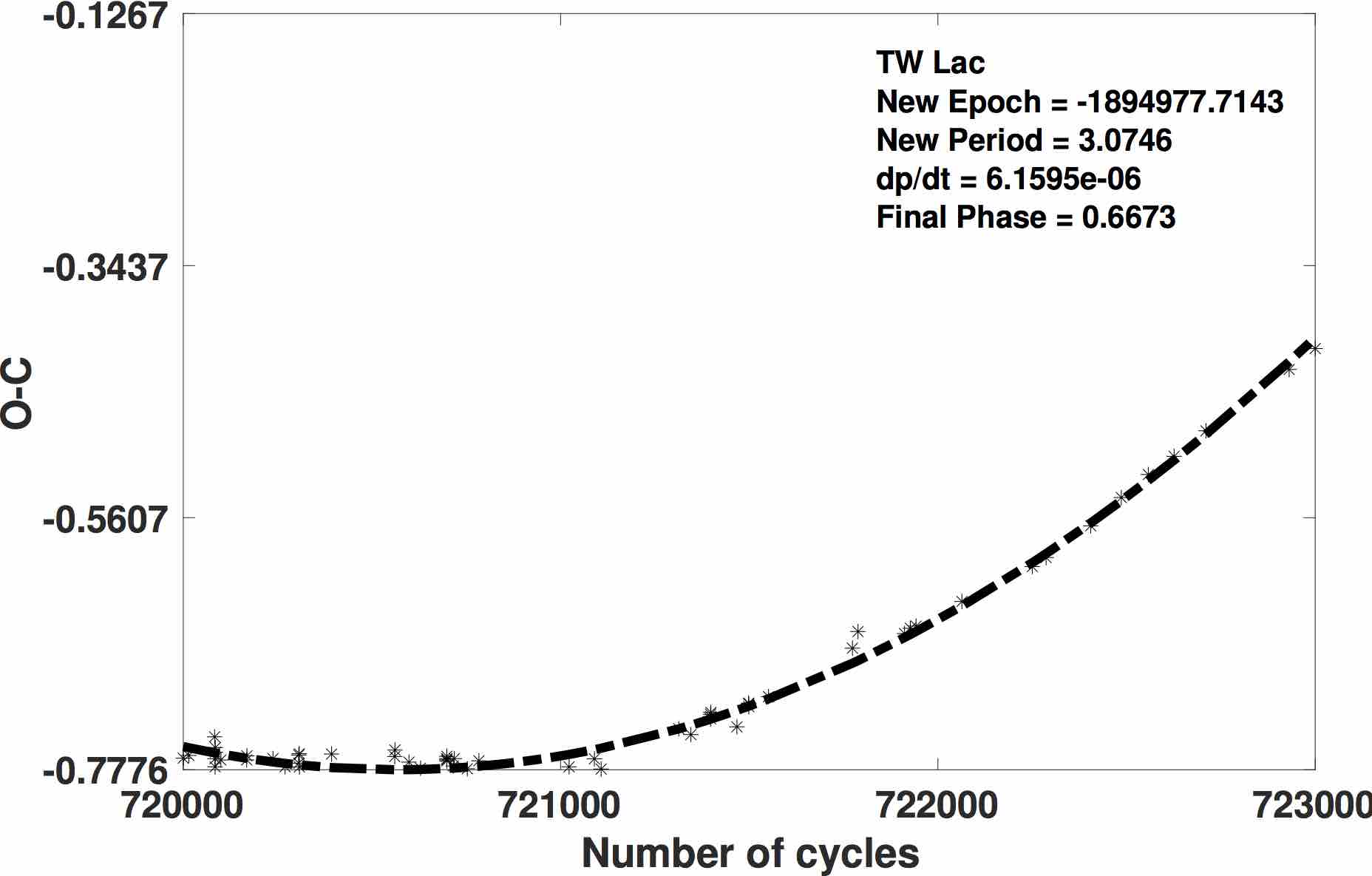}
\caption{O-C diagram of TW Lac.}
\label{fig24}
\end{figure}

TW Lac (=GSC 03987-01714=AN 123.1925, V=11.85) is a detached Algol type binary. Its variability was first studied by \cite{gaposchkin1932}, which led to extensive observations of minima for this variable. The spectral type was derived to be A3 IV by \cite{halbedel1984} and period 3$^{d}$.037518 (\cite{kreiner2004}). It was catalogued as Algol-type by \cite{budding2004} and was categorized as a candidate semi detached system for pulsation by \cite{soydugan2006}. The minima and period changes were intensively studied by many authors {\cite{whitney1957, whitney1959, wood1963, kreiner1971, agerer2003, malkov2006, erdem2007, dogru2007}}. In the current study the period study has been performed on 65 observations of ToM. The data shows a distinct sinusoidally varying period as shown in the Figure \ref{fig24} it is also observed that there is a sinusoidal component superimposed on increasing period O-C curve.The epoch derived is as follows HJD (Min I) = 2456647.435 + 3$^{d}$.074560 $\times$ E and the variation in period is dp/dt = 6.1595x10$^{-6}$ days/year.

\begin{figure}[H]
\centering
\includegraphics[scale=0.11, angle=0 ]{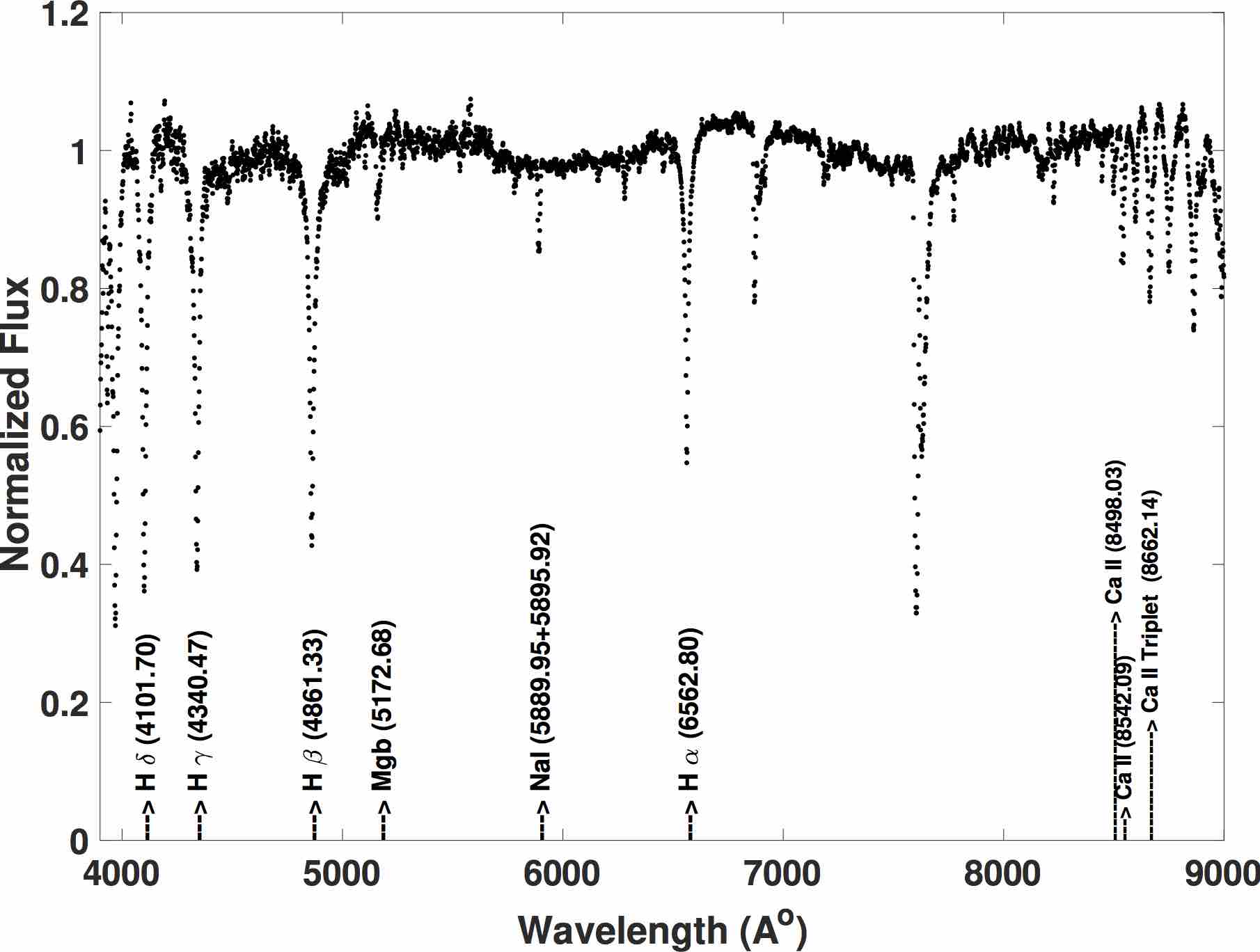}
\caption{Spectrum of TW Lac.}
\label{fig25}
\end{figure}

Though the first spectral study was done by \cite{halbedel1984}, no information was provided about the spectra except for spectral type. \cite{kaitchuck1985} have done spectroscopy of 52 short period Algols during their primary eclipse phase, and have tried to model the observed spectral line profiles to understand the nature of the accretion disk around the primary. They have found that when observed in July 1983, during eclipse which is a total eclipse, there is no emission in the observed Balmer line profiles (H$\beta$ to H$\gamma$) and the approximate maximum equivalent width of emission was less than 0.4. Their study indicates that the system is not harboring a transient disc.  We have done the spectral study for the spectra  obtained on Dec 21, 2013 at phase 0.293 calculated from the latest epoch obtained. The spectrum showing the dominant spectral lines is shown in Figure \ref{fig25} and their equivalent widths in Table \ref{T2}. It is observed that all the Balmer lines show strong absorption profile at the observed phase that is out of eclipse. The spectral type obtained using (\cite{jacoby1984}) stellar library is A8 V.

\subsection{FG Lyr}
\begin{figure}[H]
\centering
\includegraphics[scale=0.11,angle=0]{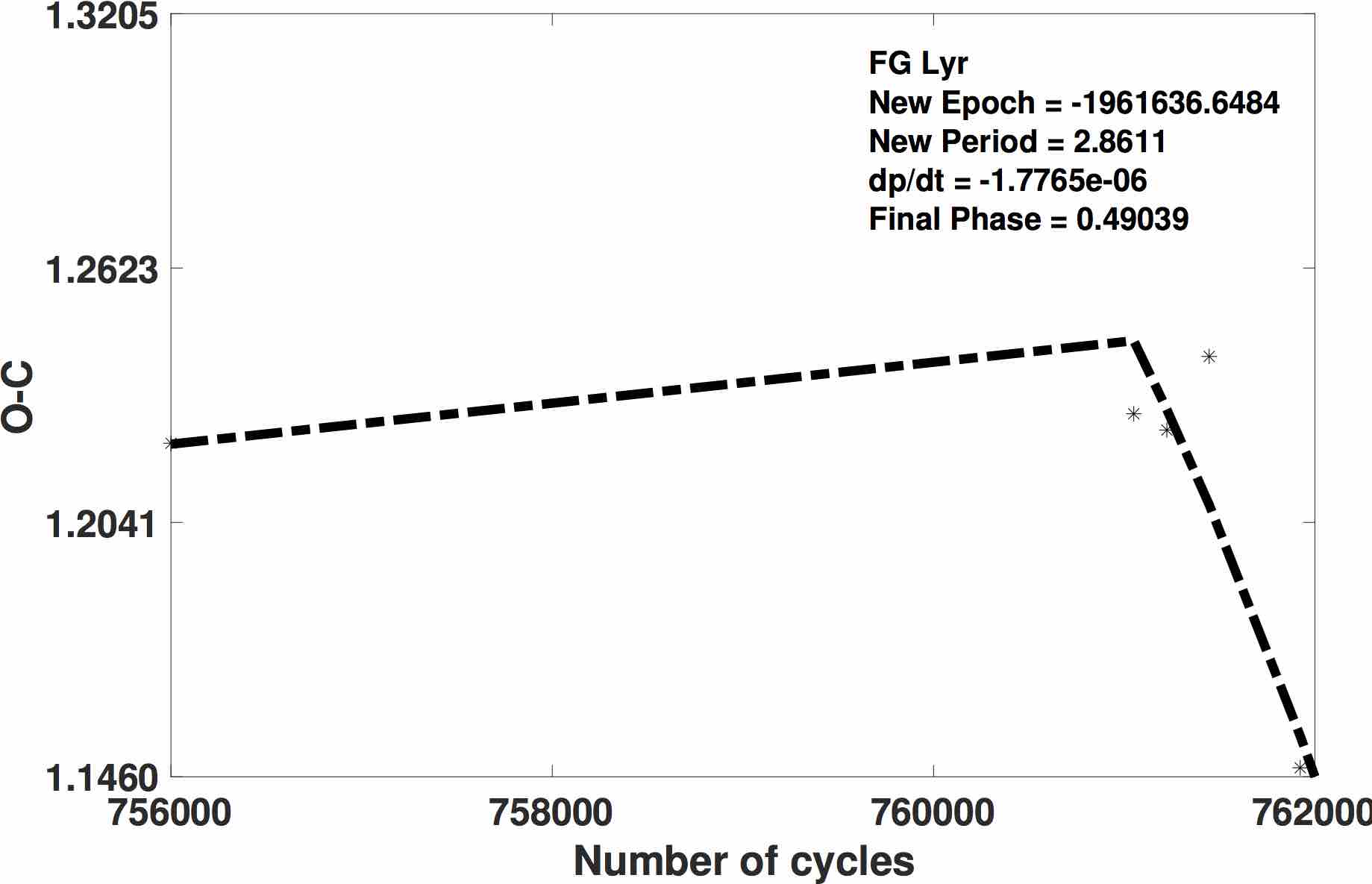}
\caption{O-C diagram of FG Lyr.}
\label{fig26}
\end{figure}

\begin{figure}[H]
\centering
\includegraphics[scale=0.11, angle=0 ]{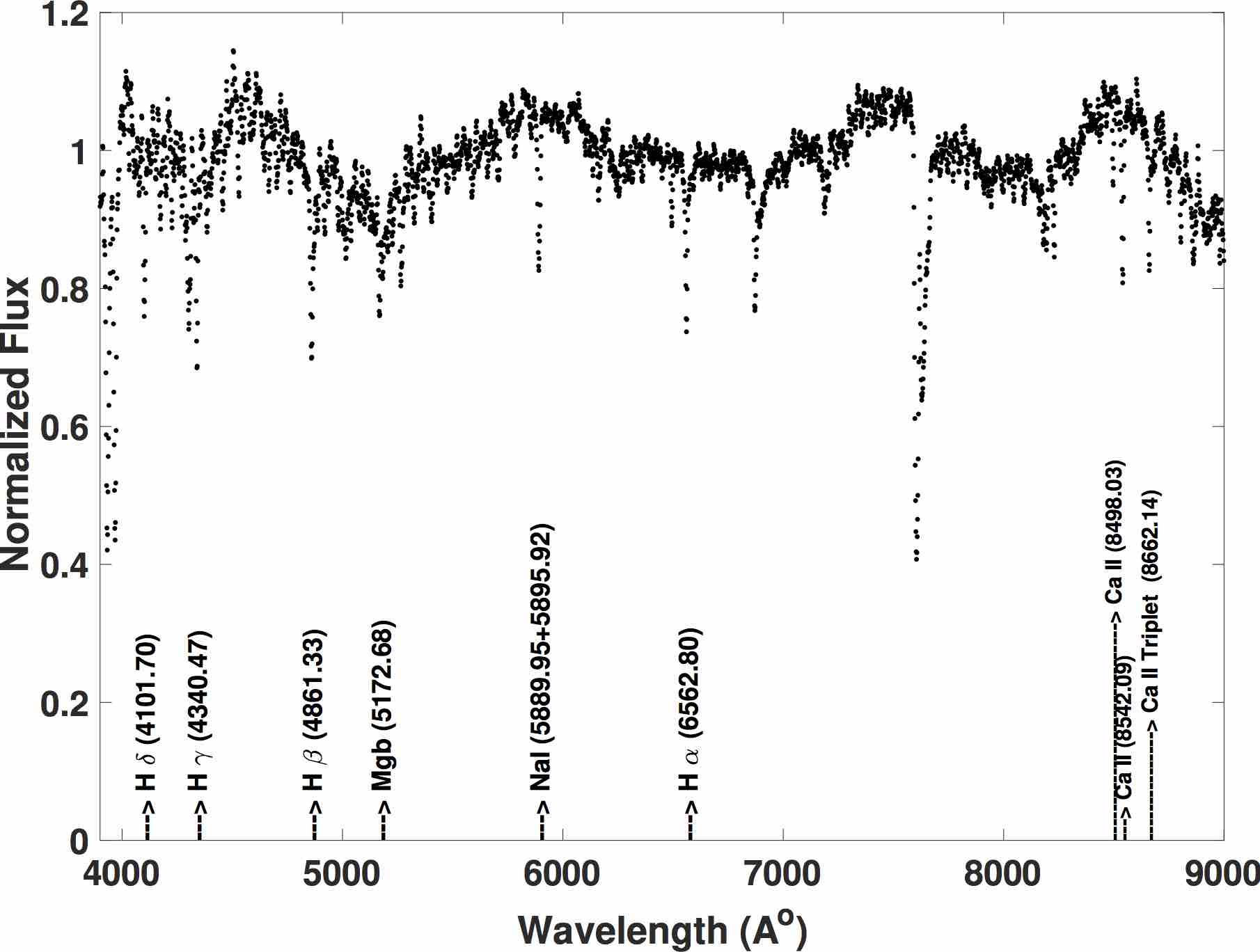}
\caption{Spectrum of FG Lyr.}
\label{fig28}
\end{figure}

FG Lyr (= AN 45.1930, V = 12.4) is an Algol-type eclipsing binary with an orbital period 2$^{d}$.87183 \cite{kreiner2004}. \cite{dworak1977} published minima from the Cracow observations of variable stars made during the years 1920-1950. Later its photoelectric minima were determined by \cite{agerer2001, hubscher2005}. No detailed study on the orbital period change of the system has been done so far. We present the first period study.  There are only six data points with 5 spread over 15 years and one point separated by 78 years in the literature. The best fit on the observed data shows a decreasing period and the O-C diagram represents the variation as shown in Figure \ref{fig26} with the rate of period decrease dp/dt  as -1.7765x10$^{-6}$ days/year. 
The new epoch derived from the present study is as follows 
HJD (Min I) = 2456770.916 + 2$^{d}$.86112 $\times$ E. 
A spectrum was obtained on April 23, 2014 for this variable at phase 0.0277 and the spectral lines identified are shown in Figure \ref{fig28} and equivalent widths are given in Table \ref{T2}. The best fit spectral model was selected (\cite{jacoby1984}) to determine the spectral class as G6 III, representing the spectral type of evolved secondary component.

\subsection{BZ Mon}
\begin{figure}[H]
\centering
\includegraphics[scale=0.11]{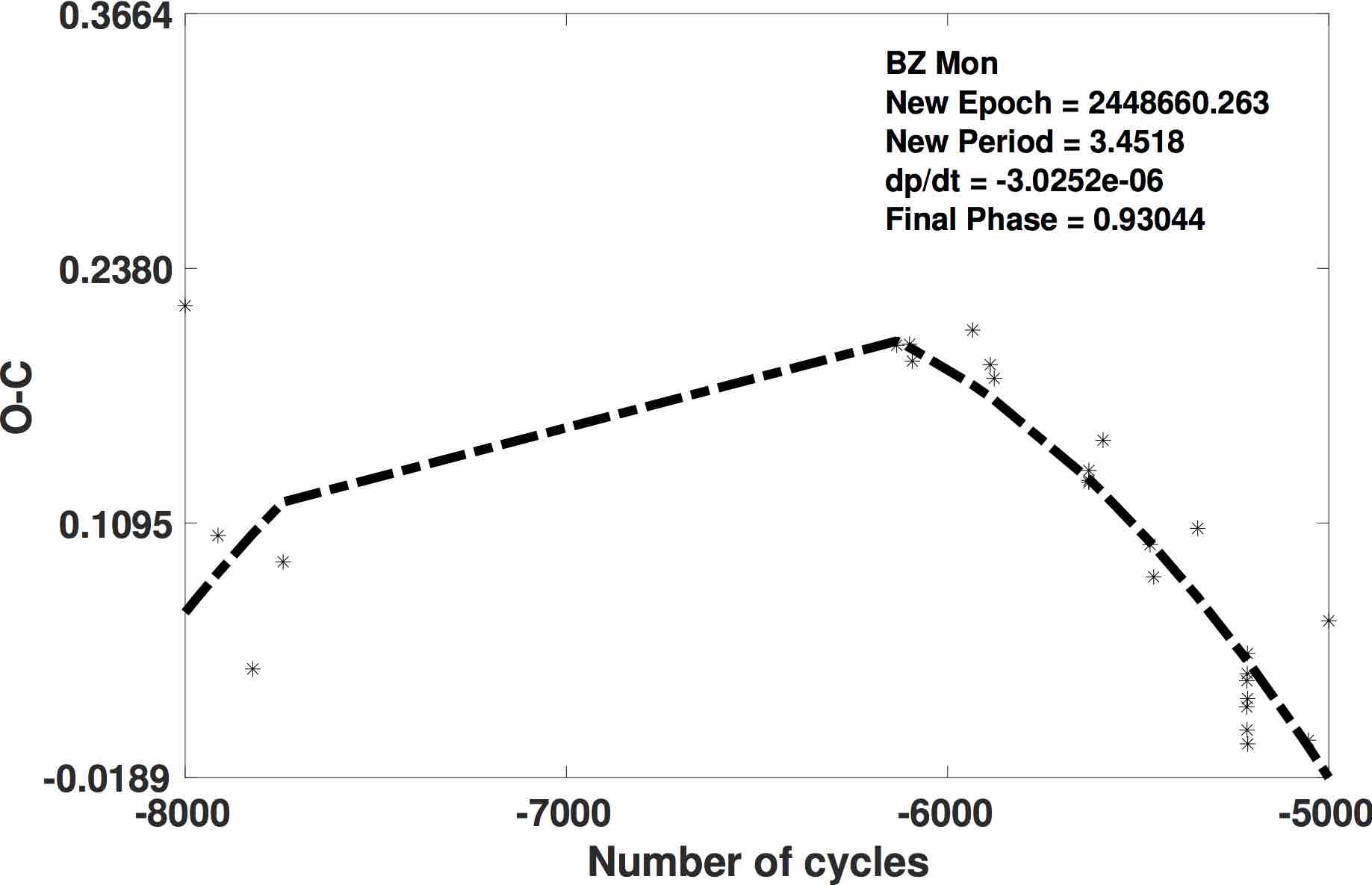}
\caption{O-C diagram of BZ Mon.}
\label{fig29}
\end{figure}
\begin{figure}[H]
\centering
\includegraphics[scale=0.11]{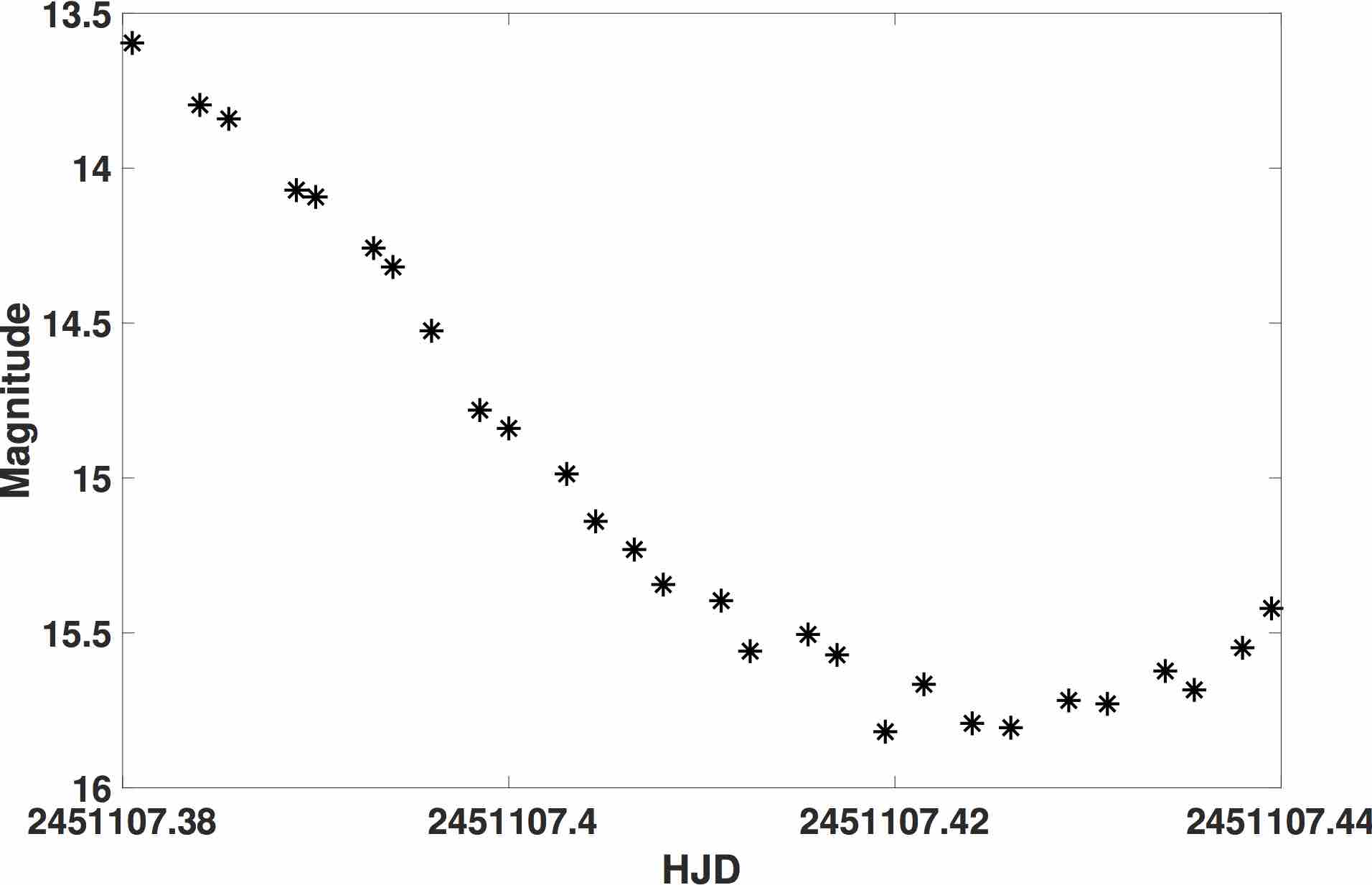}
\caption{Light curve of BZ Mon.}
\label{fig30}
\end{figure}

BZ Mon(= AN 63.1936, V = 12.10) is an Algol-type eclipsing binary with an orbital period 3$^{d}$.451721 (\cite{avvakumova2013}). The variable has a rich observational history and was first discovered by \cite{hoffmeister1936} and later basic data on light curve was given by many authors, \cite{ahnert1943, ahnert1947} with a major contribution from \cite{schaefer1980}. He has given an improved ephemeris included in GCVS Min I = HJD2443192.663+3$^{d}$.451804 \cite{zakirov2001} has conducted first photoelectric UBVR observations in 1998-99 and has determined new ephemeris, component spectral types as B5V \& G2III and also concluded that the orbital period can undergo small unexpected changes and its behavior can be represented as a parabola indicating progressive decrease in orbital period. The component parameters were determined to be M$_{h}$=4.9 M$_{\odot}$,M$_{c}$=2.7 M$_{\odot}$, R$_{h}$=3.25 R$_{\odot}$, R$_{c}$ = 5.2 R$_{\odot}$ \footnotemark and age is $\sim$ 3$\times$10$^{5}$years. Figure \ref{fig29} represents O-C variation obtained in the current study from quadratic ephemeris. The (O-C) plot shows the data points separated in two groups with no recorded observations of  minima timings for about 38 years between the two groups. The first group has a scarce set of 5 data points and the next group is recognizably spread out with 23 data points. The polynomial fit to the available data shows a decrease in the period which is in accordance with that given by \cite{zakirov2001} and the obtained dp/dt is -3.0252x10$^{-6}$ days/year. Figure \ref{fig30} represents the V light curve obtained from AAVSO. No additional work was carried out on this variable except for deriving new times of minima by \cite{zejda2004, diethelm2004, kreiner2004}. It is further listed in eclipsing binary catalogues by \cite{budding2004, malkov2006, avvakumova2013}.
\footnotetext{h - hotter component, c - cooler component}

\begin{subfigures}
\begin{figure}[H]
\centering
\includegraphics[scale=0.11, angle=0 ]{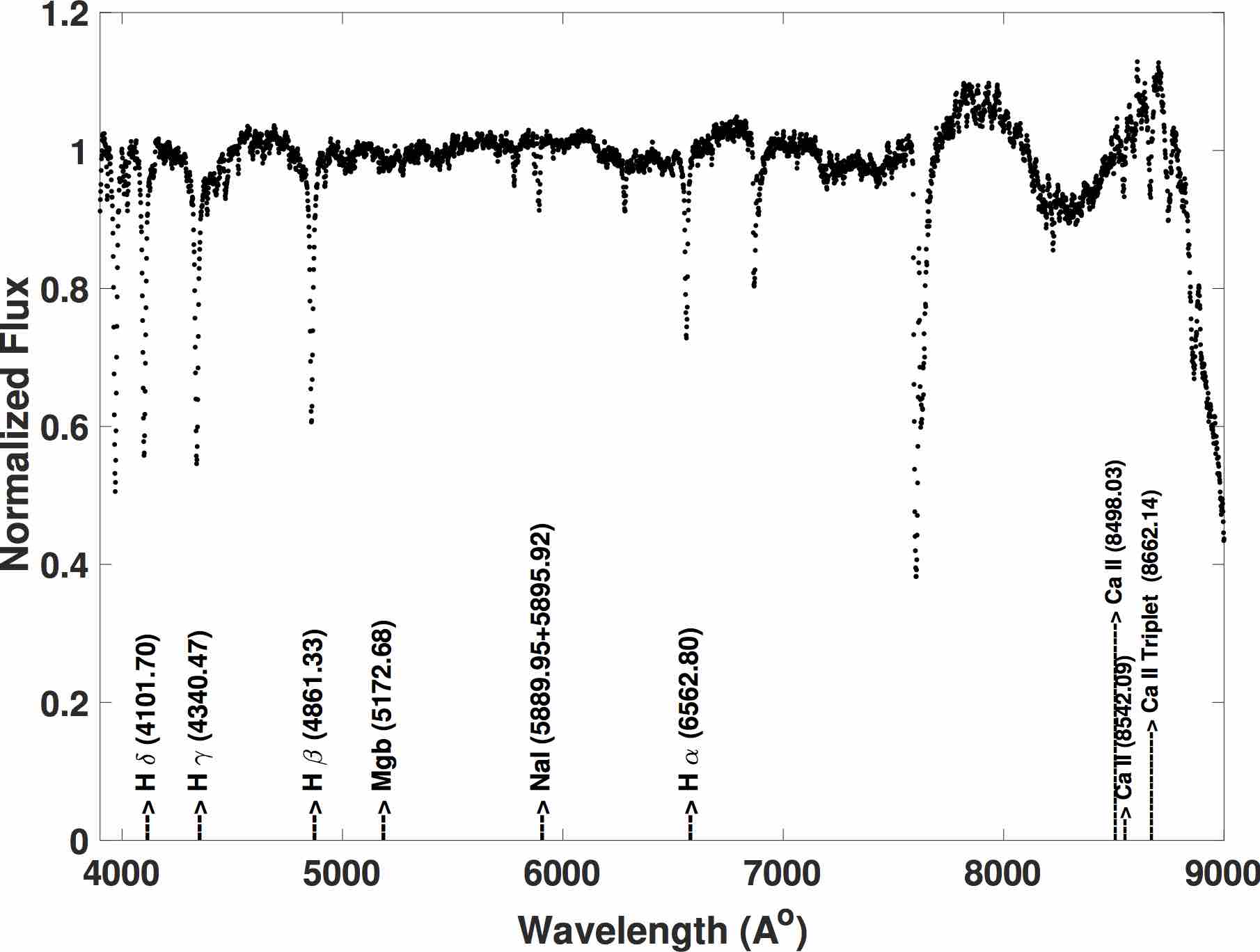}
\caption{Spectrum of BZ Mon (2013).}
\end{figure}
In the present study, the new ephemeris obtained is as follows 
HJD (Min I) = 2448660.263 + 3$^{d}$.451800 $\times$ E. \\
In the present work two spectra were obtained for this variable on Feb 20, 2013 and Feb 20, 2014 at phases 0.070 (closer to primary minima), 0.807 (closer to maxima) respectively. These phases were calculated through the new epoch derived from period study. The spectral lines identified are shown in Figure \ref{fig31} for the observed phases.There is an increase in the equivalent widths of Balmer lines near maxima compared to the spectrum obtained within the eclipse which shows spectral shallow due to filled-in absorption (as shown in Table \ref{T2}). This can be attributed to the model defined by \cite{zakirov2001} where the period decrease was explained to be the result of non conservative mass flow and mass loss by the system. \cite{avvakumova2013} have derived the spectral type to be B7 V. The best fit spectral model was selected (\cite{jacoby1984}) to determine the spectral class as A7 V, which has to be further validated.

\begin{figure}[H]
\centering
\includegraphics[scale=0.11, angle=0 ]{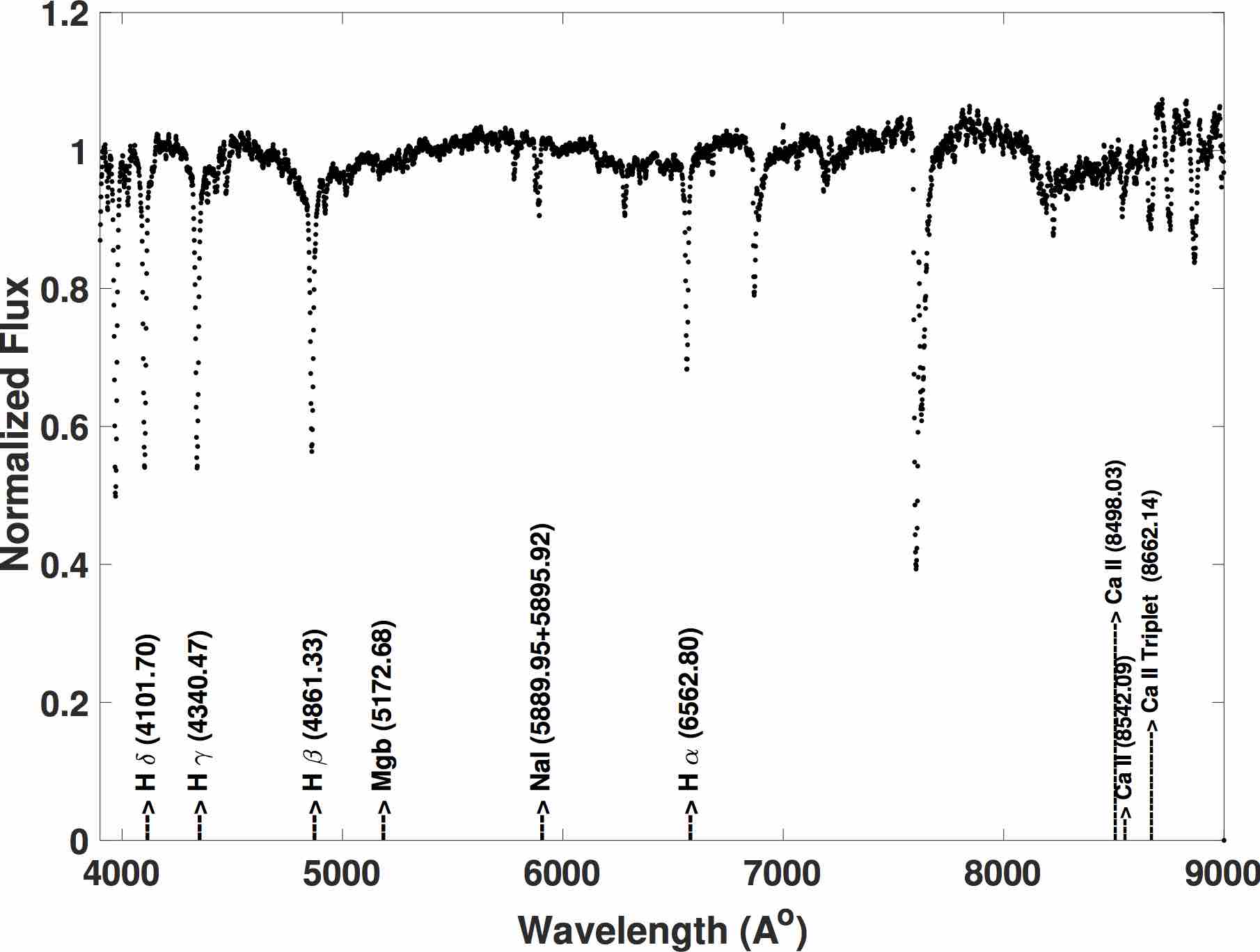}
\caption{Spectrum of BZ Mon (2014).}
\end{figure}
\label{fig31}
\end{subfigures}

\subsection{CH Mon}
\begin{figure}[H]
\centering
\includegraphics[scale=0.11, angle=0 ]{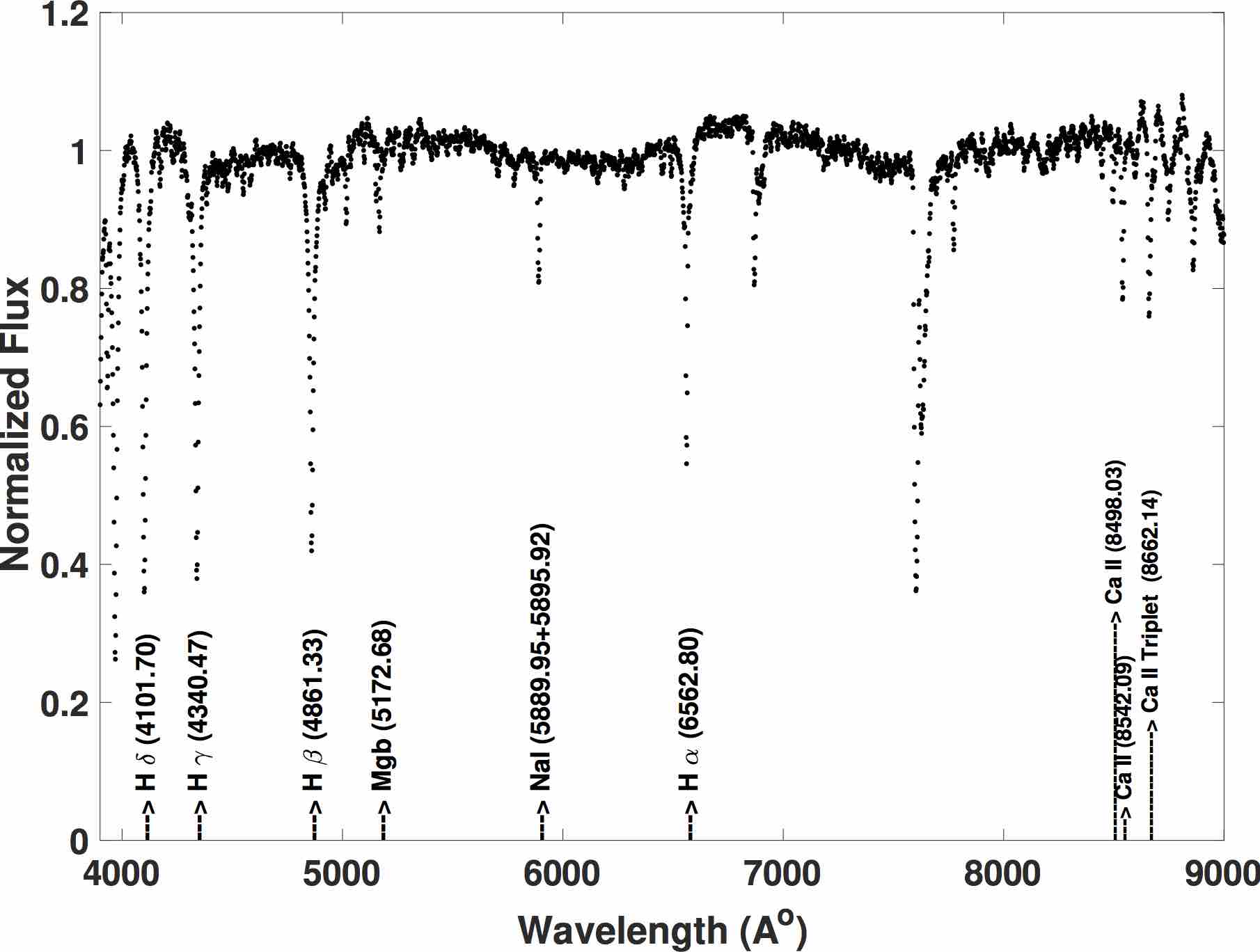}
\caption{Spectrum of CH Mon.}
\label{fig32}
\end{figure}

CH Mon(=AN 73.1936, GSC 00160-00530,B=13.1) was first reported as a variable by \cite{hoffmeister1936} and was further catalogued as an Algol type by \cite{ahnert1947, kinnunen2000, budding2004} and \cite{malkov2006} . The period of the variable was determined to be 6$^{d}$.922312 by \cite{avvakumova2013} and no further study on times of minima or period variation was carried out in the literature. For the first time we obtained a spectra for CH Mon on Nov 12, 2013 at phase 0.014. It is again one of the least studied Algol type binaries. The dominant spectral lines are shown in Figure \ref{fig32} and their equivalent widths calculated are as given in Table \ref{T2}.  The H$\alpha$ absorption profile is observed to be relatively less prominent than the other Balmer lines which could be due to fill in effect.The spectral type was determined to be A8 V (\cite{jacoby1984}) based on minimum res$^{2}$ corresponding to evolved secondary component.

\subsection{FW Mon}
\begin{figure}[H]
\centering
\includegraphics[scale=0.11, angle=0 ]{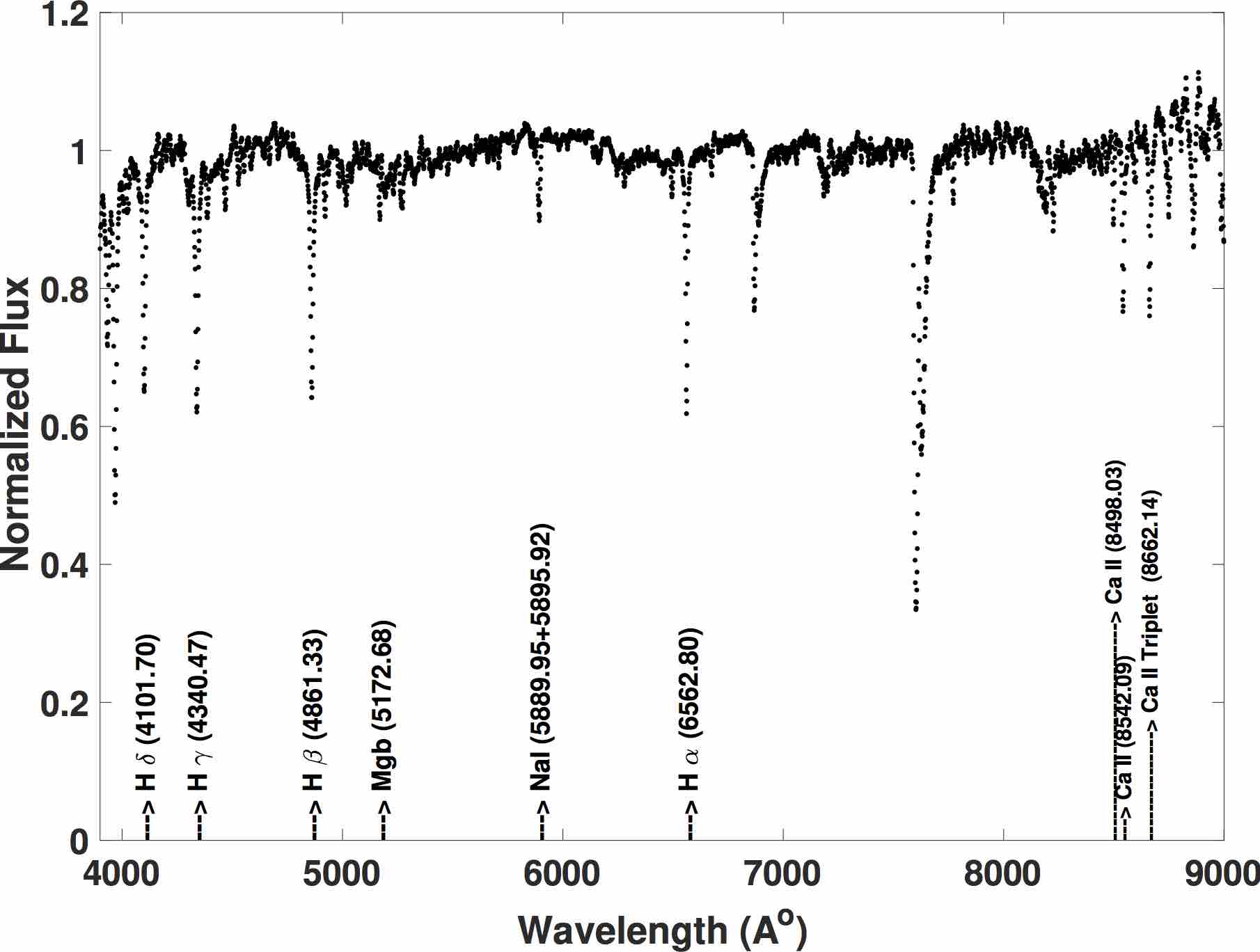}
\caption{Spectrum of FW Mon.}
\label{fig33}
\end{figure}
FW Mon (= GSC 04845-02526=4845-2526-1, V = 9.97) is an Algol-type eclipsing binary with an orbital period 3$^{d}$.8735900 (\cite{kukarkin1971}). \cite{chang1948} was the first to study the variability of FW Mon and report the spectral types of brighter and fainter components as B5 and F2. \cite{brancewicz1980} have presented physical parameters of FW Mon. \cite{srivastava1993} have done photoelectric observations and no period variation was found. Later on \cite{budding2004} has presented minima of FW Mon in the catalogue of Algol type binary stars. In the current study we performed period variation study for 15 data points spanning over 31 years and one point separated by 42 years. The data points show a large scatter but the best polynomial fit gives a decreasing period, however additional observations are required to understand the change clearly. The rate of decrease of period dp/dt obtained is -1.6204x10$^{-7}$ days/year and the new epoch obtained is HJD (Min I) = 2448630.672 + 3$^{d}$.873589 $\times$ E.  
One spectra was obtained on March 18, 2013 at phase 0.003 derived using the latest times of minima in the literature. Figure \ref{fig33} displays the identified spectral lines and the equivalent widths obtained are given in Table \ref{T2}. The spectral class of the variable is determined to be A9 V from (\cite{jacoby1984}) our spectral observation.

\subsection{HP Mon}

\begin{figure}[H]
\centering
\includegraphics[scale=0.11]{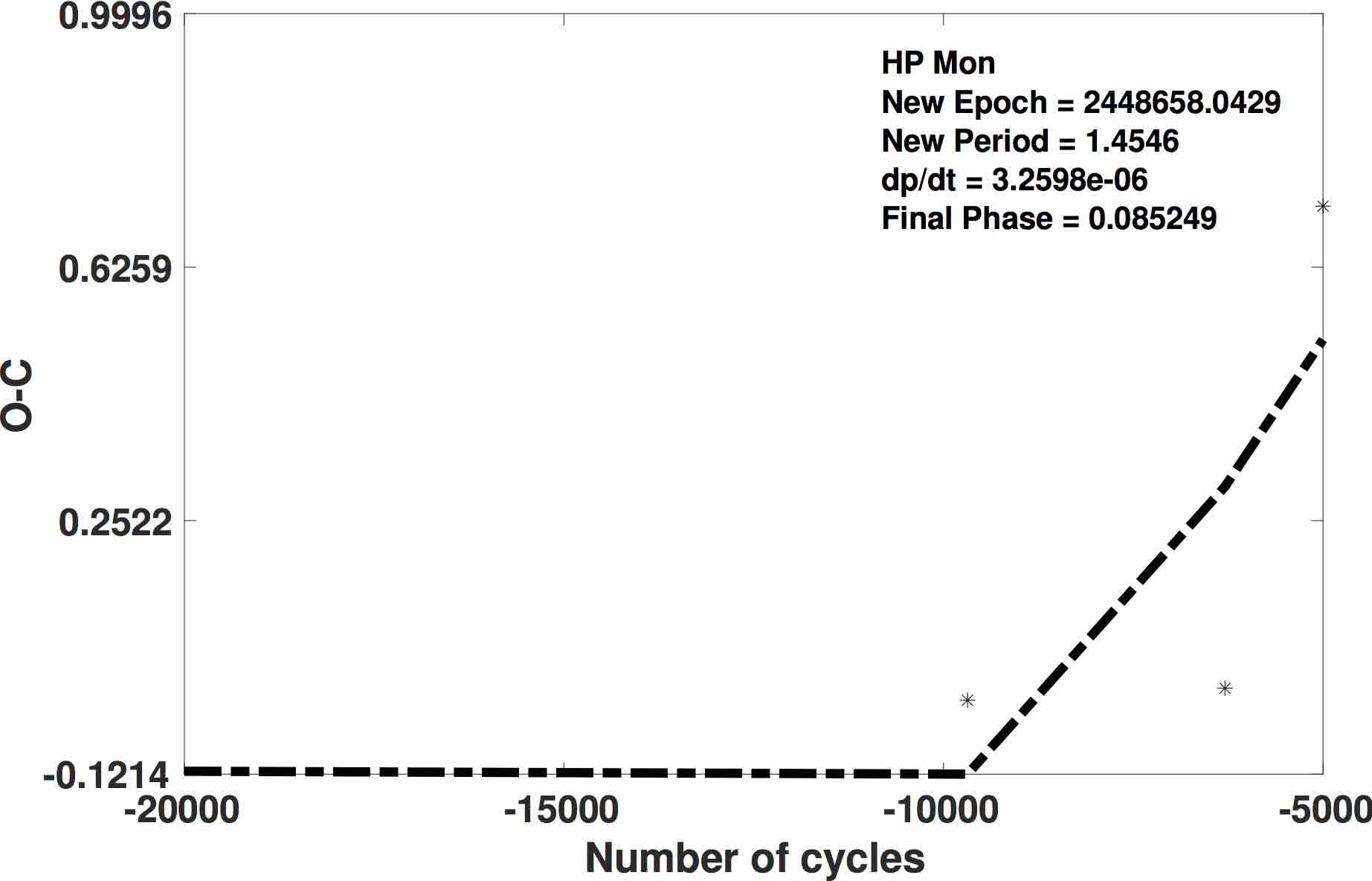}
\caption{O-C diagram of HP Mon.}
\label{fig34}
\end{figure}

\begin{subfigures}
\begin{figure}[H]
\centering
\includegraphics[scale=0.11, angle=0 ]{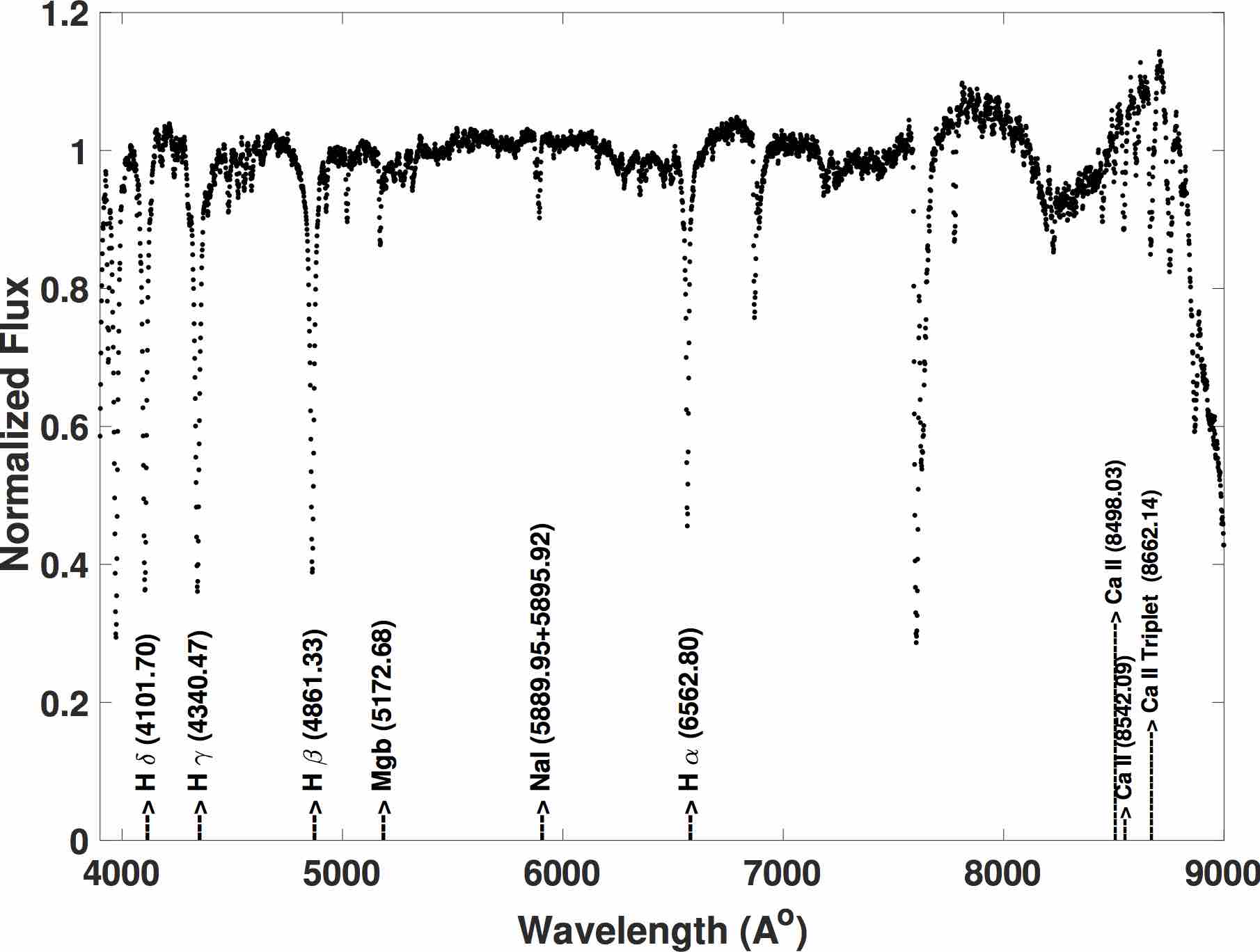}
\caption{Spectrum of HP Mon (2013).}
\end{figure}

HP Mon (= 2MASS J07103072-0533003) is an Algol-type eclipsing binary with an orbital period 1$^{d}$.45461. It was listed in the catalogue of Algol type binary by \cite{budding2004} and \cite{malkov2006}.  Times of minima obtained through photoelectric observations were presented by \cite{krajci2006}and \cite{hubscher2012}. No period variation study was carried out so far, we present the first period study. This is another least studied Algol with only four data points available and spread across 70 years. The best fit curve shows increase in the period. The O-C variation is shown in Figure \ref{fig34} and dp/dt obtained is 3.2598x10$^{-6}$ days/year.
The new epoch obtained in the current study is as follows 
HJD (Min I) = 2456344.174 + 1$^{d}$.454641 $\times$ E. 
Two spectra were obtained for this variable on Feb 20, 2013 and Feb 20, 2014 at phases 0.342, 0.261 respectively for the first time. The equivalent widths of the prominent spectral lines (Figure \ref{fig35}) are tabulated in Table \ref{T2}.  The best fit spectral model was selected from (\cite{jacoby1984}) and spectral class is determined to be A5 V.
\begin{figure}[H]
\centering
\includegraphics[scale=0.11, angle=0 ]{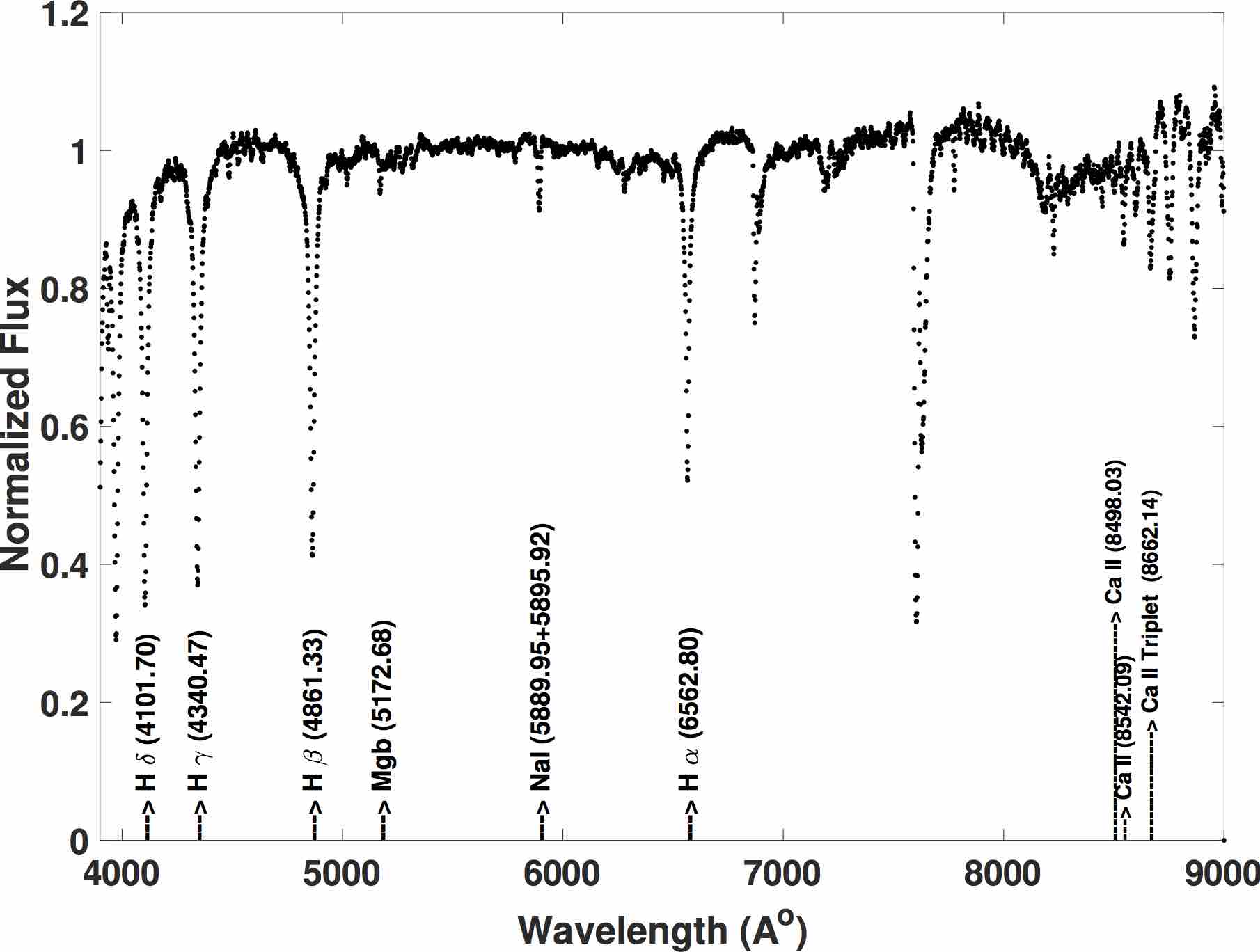}
\caption{Spectrum of HP Mon (2014).}
\end{figure}
\label{fig35}
\end{subfigures}

\subsection{RV Oph}
\begin{figure}[H]
\centering
\includegraphics[scale=0.11]{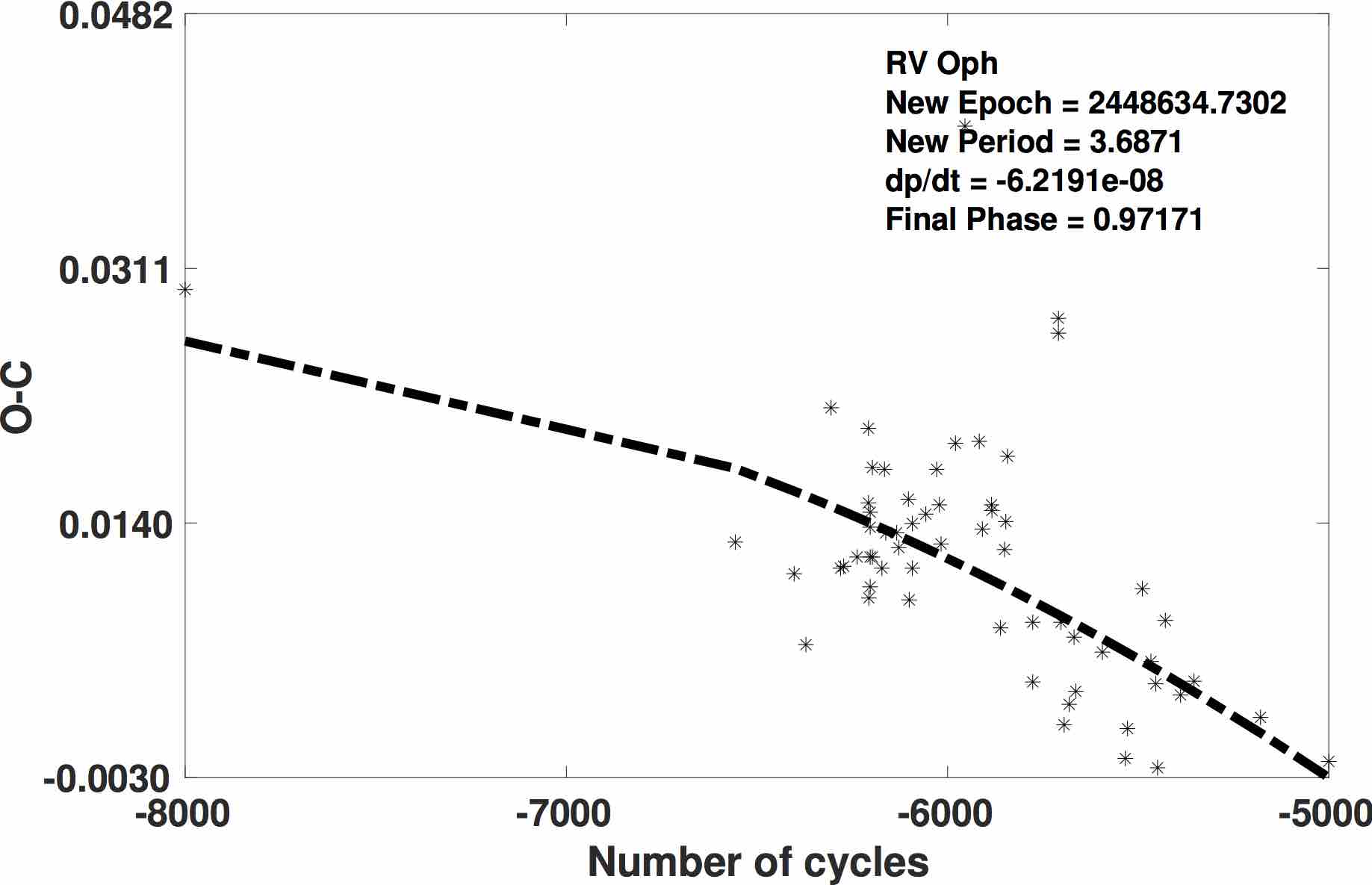}
\caption{O-C diagram of RV Oph.}
\label{fig36}
\end{figure}

\begin{figure}[H]
\centering
\includegraphics[scale=0.11, angle=0 ]{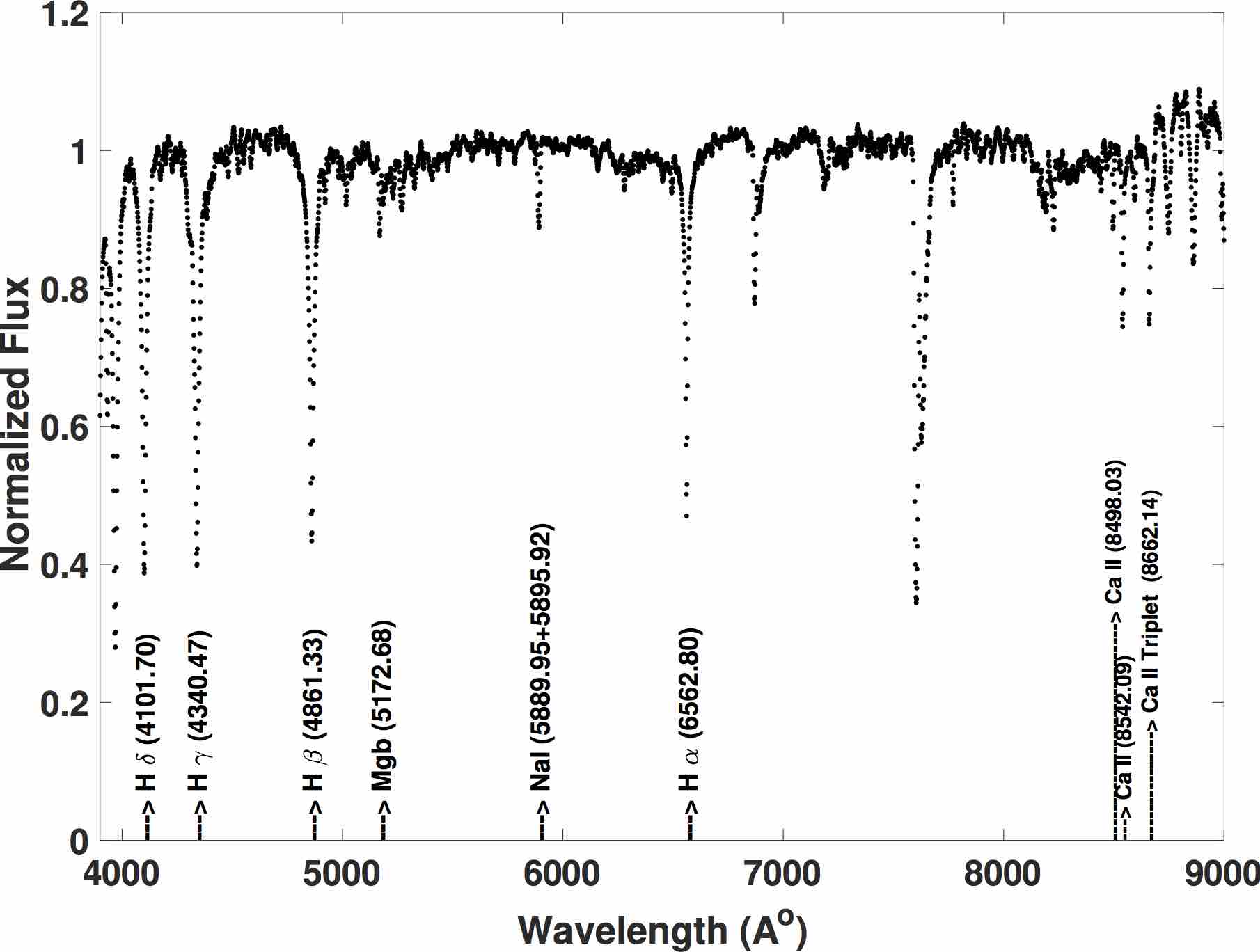}
\caption{Spectrum of RV Oph.}
\label{fig37}
\end{figure}

RV Oph (= 2MASS J16490246-1927524, V = 8.00) is an Algol-type eclipsing binary with an orbital period 3$^{d}$ .687125. The variability of this star was first discovered by Mrs.Fleming (\cite{pickering1904}) and has been further observed by \cite{dugan1916}. \cite{walter1970} carried out photoelectric observations of the variable in B and V bands. Compared to the previous observations the asymmetry was reduced and a model of gas stream was given explaining characteristics of light curves. \cite{mezzetti1980} presented revised photometric light curves of RV Oph. From the photometric studies by \cite{liao2010} the light curve was found to be distinctly asymmetrical near primary minimum and irregularities in the phases out of eclipses were observed indicating influences of gas streams. The (O-C) has been plotted for 64 data points spanning over 45 years and one point separated by about 40 years.The data points show a large scatter but the best fit shows a decreasing period in Figure \ref{fig36} with dp/dt = -6.2191x10$^{-8}$ days/year. The new epoch deduced is HJD (Min I) = 2448634.730 + 3$^{d}$.6871 $\times$ E. No spectroscopic observations were done till date. Thus, the first spectrum of the variable with a single spectrum observed on March 18, 2013 at phase 0.324 as derived from latest epoch in the literature. The spectral lines identified are given in Figure \ref{fig37} and the equivalent widths are given in Table \ref{T2}. It is observed that there is prominent fill in effect in H$\alpha$ line when compared to other Balmer lines. \cite{kreiner1971} has given the spectral type of the variable as K0 whereas in the current study the best fit spectral model was selected from \cite{jacoby1984} to determine the spectral class as F0 V which can be validated by further observations.  

\subsection{FH Ori}

\begin{figure}[H]
\centering
\includegraphics[scale=0.11, angle=0 ]{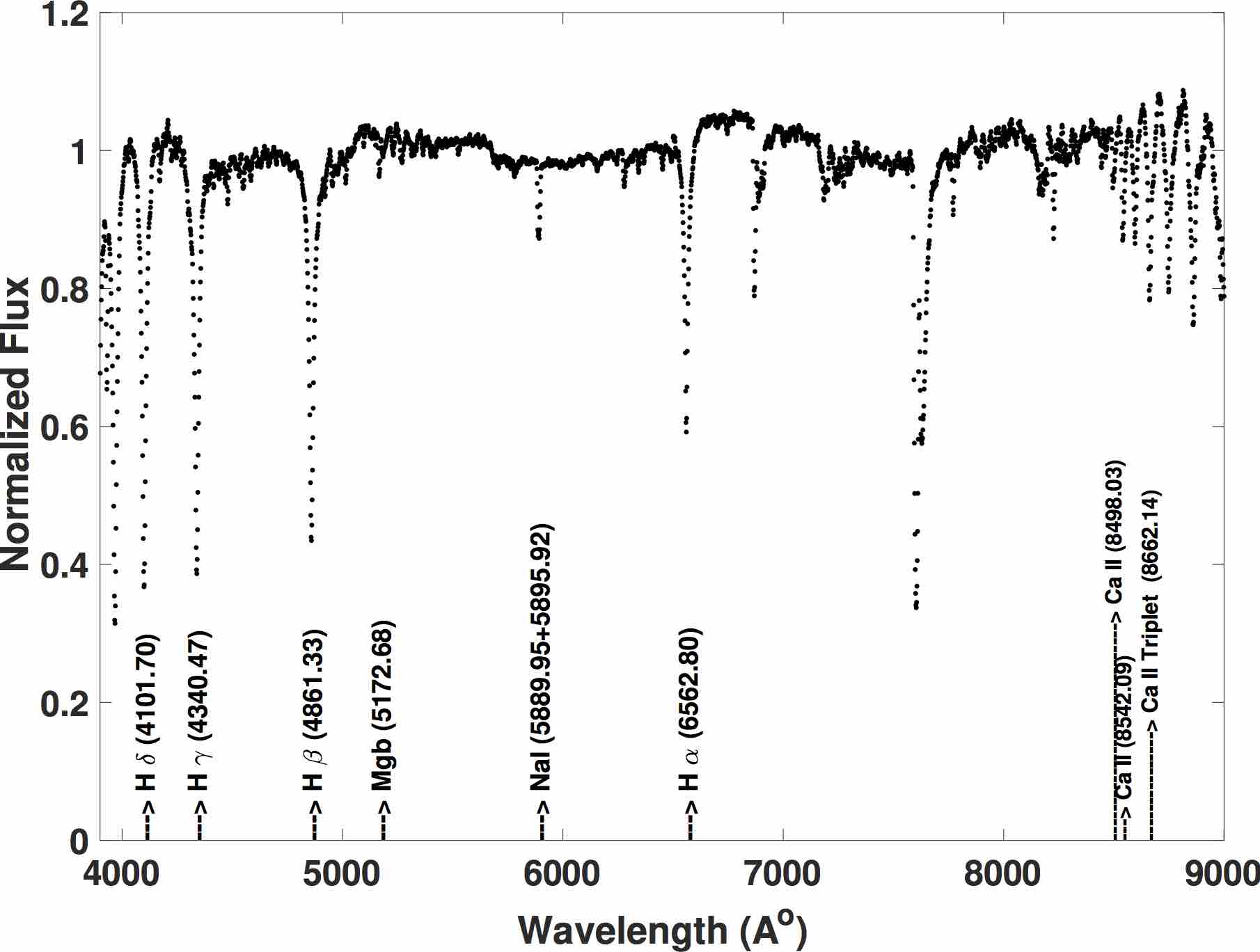}
\caption{Spectrum of FH Ori.}
\label{fig38}
\end{figure}

FH Ori(=GSC 00109-02559,TYC 109-2559-1,V=11.37) is a well studied Algol system with period of 2$^{d}$.151110 (\cite{drake2014}). The variability of FH Ori was first detected by \cite{hoffmeister1929, guthnick1934} and the first photoelectric observations were done by \cite{zakirov1994} showing that FH Ori is a typical EA-type system. Many times of minima have been published (\cite{kreiner1971, dworak1977, agerer1998, qian2001}) to study period variations which indicated secular decrease in the orbital period. This was attributed to variable magnetic coupling and gravity coupling between the two components. They also determined that it is a variable with apsidal motion with high relative velocity and the orbital period is accounted to mass transfer from primary to secondary and spectroscopic studies were recommended for detailed results.  We present first spectral study showing dominant profiles. The spectrum was obtained on Oct 14, 2013 at phase 0.2491. The prominent spectral lines are given shown in Figure \ref{fig38} and their equivalent widths in Table \ref{T2}. The spectral type determined by \cite{avvakumova2013} is A1V and by using the (\cite{jacoby1984}) stellar library we determined it to be A2 V.

\subsection{Z Ori}

\begin{figure}[H]
\centering
\includegraphics[scale=0.11, angle=0 ]{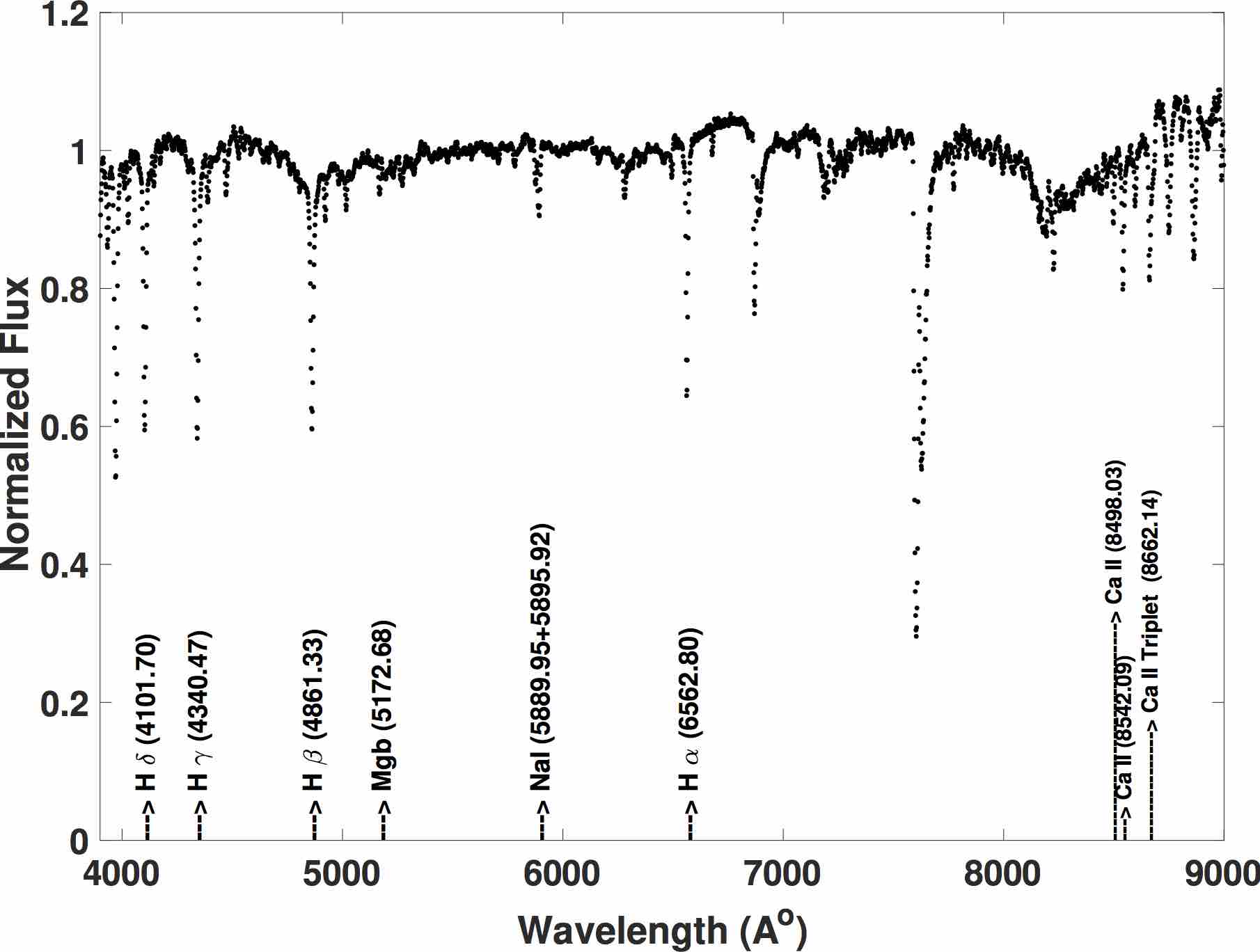}
\caption{Spectrum of Z Ori.}
\label{fig39}
\end{figure}

Z Ori (= GSC 00728-01172, TYC 728-1172-1, V=9.98) is one of the least studied Algol type binary. Its variability was first given by \cite{gaposchkin1932} and a spectrographic observations were carried out by \cite{struve1947} to determine the spectral type. The latest period available in the literature is 5$^d$.203265 (\cite{avvakumova2013}). \cite{crawford1955} categorized it to be a binary with subgiant secondary and spectral type as B5. Study for period variation was carried out by \cite{prikhod1962} who showed that the variation is cyclic. Very few minima are available in the literature to do period variation studies. One spectrum of Z Ori was obtained on Mar 21, 2014 at phase 0.9913 obtained from latest epoch available in the literature. The prominent spectral lines in the spectrum are shown in Figure \ref{fig39} and their equivalent widths are given in Table \ref{T2}. The spectrum displays absorption profiles of all the Balmer lines which could be due to evolved secondary component as seen from the phase. H$\alpha$ line is weaker relative to other Balmer lines.

\begin{figure}[H]
\centering
\includegraphics[scale=0.11, angle=0 ]{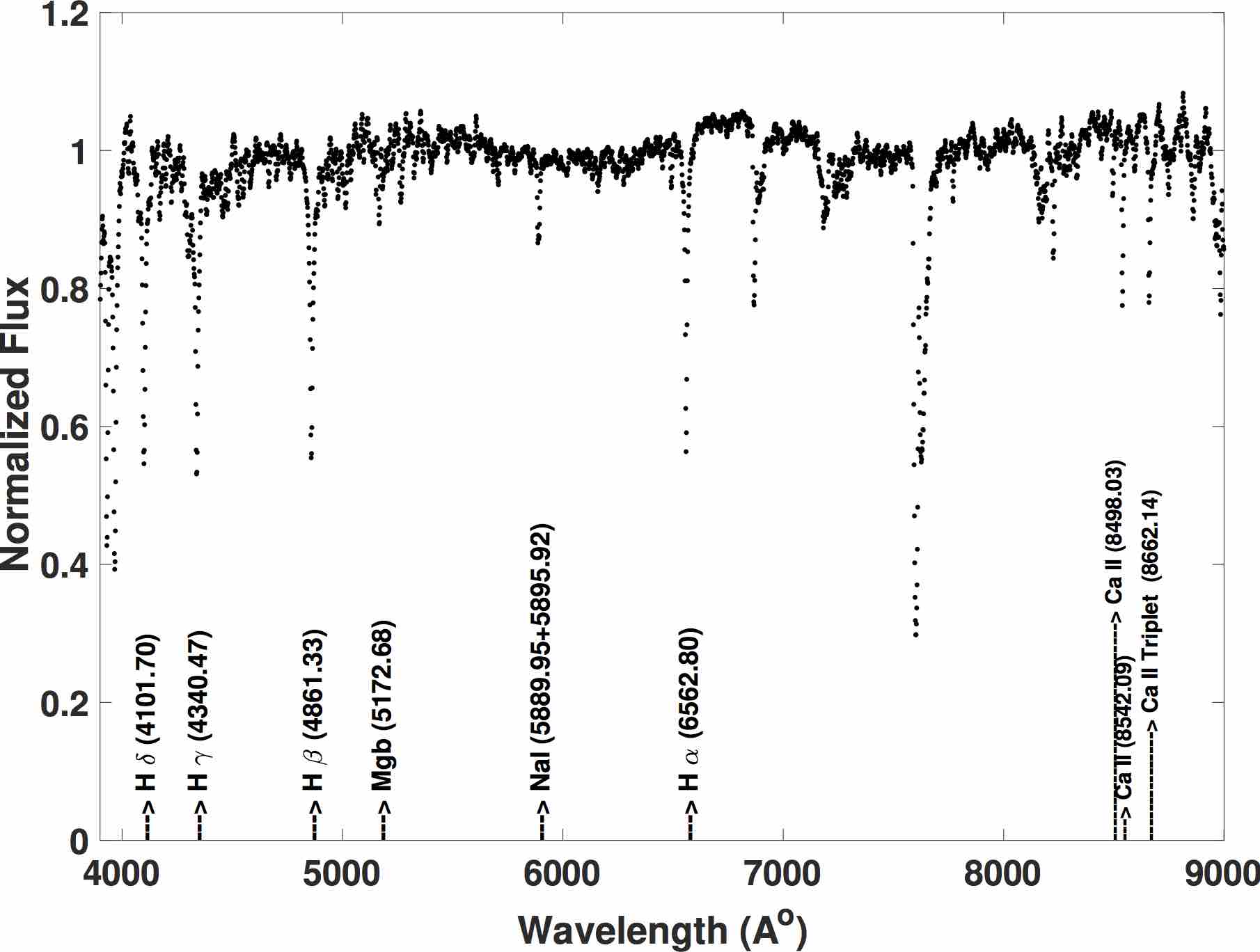}
\caption{Spectrum of V0640 Ori.}
\label{fig40}
\end{figure}
\subsection{V0640 Ori}

V0640 Ori (= GSC 05348-00080, TYC 5348-80-1, V=10.94) is one of the least studied Algol with a period of 2$^{d}$.020740 (\cite{avvakumova2013}).  It was catalogued as an Algol type binary by \cite{budding2004} and \cite{malkov2006}. Times of minima were given by many authors (\cite{diethelm2003, diethelm2004, dvorak2004, dvorak2005} and \cite{locher2005}). The times of minima in the literature are very few to carry out O-C studies. We obtained one spectrum for V0640 Ori on Oct 13, 2013 at phase 0.013 calculated with latest epoch in the literature.  The dominant spectral lines are given in Figure \ref{fig40} and equivalent widths are given in Table \ref{T2}. The spectrum displays absorption profiles of all the Balmer lines which could be due to evolved secondary as seen from the phase. H$\delta$ line shows weaker absorption profile relative to other Balmer lines. The best fit spectral model was found to be (\cite{jacoby1984}) A3 III.

\subsection{CK Per}
\begin{subfigures}
\begin{figure}[H]
\centering
\includegraphics[scale=0.11, angle=0 ]{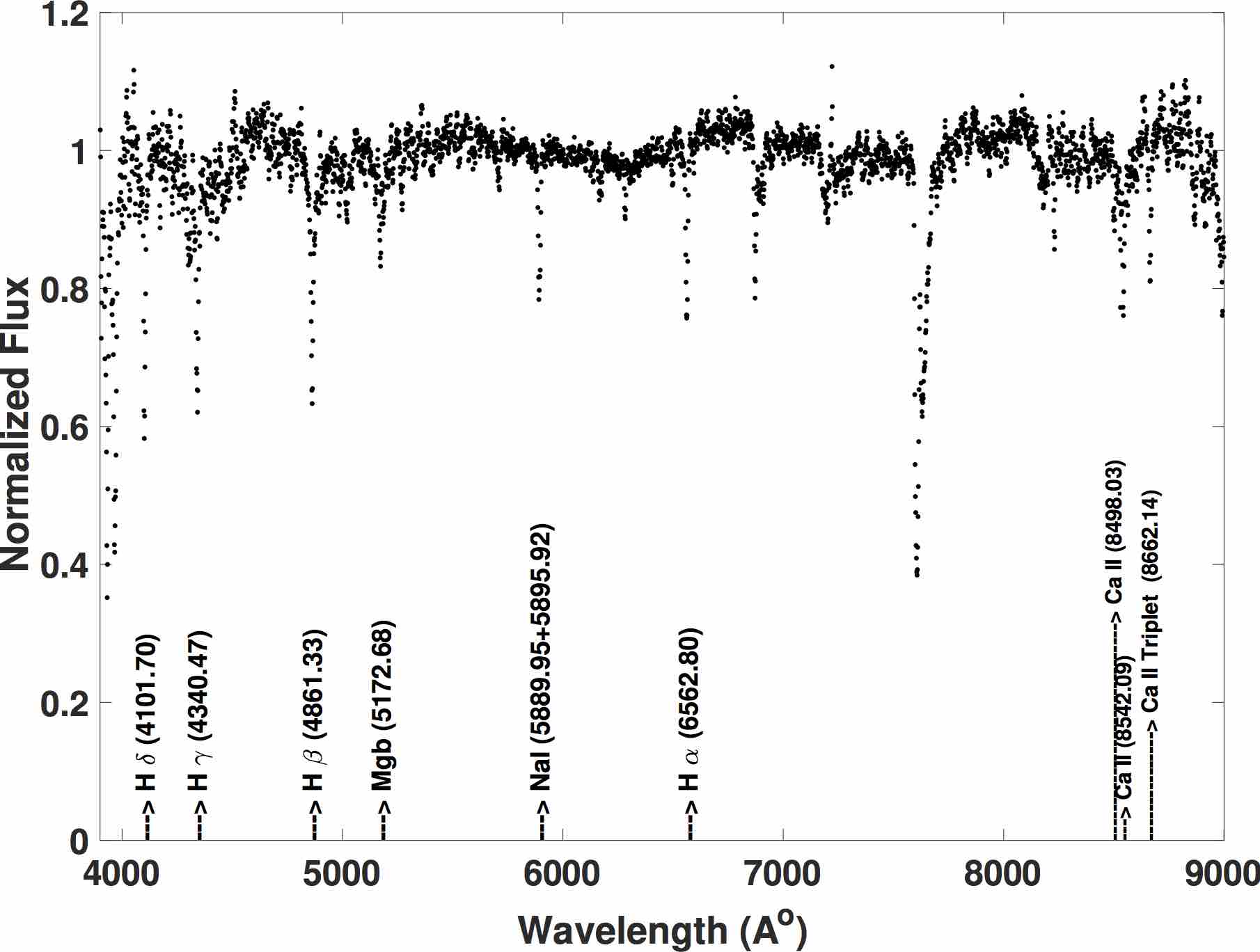}
\caption{Spectrum of CK Per (Oct,2013).}
\end{figure}

\begin{figure}[H]
\centering
\includegraphics[scale=0.11, angle=0 ]{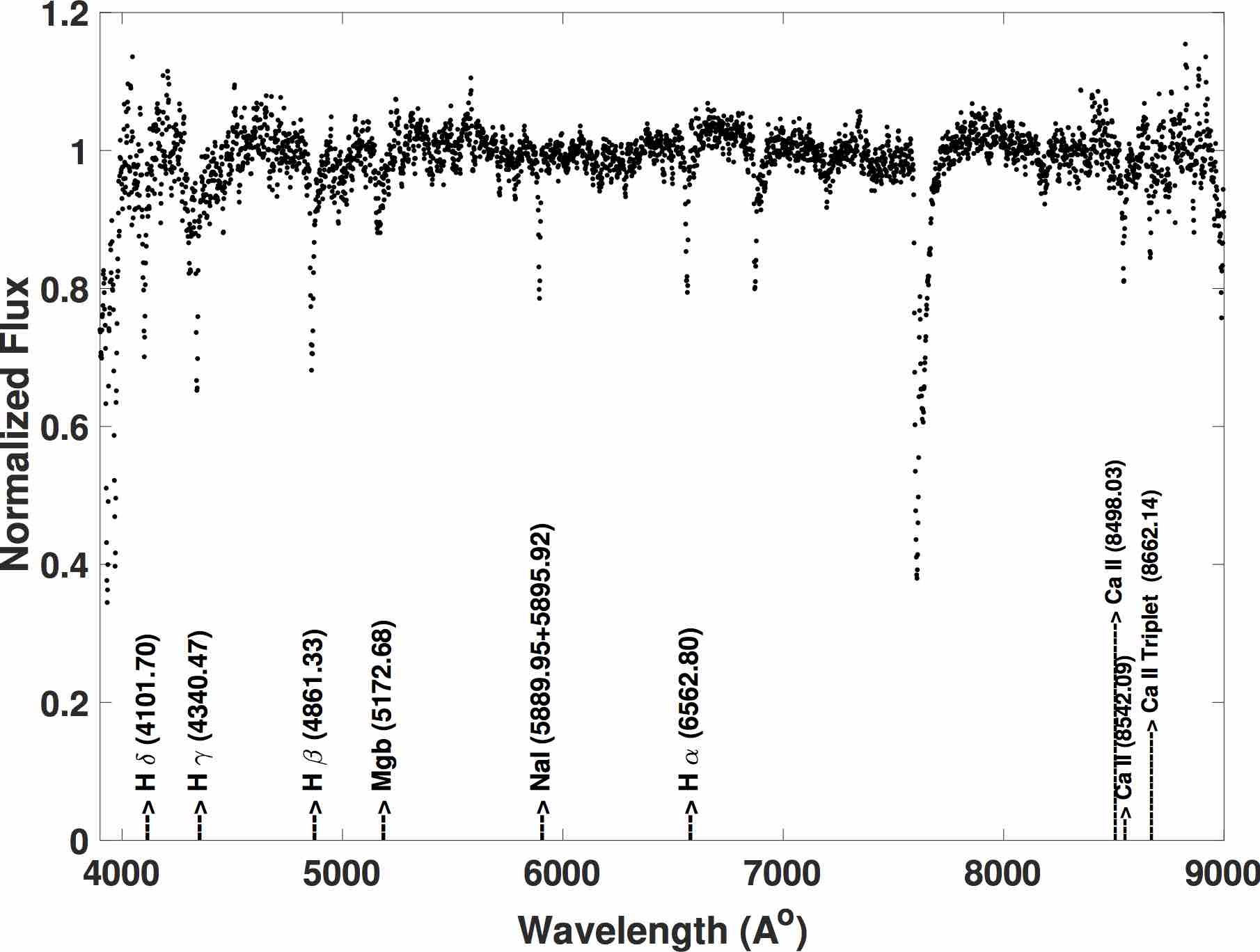}
\caption{Spectrum of CK Per (Nov,2013).}
\label{fig41}
\end{figure}
\end{subfigures}

CK Per (AN 10.1940,2MASS J02050675+5610247, V=15.2) is one of the least studied Algol system. It was catalogued as a variable star by \cite{ahnert1947}. The period of the variable is 2$^{d}$.372800 (\cite{kukarkin1971}) and was further catalogued as an Algol type by \cite{budding2004, malkov2006}. Two spectra of CK Per were obtained on Oct 13 \& Nov 20, 2013 at phases 0.9948 \& 0.015 respectively for the first time.  The dominant spectral lines are in Figure \ref{fig41} and equivalent widths are listed in Table \ref{T2}. The H $\delta$ line shows stronger absorption profile relative to other Balmer lines and also distinctly stronger absorption profile of Na line in one spectrum obtained in  Oct 2013 where as it shows relatively weaker absorption profile in Nov 2013 near primary minima. This variable will be an interesting object for period study. Using (\cite{jacoby1984}) stellar library, the spectral type is determined as F8 V.

\subsection{Z Per}
\begin{figure}[H]
\centering
\includegraphics[scale=0.11]{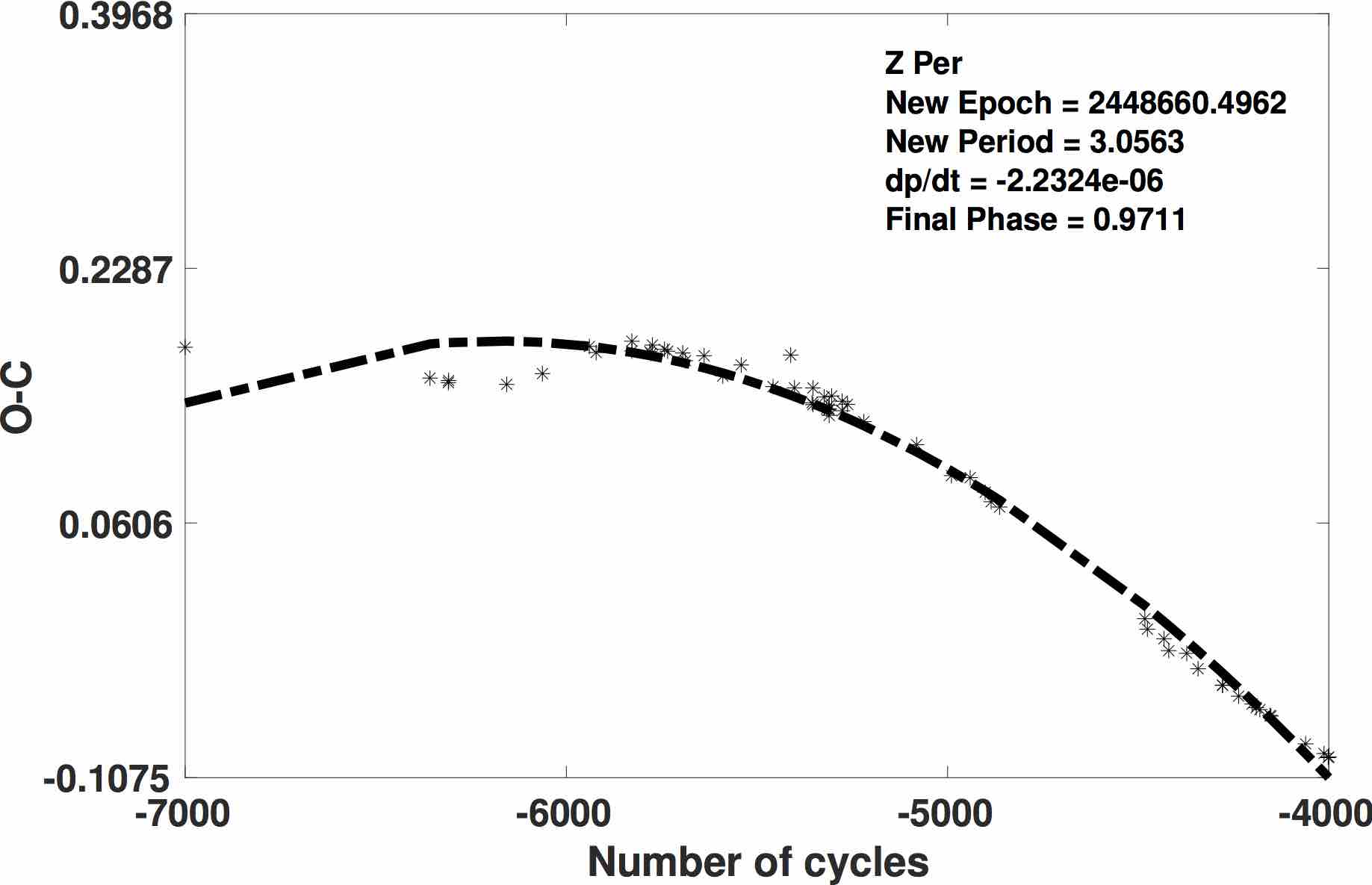}
\caption{O-C diagram of Z Per.}
\label{fig42}
\end{figure}

\begin{figure}[H]
\centering
\includegraphics[scale=0.11]{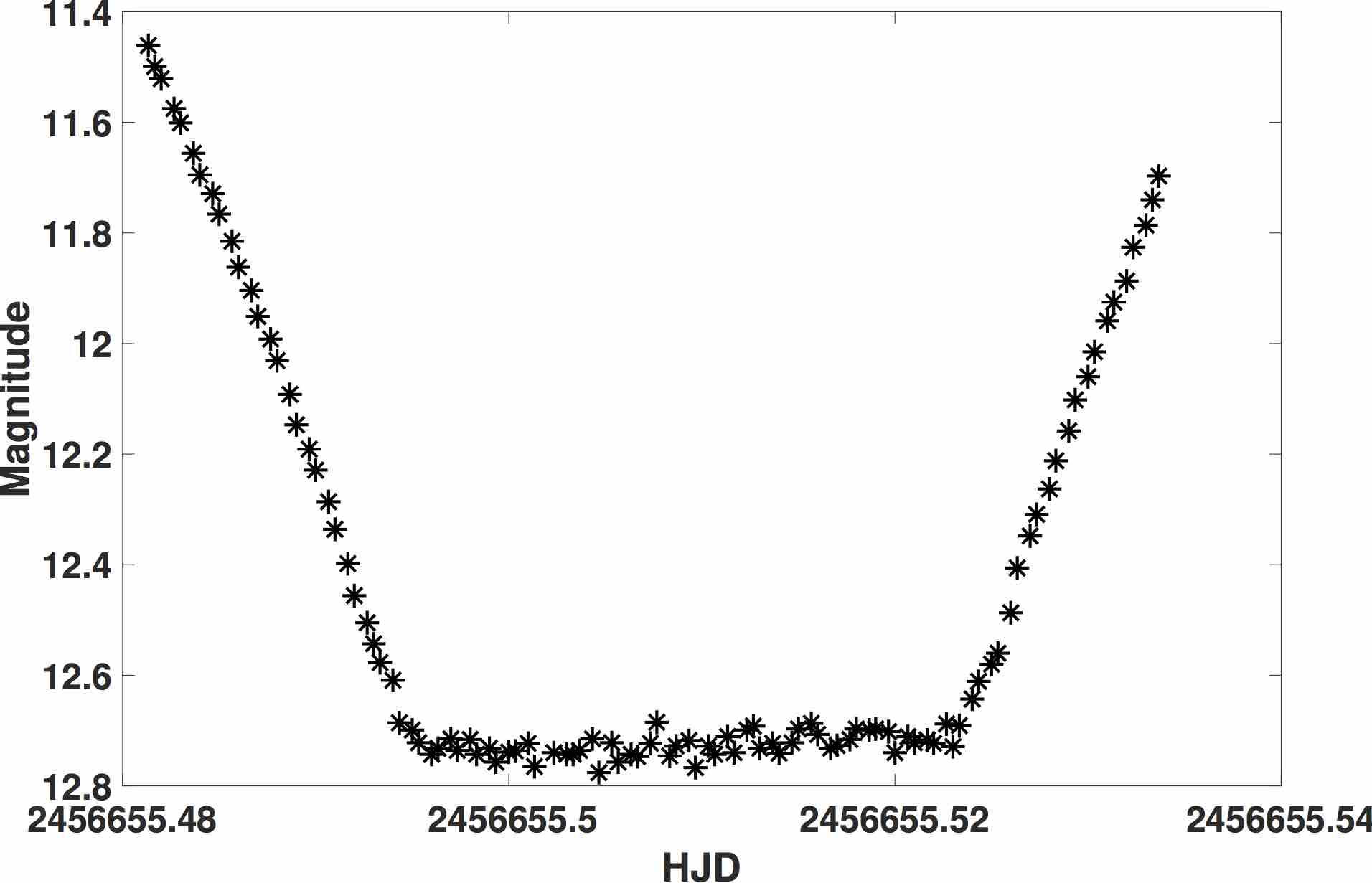}
\caption{Light curve of Z Per.}
\label{fig43}
\end{figure}

Z Per (= GSC 04147-01115 = TYC 4147-1115-1, V = 10.60) is an Algol-type eclipsing binary with an orbital period of 3$^{d}$.056306.  The orbital period of this Algol has been given by many authors (\cite{wood1963, mallama1987}). These are about 61 ToM spread about 54 years and the best fit is distinctly representing a decreasing period. Many ToM were given in the literature and the variable was observed to have a decreasing period since the first AAVSO observations (\cite{kreiner1971, samolyk1997, samolyk2010}). Along with the secular decrease in orbital period, several sudden jumps were noticed in the orbital period of Z Per within a time interval of 94 years i.e, from early 1901 and late 1995. A sinusoidal variation is evident (Figure \ref{fig42}) and could be an unseen third star in the system with dp/dt = -2.2324x10$^{-6}$ days/year. The O-C residuals obtained in the current study is as follows HJD (Min I) = 2456343.109 + 3$^{d}$.056283 $\times$ E. The light curve obtained from the data available in the literature is shown in Figure \ref{fig43} The spectral type of this variable is given as A0+G2 IV (\cite{budding2004, hoffman2006}).

\begin{subfigures}
 \begin{figure}[H]
\centering
\includegraphics[scale=0.11, angle=0 ]{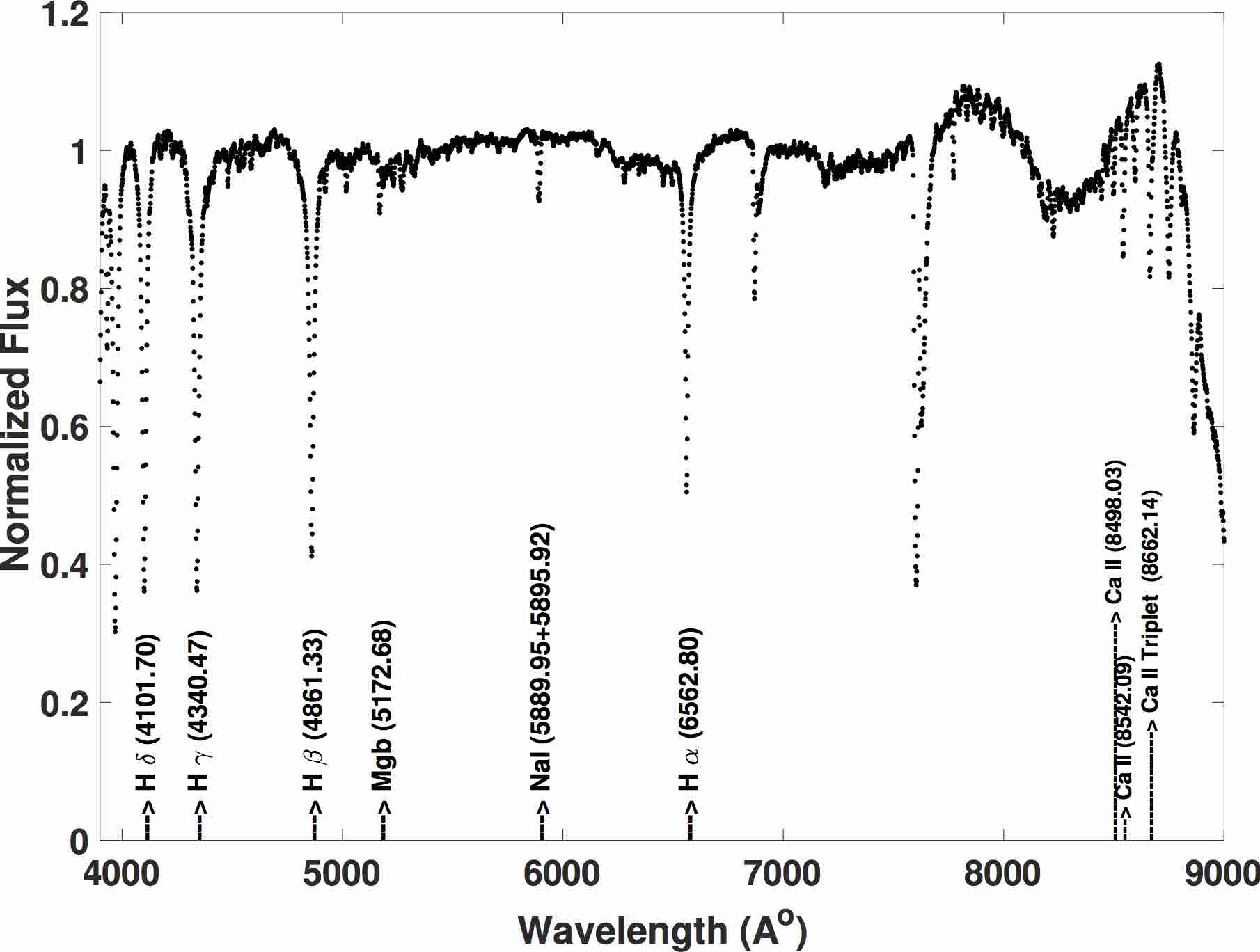}
\caption{Spectrum of Z Per (Feb, 2013).}
\end{figure}

\begin{figure}[H]
\centering
\includegraphics[scale=0.11, angle=0 ]{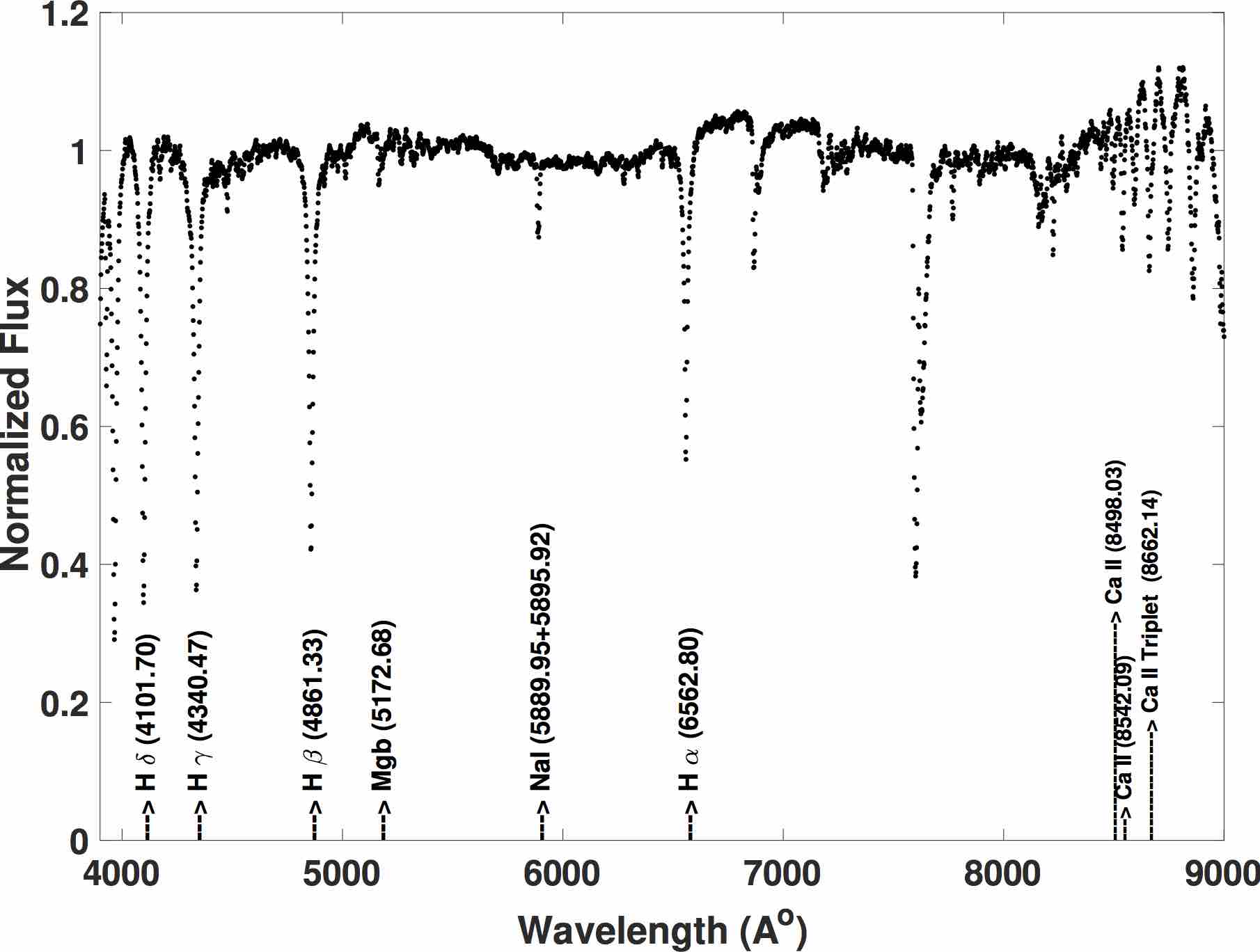}
\caption{Spectrum of Z Per (Oct, 2013).}
\end{figure}

We present the first spectral study of Z Per in this study. Three spectra were obtained for Z Per on Feb 20, 2013 at phase 0.015, Oct 13, 2013 at phase 0.016 and Feb 20, 2014 at phase 0.443. The dominant spectral lines obtained are shown in Figure \ref{fig44}  and from  the best fit spectral model (\cite{jacoby1984}) the spectral class was determined as A7 V. 

\begin{figure}[H]
\centering
\includegraphics[scale=0.11, angle=0 ]{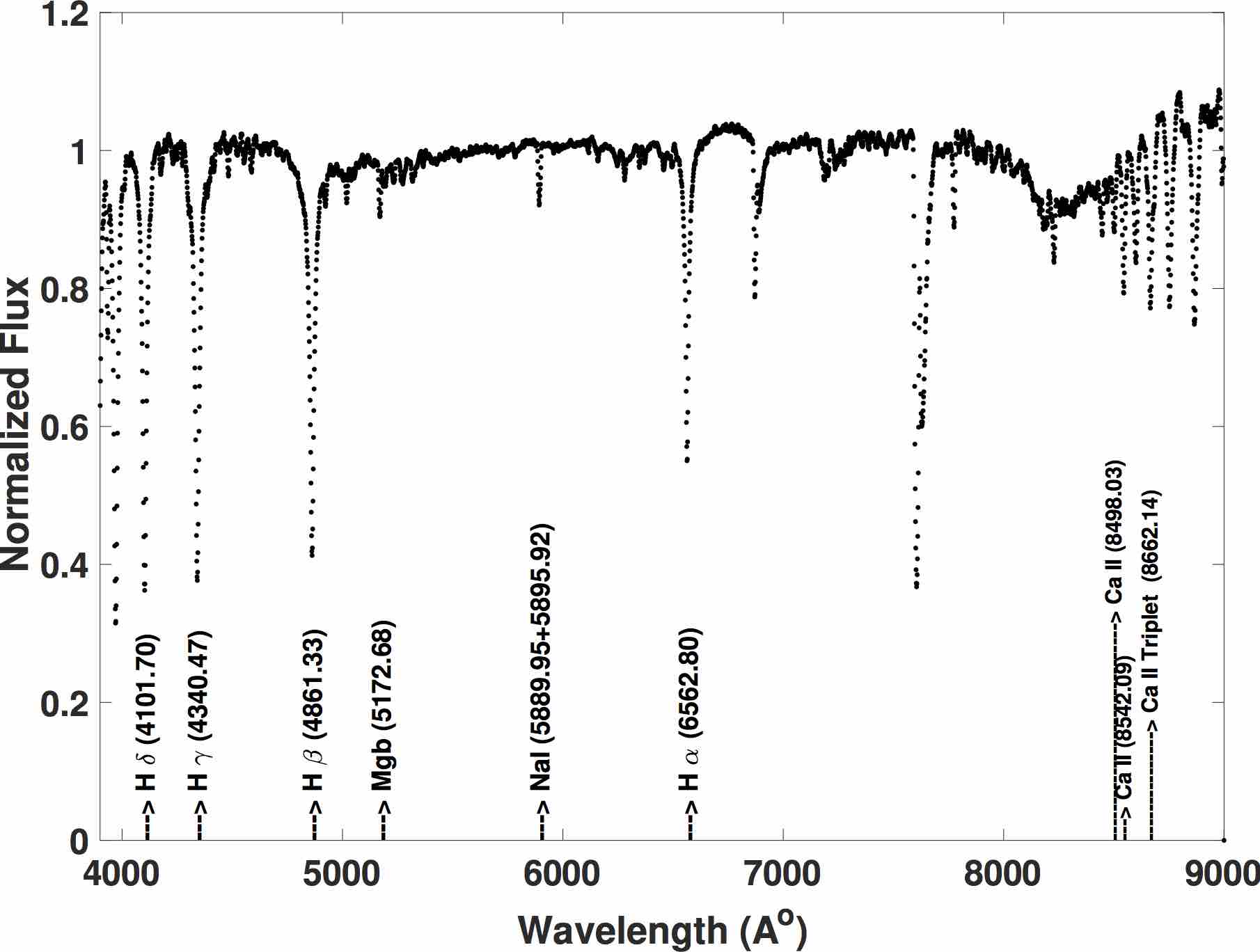}
\caption{Spectrum of Z Per (Feb, 2014).}
\end{figure}
\label{fig44}
\end{subfigures}

\subsection{XY Pup}

\begin{figure}[H]
\centering
\includegraphics[scale=0.11, angle=0 ]{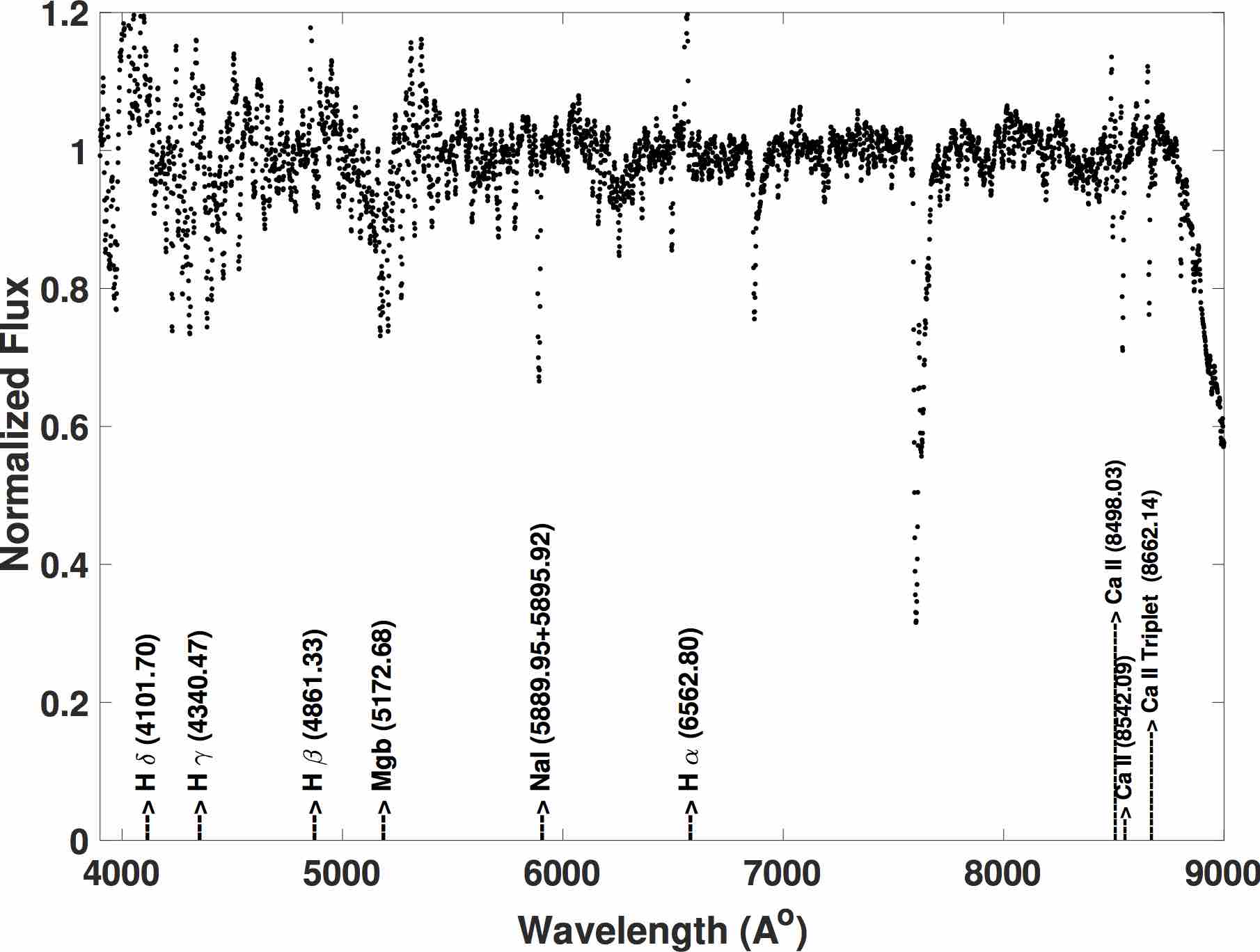}
\caption{Spectrum of XY Pup.}
\label{fig45}
\end{figure}

XY Pup (=GSC 05421-00222, TYC 5421-222-1) was first discovered by \cite{hoffmeister1929}. In the first spectral studies its spectral type was given as A3. \cite{popper1962, deLoore1984} also found two components at each of the D lines though not well separated. They also detected H$\alpha$ absorption line to be blended with strong double emission attributed to the presence of circumbinary gaseous envelope photospheric matter. \cite{brancewicz1980}, classified the variable to be Algol type with spectral type A3e+K3 IV. The ToM were given in the literature by \cite{dworak1977, dvorak2004, kreiner2004}. Extensive photometry was performed in the visible, IR \& Radio by \cite{woodsworth1977, kilkenny1985}. \cite{popper1989} derived radial velocities from photometric analysis of Na D lines for cooler component and found evidence for circumstellar matter. The H$\alpha$ line is in emission. It was also suggested that the velocities represented Pseudo elliptical motion. A spectrum was observed on Feb 20, 2013 and the phase obtained as 0.1956 using the epoch in the literature. The dominant spectral lines obtained are shown in Figure \ref{fig45} and the equivalent width of the spectral lines are given in Table \ref{T2}, in which sodium line shows greater equivalent width than all the other Algols under current study. As seen from the Table \ref{T2} the variable is distinctly characterized by emission spikes blended with absorption profile in Balmer lines which can be due to central emission (\cite{vesper2001}) and the double peaked emission is also suggestive of accretion disc. 

\subsection{AC Tau}

\begin{subfigures}
\begin{figure}[H]  
\centering
\includegraphics[scale=0.11, angle=0 ]{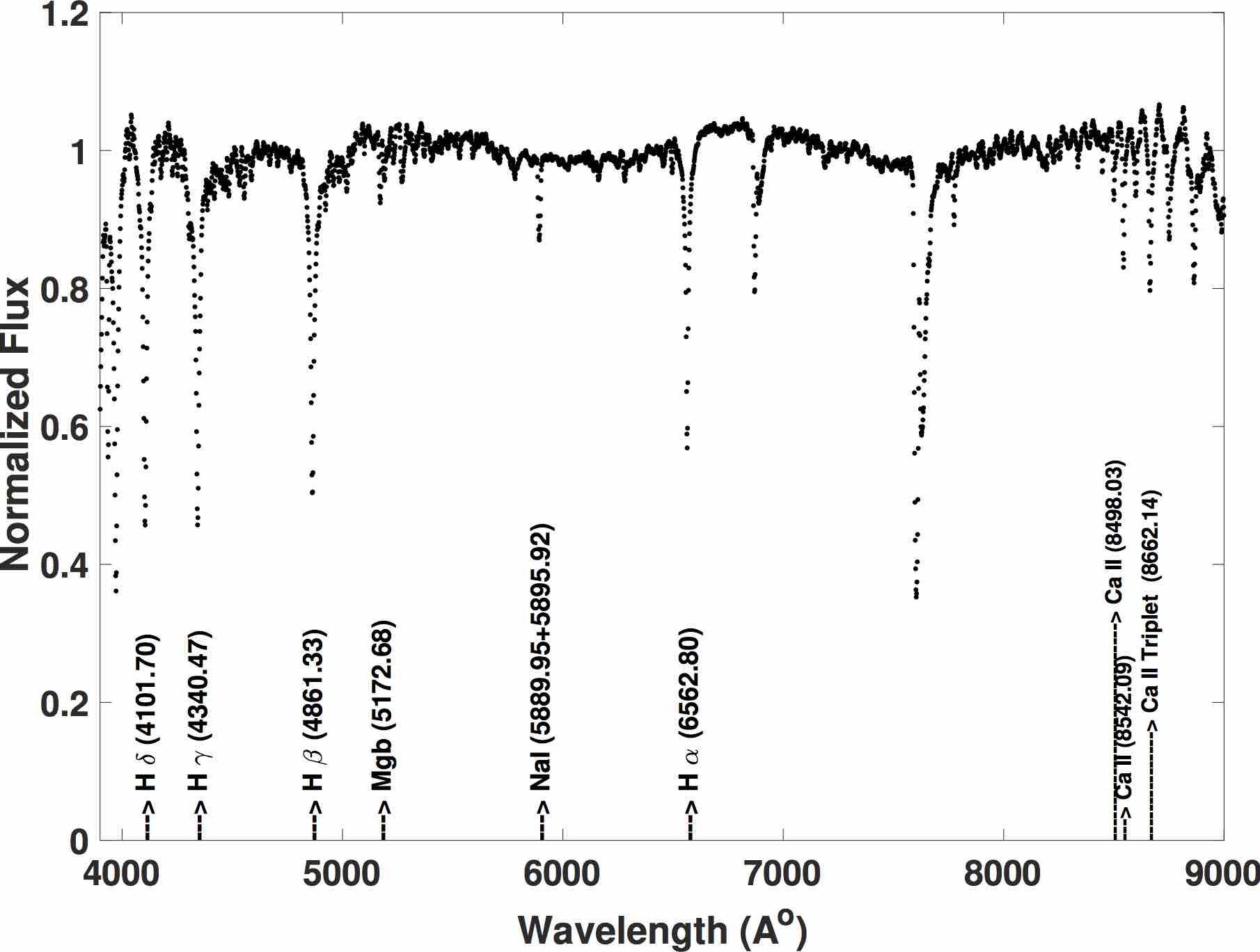}
\caption{Spectrum of AC Tau (Nov, 2013).}
\end{figure}

AC Tau(=AN 41.1929,GSC 00082-00147, V=11.09) is one of the least studied. It was catalogued as a variable star by \cite{hoffmeister1929, webbink2002}. The period of the variable is 2$^{d}$.043300 as given by \cite{samus2009}. The spectral type was derived as F0+K6  by \cite{baldwin1997}. The ROTSE data has given the O-C as -0.976 however the presence of third body for period variation was not asserted. Many ToM have been recorded by \cite{wood1963, kreiner1976, safar2000, nelson2003, kreiner2004, locher2005, samolyk2008}. Qian (\cite{qian2000}) has compiled various times of light minimum and studied the changes in its orbital period which indicated cyclic variation with a secular increase. Pulsations have been defined and studied by \cite{soydugan2006, zhang2013} suggesting the delta scuti nature of the pulsations of the primary and a possible relation between pulsation and orbital period. In the current study we present one spectrum for AC Tau which was obtained on Dec 21, 2013 at phase 0.938 derived using latest epoch in the literature.

\begin{figure}[H]
\centering
\includegraphics[scale=0.11, angle=0 ]{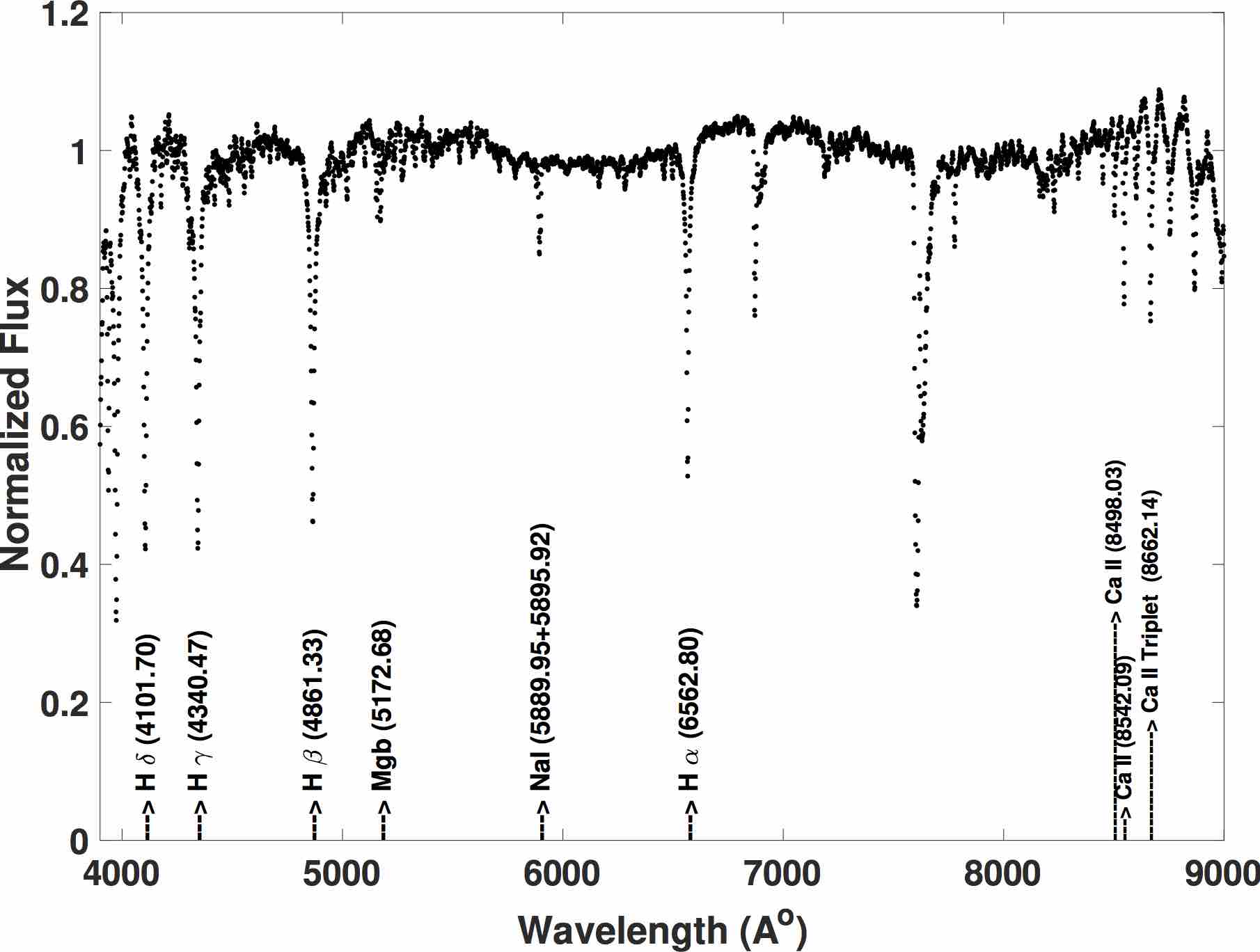}
\caption{Spectrum of AC Tau (Dec, 2013).}
\end{figure}
\label{fig47}
\end{subfigures}

We present the dominant spectral lines in Figure \ref{fig47} and their equivalent widths in Table \ref{T2}. The spectrum displays absorption profiles of all the Balmer lines. The spectral class determined in this current study using best fit spectral model (\cite{jacoby1984}) is F9 V.

\subsection{RW Tau}

\begin{figure}[H]
\centering
\includegraphics[scale=0.11]{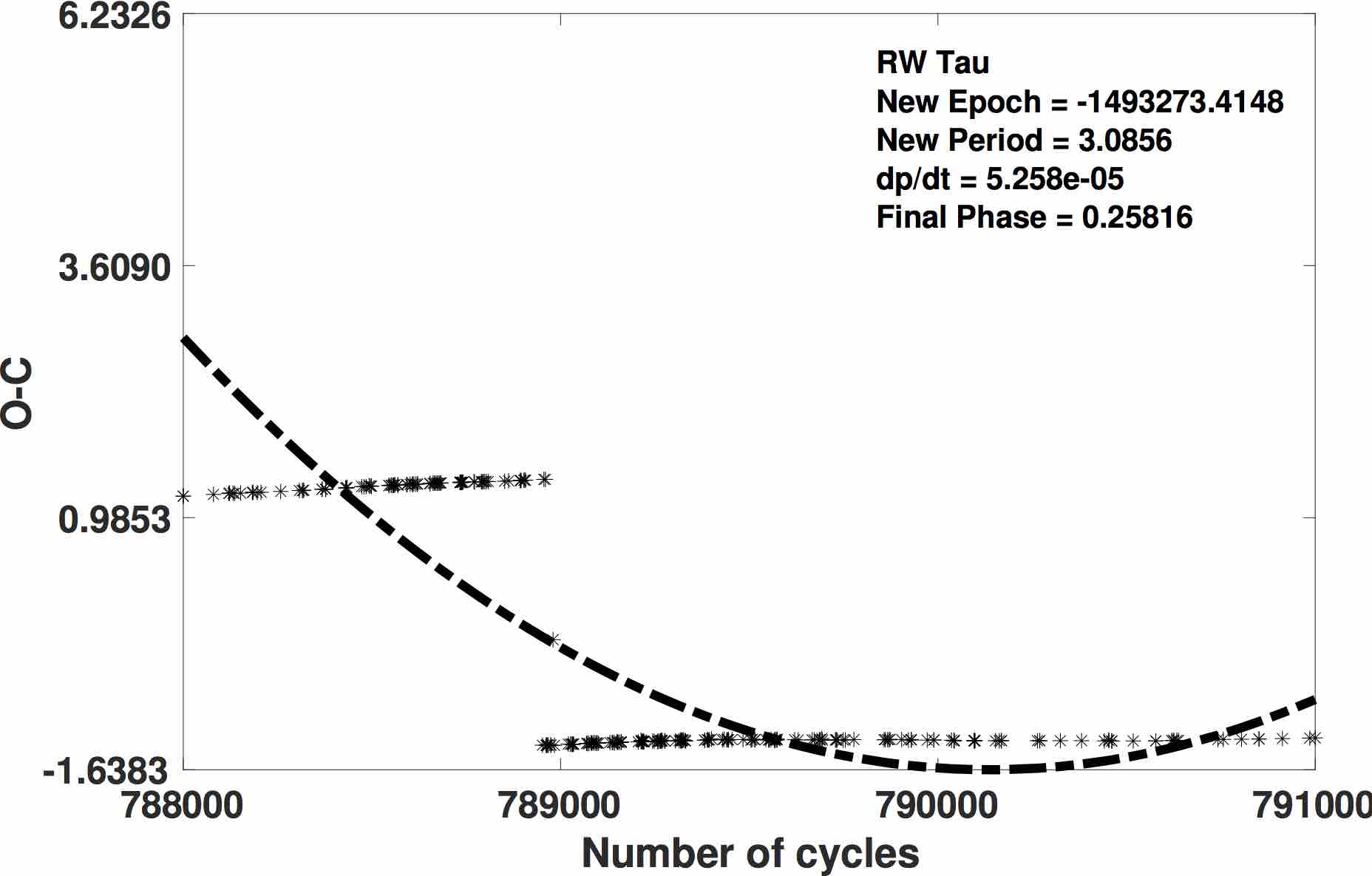}
\caption{O-C diagram of RW Tau.}
\label{fig48}
\end{figure}

RW Tau(=GSC 01826-00031=AN 102.1905, V=8.08) is a well studied Algol system with an orbital period of 2$^{d}$.76876 belonging to spectral type B9V (\cite{abt2008}). It was first catalogued by \cite{shapley1913} and many ToM are given in the literature by various authors. The O-C variation obtained is shown in Figure \ref{fig48} and the derived dp/dt = 5.25816x10$^{-8}$ days/year which represents a constant period. The new epoch obtained is as follows HJD (Min I) = 2456582.117 + 3$^{d}$.085628 $\times$ E. Accretion disks were first observed in RW Tau by \cite{wyse1934} and \cite{struve1948, struve1949} studied the rings in Algol system and suggested that the rings were due to accretion of gas from the secondary star. 

\begin{figure}[H]
\centering
\includegraphics[scale=0.11]{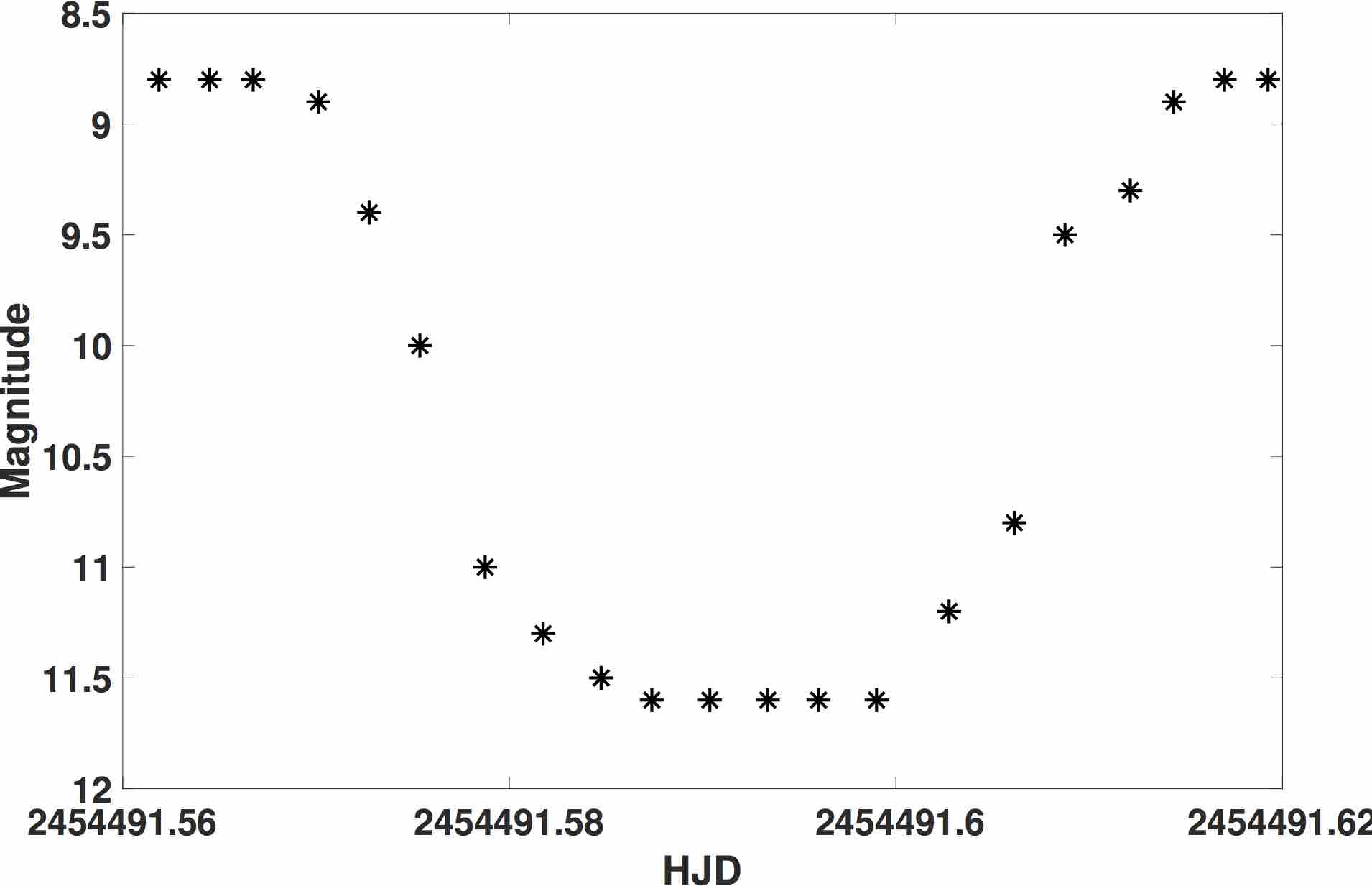}
\caption{Light curve of RW Tau.}
\label{fig49}
\end{figure}
\begin{figure}[H]
\centering
\includegraphics[scale=0.11, angle=0 ]{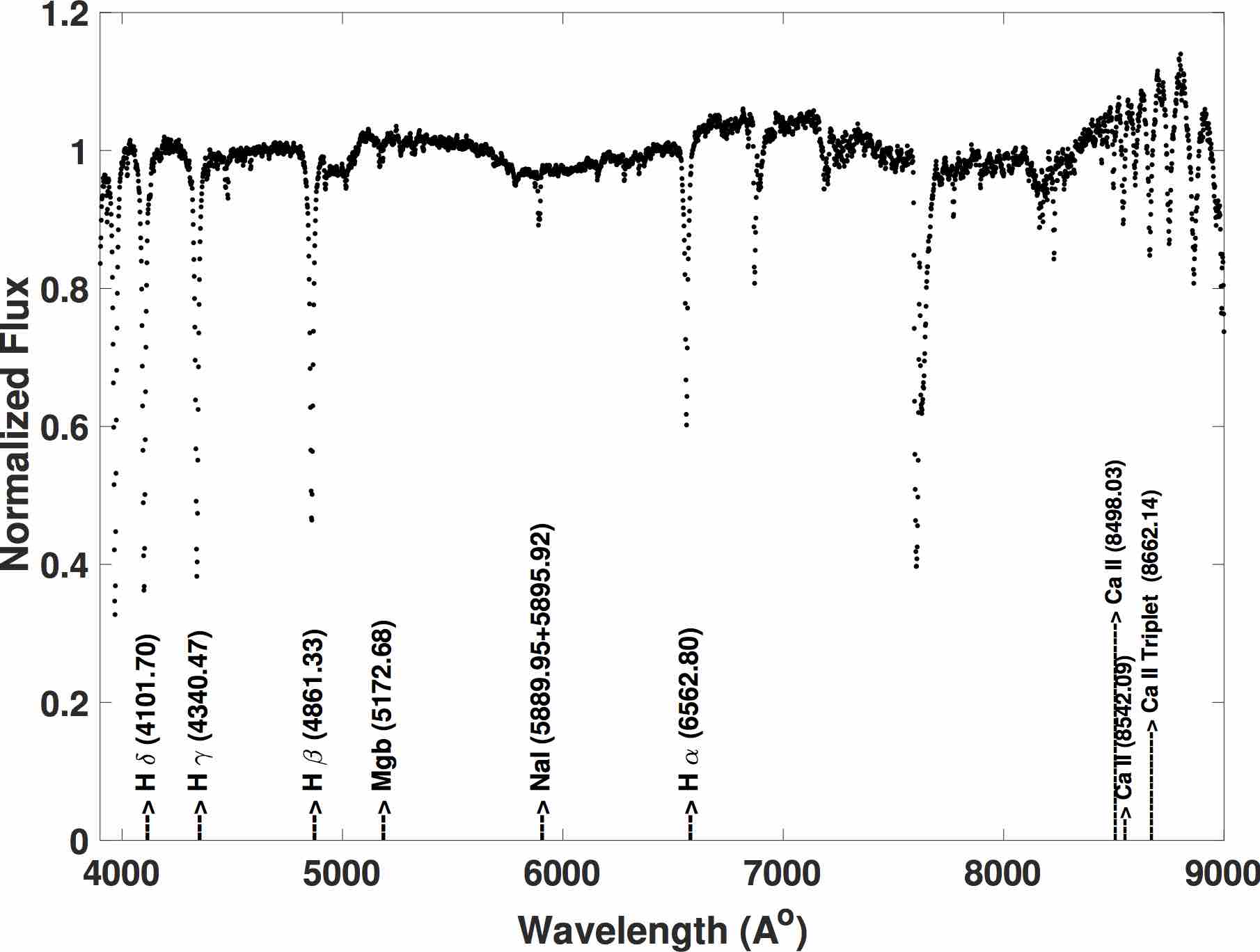}
\caption{Spectrum of RW Tau.}
\label{fig50}
\end{figure}
\cite{kaitchuck1982} have done detailed study and computed disk radii from measured times of disappearance and reappearance of each doppler component during eclipse. RW Tau is well known for its transient disks and \cite{lubow1975} have calculated the gas stream trajectories. \cite{kaitchuck1985} showed emission line widths to be almost constant during eclipse and twice as broad as expected from rotational broadening in a disk. One spectra on Oct 16, 2013 was observed for RW Tau at phase 0.115.  
The prominent spectral lines are shown in Figure \ref{fig50}  and their equivalent widths are given in Table \ref{T2}. The spectral type as obtained from the best fit spectral model using \cite{jacoby1984} is B9 III which is similar to that obtained by \cite{joy1949}. As per the available models \cite{lubow1975} and \cite{kaitchuck1985}, RW Tau is observed to be showing primary radius larger than  $\omega_{disk}$. This implies that a stable accretion disk could not form but possibility of circumstellar bulges exists. The light curve obtained from the data available in the literature is shown in Figure \ref{fig49}. The system may have circumbinary gaseous envelop.

\subsection{AF UMa}
\begin{figure}[H]
\centering
\includegraphics[scale=0.11]{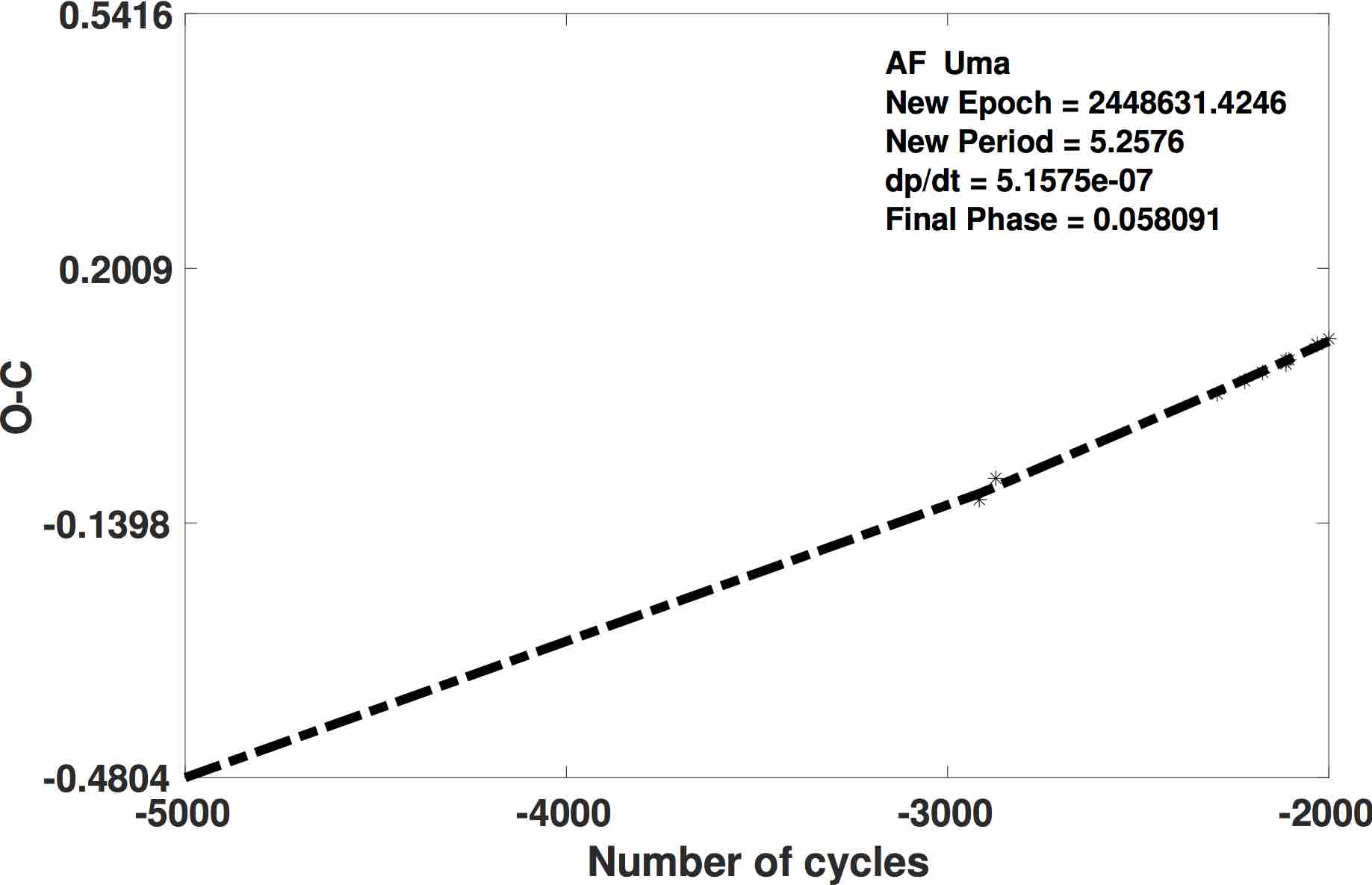}
\caption{O-C diagram of AF UMa.}
\label{fig51}
\end{figure}

\begin{figure}[H]
\centering
\includegraphics[scale=0.11]{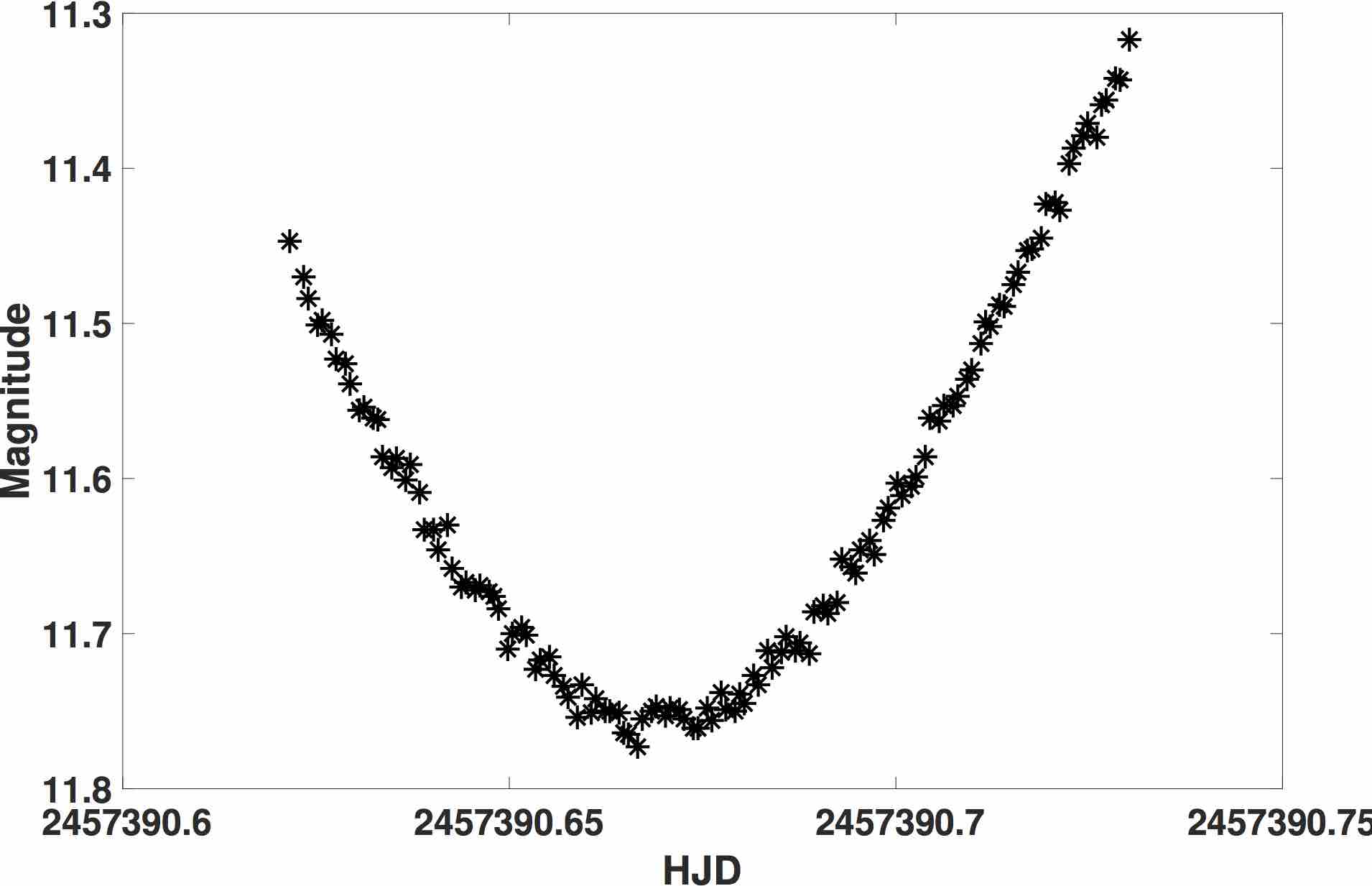}
\caption{Light curve of AF UMa.}
\label{fig52}
\end{figure}
AF UMa (= GSC 04147-01115 = TYC 4147-1115-1, V = 10.60) is an Algol-type eclipsing binary with an orbital period 5$^{d}$.25755. No detailed study on the orbital period changes of the system has been done so far. \cite{brancewicz1980} presented geometric and physical parameters of AF UMa. Many times of minima were reported in the literature by \cite{malkov2006, samolyk2008, samolyk2010, samolyk2011, brat2011}. The photoelectric observations were carried out and the minima were given in IBVS by  \cite{hubscher2006, hubscher2012, hubscher2015, hubscher2016}. The spectral type of the variable is A0 (\cite{brancewicz1980}).  The O-C diagram is shown in Figure \ref{fig51}, which has been plotted for only 11 data points spanning over 28 years and one point separated by 57 years.  The parabolic relation shows a tight fit for the observed data, indicating an increase in the period. However, this information can be authenticated with further observations. Using the quadratic term found in the O-C analysis, the increase rate in the period of AF UMa is derived to be about dp/dt = 5.1575x10$^{-7}$  days/year. 
\begin{subfigures}
\begin{figure}[H]
\centering
\includegraphics[scale=0.11, angle=0 ]{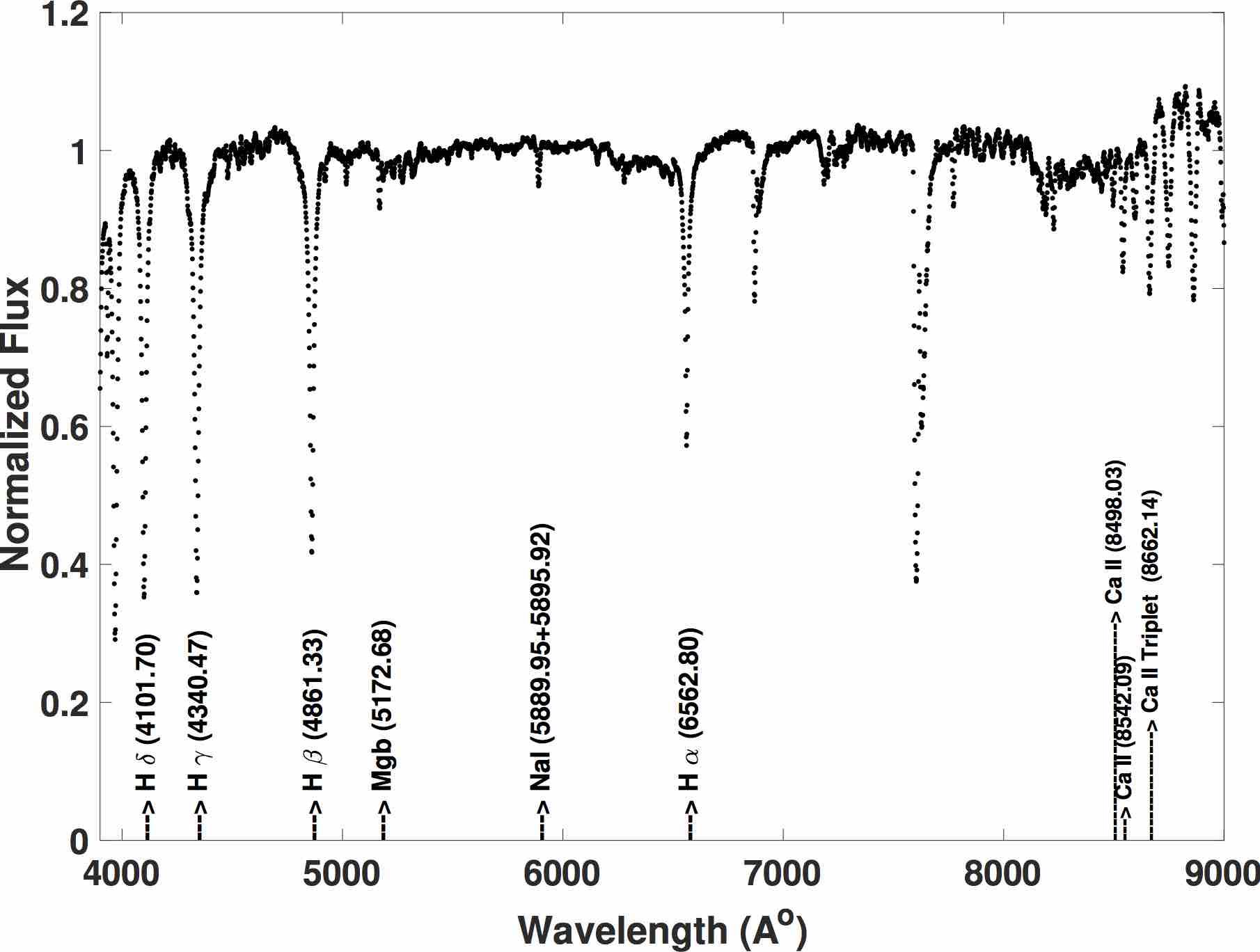}
\caption{Spectrum of AF UMa (Mar, 2013).}
\end{figure}

\begin{figure}[H]
\centering
\includegraphics[scale=0.11, angle=0 ]{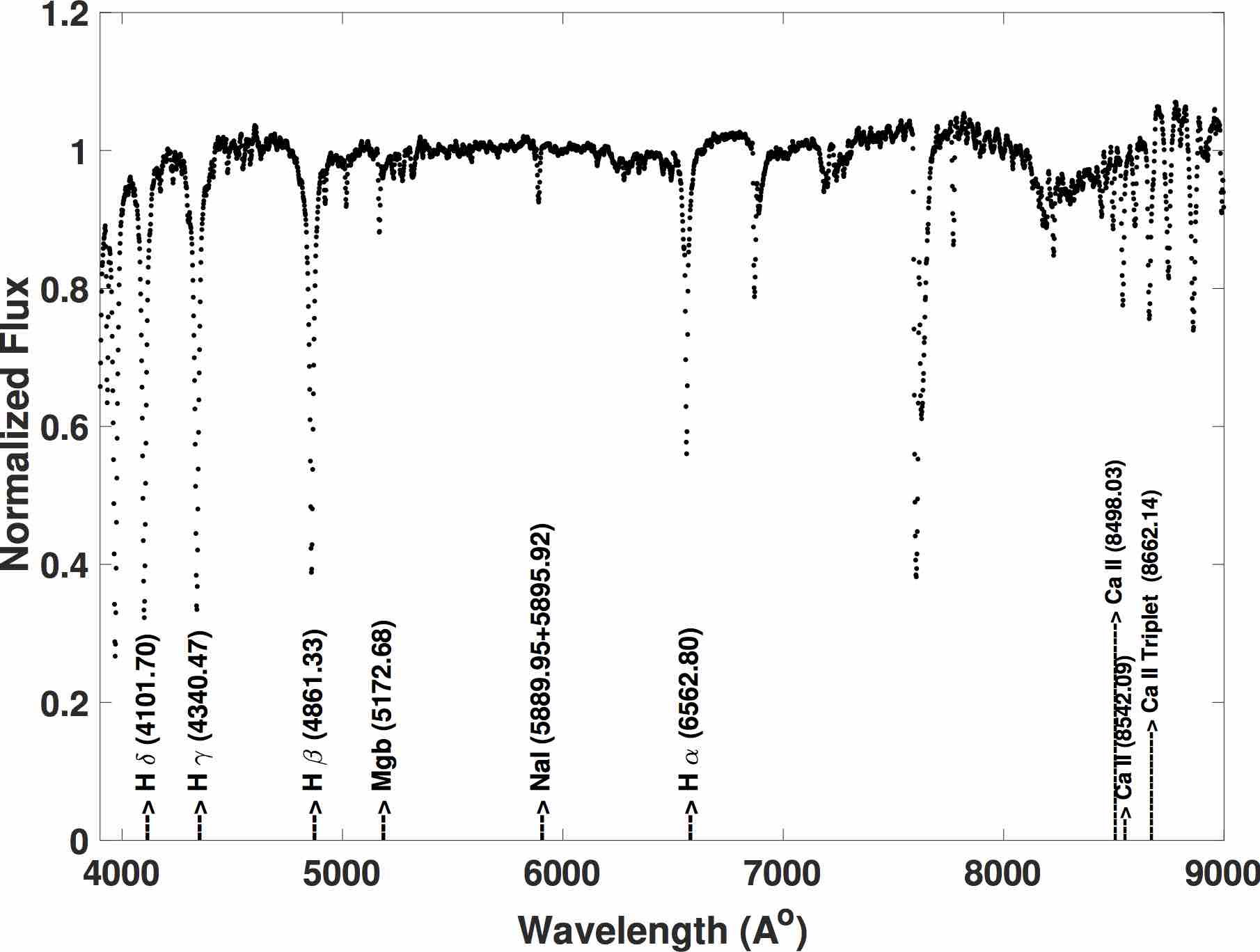}
\caption{Spectrum of AF UMa (Mar, 2014).}
\end{figure}
\label{fig53}
\end{subfigures}

The new ephemeris is obtained to be as HJD (Min I) = 2448631.4246 + 5$^{d}$.2576 $\times$ E. 
Figure \ref{fig52} shows the V-light curve from the observational data taken from AAVSO. Two spectra of AF UMa were observed on Mar 18, 2013 and Mar 21, 2014. The spectra obtained were at phases 0.942 and 0.940 respectively which were calculated using the new epoch derived from period study. The spectral lines identified are shown in Figure \ref{fig53}. It is observed that there is an increase in the equivalent widths of the Balmer lines (as shown in Table \ref{T2}) especially in the H$\alpha$ absorption line. AF UMa falls in to category 1A (Figure \ref{fig55}) of model given by \cite{vesper2001} which signifies a characteristic of symmetric disk or bulge but no direct indications of stream disk interaction. As per the models available \cite{lubow1975, kaichuck1985}, the position of AF UMa is observed to be showing primary radius greater than $\omega_{disk}$ but smaller than $\omega_{min}$. 
This implies that the accretion stream would have interacted with the primary component and formed into a stable disk as evident from the absorption line profiles observed in the spectra. The best estimate of spectral type for AF UMa based on minimum res$^{2}$ and visual inspections is derived to be A7 V using \cite{jacoby1984} stellar library. The spectral type derived is varying from that reported earlier which could be due to phase dependance.

\subsection{VV Vul}

\begin{figure}[H]
\centering
\includegraphics[scale=0.11, angle=0 ]{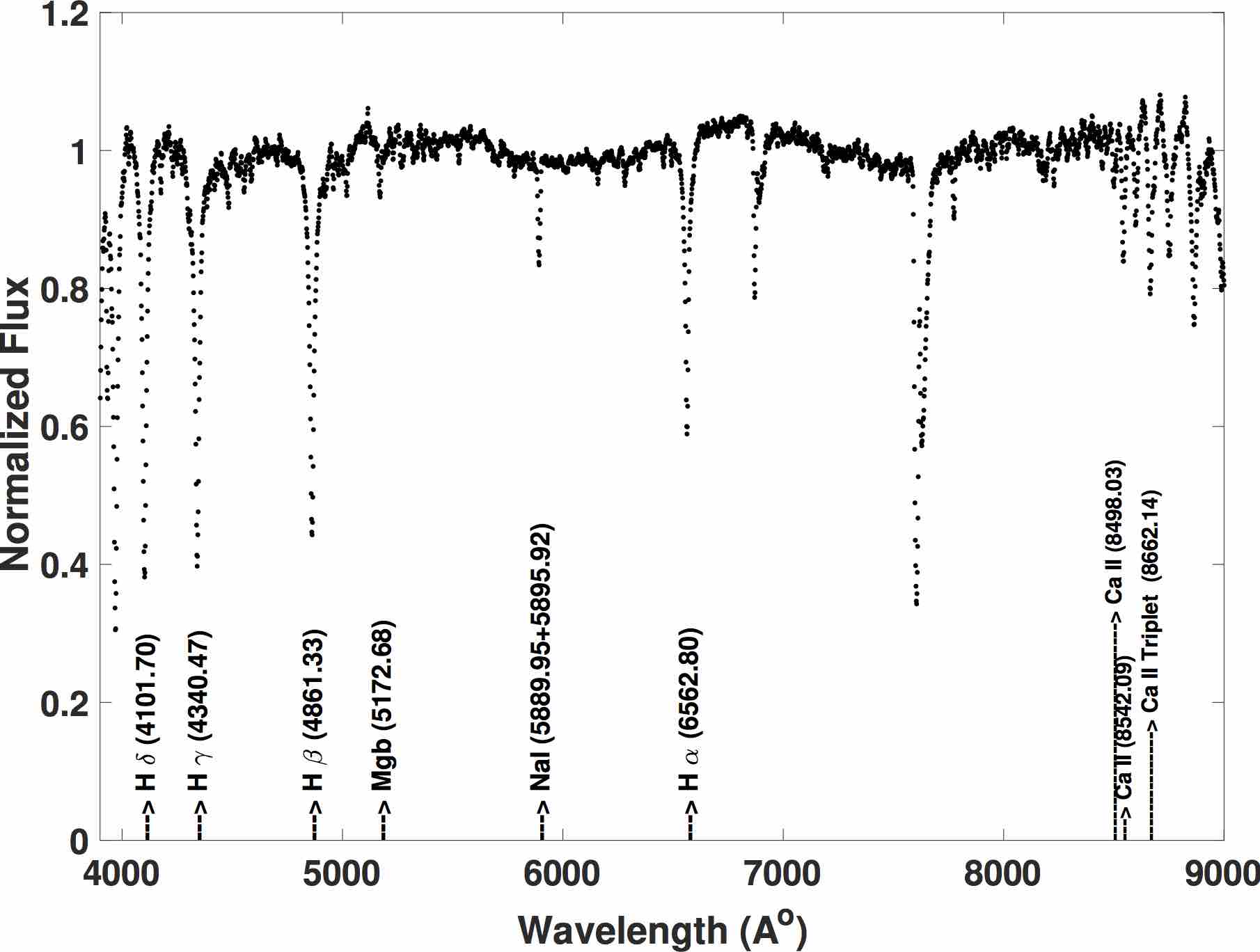}
\caption{Spectrum of VV Vul.}
\label{fig54}
\end{figure}

\begin{figure}[H]
\centering
\includegraphics[scale=0.11, angle=0 ]{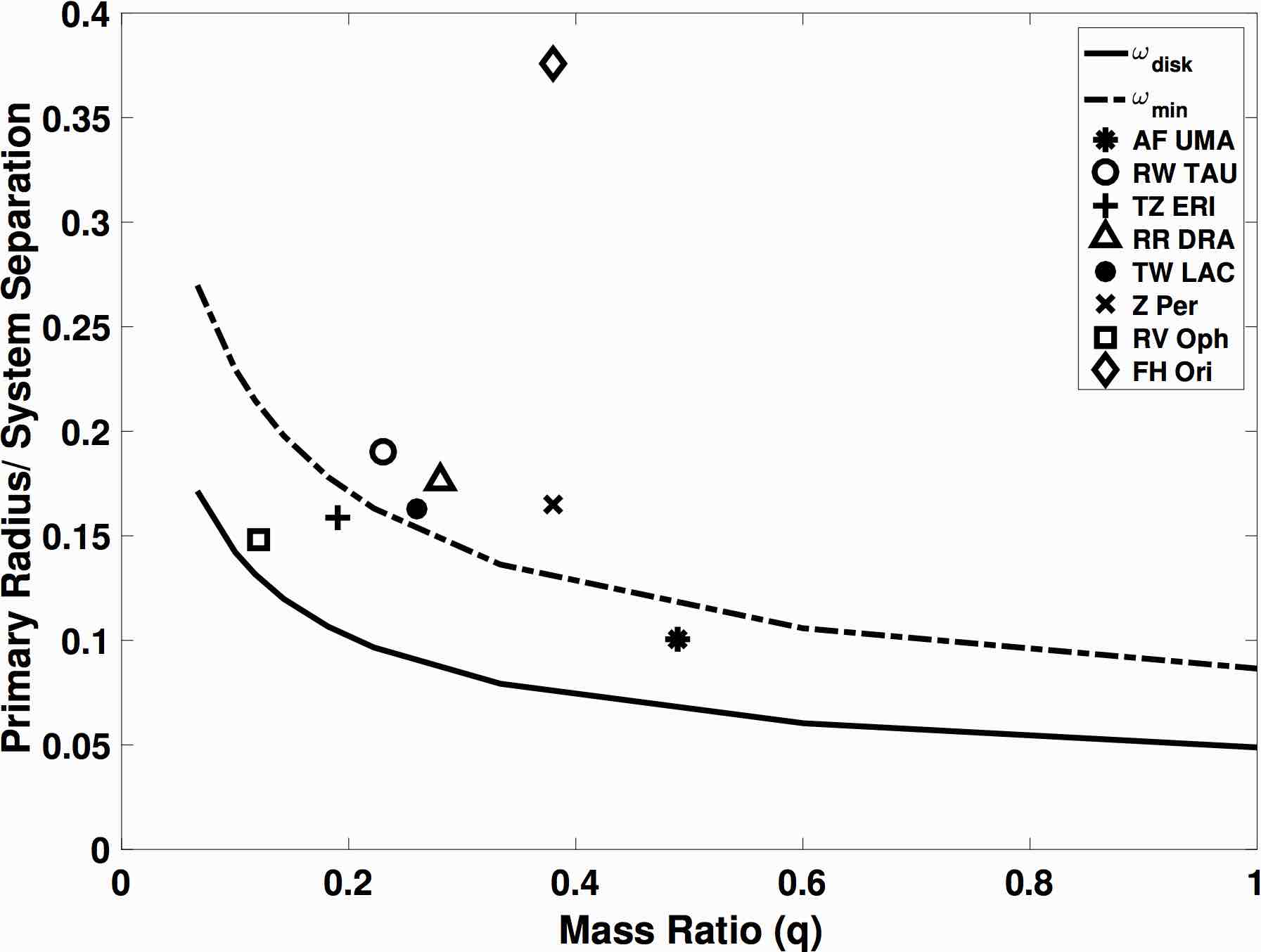}
\caption{A plot of primary radius vs mass ratio as per the model given by \cite{lubow1975, kaichuck1985, vesper2001}.}.
\label{fig55}
\end{figure}

VV Vul (=TYC 2180-593-1, V=12.40) was first discovered by \cite{wolf1904} while searching for variables in \cite{halbedel1984} derived the spectral class to be A2/3 V and the period was derived as 3$^{d}$.411400 by \cite{kreiner2004}. Few times of minima were recorded between 1904 \& 2009 and were catalogued by \cite{budding2004, malkov2006} and the period changes were determined  by \cite{qian2000, qian2002}. O-C studies done by Qian showed a secular increase in the period and from the assumed parameters gave a high rate of mass transfer. In the current study a spectrum for VV Vul was obtained at phase 0.868 on Nov 20, 2013.  The dominant profiles in the spectra are shown in Figure \ref{fig54} and equivalent widths derived are given in Table \ref{T2}.

\section{Conclusions}
\begin{figure}[H]
\centering
\includegraphics[scale=0.11, angle=0 ]{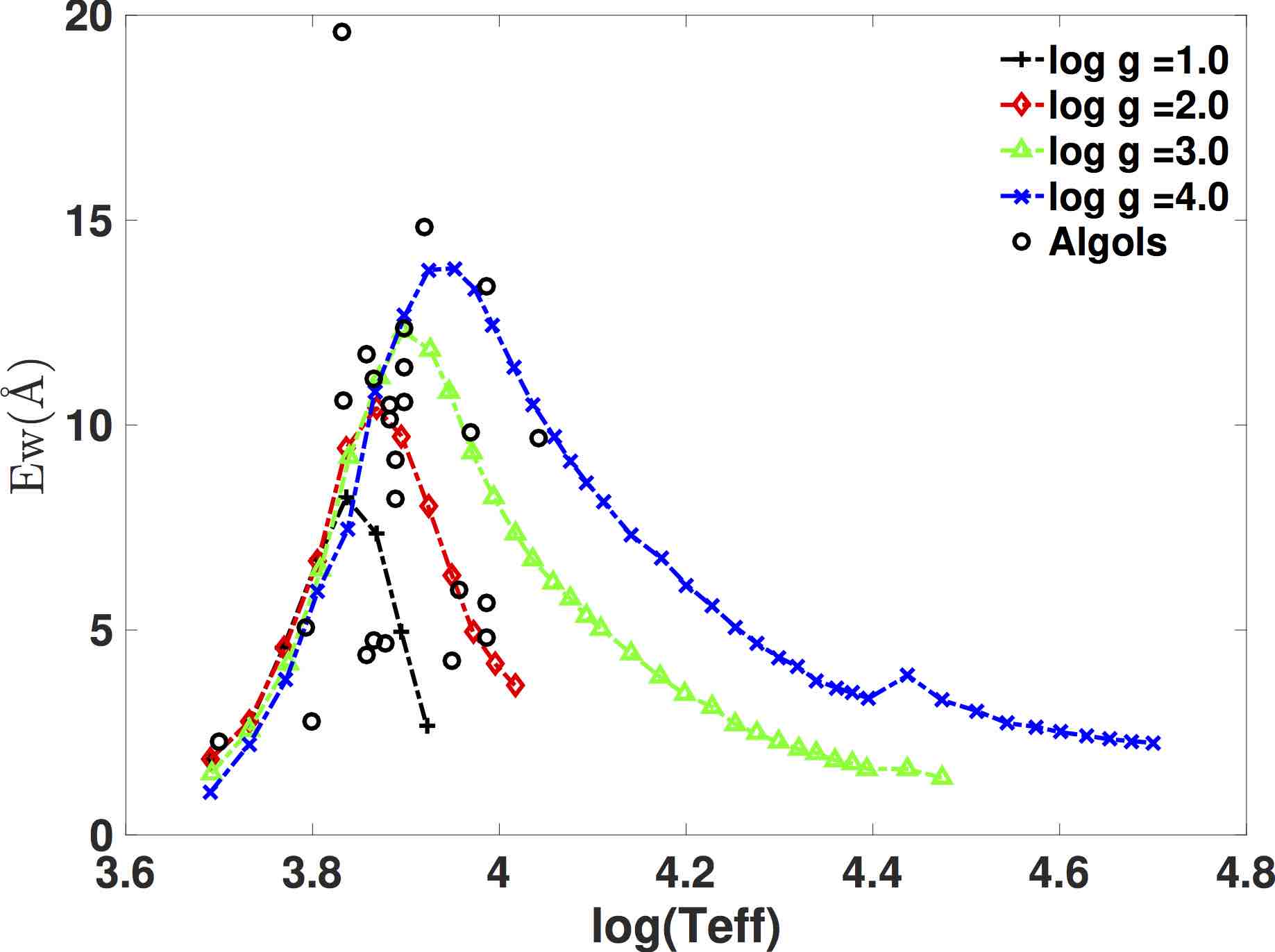}
\caption{Equivalent width of H$\gamma$ as a function of the effective temperature adopted from Martin (\cite{martin2005}) and Algols in study overplotted to derive log g.}
\label{fig56}
\end{figure}

The dominant spectral lines in all the Algols were observed to be of H$\delta$, H$\gamma$, H$\beta$, H$\alpha$, Na I, Ca II \&  Ca triplet with absorption profile except for XY Pup in which emission spike in H lines and absorption in Na line is observed to be dominant. These systems are listed in Table \ref{T1} in alphabetical order along with details of observations and we report the equivalent widths (EWs) obtained and observed features in Table \ref{T2}. For the variables AF UMa, RW Tau, TZ Eri, RR Dra, TW Lac, Z Per, RV Oph \& FH Ori we tried to examine the systems with the model of \cite{lubow1975} who established two parameters $\omega_{min}$ and $\omega_{disk}$ that indicate the characteristics of accretion disk. They signify the stream trajectory and disk formation in the orbit plane as a function of mass ratio. It was observed that none of the systems had parameters below the $\omega_{min}$ line which signifies that the systems stream misses the primary component and could form a stable accretion disk. The spectral profiles obtained indicated clear absorption lines. \\
We checked the physical parameters (T$_{eff}$ and log g) for the Algols in this study and this task was accomplished by fitting model atmosphere spectra \cite{jacoby1984} to the observed spectra and the spectral types were derived. Using the equivalent widths obtained for H$\gamma$ line we tried to derive log g values for the Algols as seen in Figure \ref{fig56}, as H$\gamma$ line is considered sensitive to gravity and temperature thus making it an effective tool in determining the fundamental stellar parameters (\cite{martin2005}). \\

 The G-band extended around 4300 \AA \hspace{0.5mm} is used as an approximate measure of CH abundance, make an appearance around F3 and become very strong at late G-type to K-type (\cite{morgan1943}). The variables V 769 Aql, XY Cet, RR Dra, SX Gem, FG Lyr, CH Mon, FW Mon, Z Ori, V 640 Ori, CK Per, Z Per for which the spectra were obtained at phases close to zero, indicate relatively weak G-band. High resolution spectra during totality phase are needed to confirm and derive carbon abundance (\cite{parthasarathy1983}). This was further validated by comparing the G-band absorption profiles with that of H$\gamma$ absorption lines. For many Algols in our sample there are no recent observed epochs of primary minima, therefore the phases given in Table \ref{T1} may not be accurate. 
 The H-alpha line in the spectrum of several systems in our sample is relatively weak and or partly filled in which indicates mass transfer and or circumstellar gas. Monitoring the H-alpha profile of these systems at high resolution can reveal more information. Wavelength region more than 8000 \AA\hspace{0.5mm} and in the Ca II triplet lines the blending of the primary star spectrum with the flux of the late-type secondary can be significant. The late-type secondary components in Algol-type systems seems to have strong chromospheric activity than single stars of similar spectral type therefore the Ca II IR triplet strengths can be weaker (\cite{parthasarathy1979}). High resolution and high signal to noise ratio spectra at 0.25 and 0.75 phases of most of the system in our sample may reveal the H-alpha line of the secondary component (see for example \cite{parthasarathy2017} and \cite{parthasarathy2018}). Algol-type semidetached systems in which the primary minimum showing long duration  flat bottom totality phase similar to that of Z Per (see Figure \ref{fig44}) are well suited for obtaining uncontaminated high resolution and high signal to noise ratio spectra of the late-type  evolved low mass secondary component to derive its detailed chemical composition including, C, N, O and S-process elements and may be even C12/C13 isotope ratio. Such a study will enable us to understand the nucleosynthesis, mixing, mass-transfer, mass-loss and evolution process experienced by Algol-type close binaries.

\section{Acknowledgement} 
We are thankful to Prof. G. C. Anupama for helping us to get the observational data with the Himalayan Chandra Telescope (HCT). We are very grateful to Dr. Oleg Malkov for his encouraging comments.
\nocite{*}
\RaggedRight
\bibliographystyle{spr-mp-nameyear-cnd}
\bibliography{references}

\end{document}